\documentclass{PoS}

\usepackage{graphicx}              
\usepackage{amsmath}               
\usepackage{amsfonts}              
\usepackage{amsthm}   		         
\usepackage{psfrag}
\usepackage{calc}

\usepackage{amsmath}               
\usepackage{amsfonts}              
\usepackage{amsthm}   		         

\title{Group field theories}

\ShortTitle{Group field theories}

\author{\speaker{Thomas Krajewski}\\
       Centre de Physique Th\'eorique\\
       Aix-Marseille Universit\'e\\
       F-13288 Marseille cedex 9 (France)\\
       E-mail: \email{krajew@cpt.univ-mrs.fr}}

\abstract{Group field theories are particular quantum field theories defined on $D$ copies of a group which reproduce spin foam amplitudes on a space-time of dimension $D$. In these lecture notes, we present the general construction of group field theories, merging ideas from tensor models and loop quantum gravity. 
This lecture is organized as follows. In the first section, we present basic aspects of quantum field theory and matrix models. The second section is devoted to general aspects of tensor models and group field theory and in the last section we examine properties of the group field formulation of $BF$ theory and the EPRL model. We conclude with a few possible research topics, like the construction of a continuum limit based on the double scaling limit or the relation to loop quantum gravity through Schwinger-Dyson equations.}

\FullConference{3rd Quantum Gravity and Quantum Geometry School \\
		 February 28 - March 13, 2011\\
		 Zakopane, Poland}

\PoS{PoS(QGQGS 2011)005}

\begin{document}




\tableofcontents

\bigskip


\section{Introduction}

Despite decades of efforts, constructing a quantum theory of gravity remains one of the most tantalizing open problem in fundamental physics. Indeed, the gravitational field encodes the geometry of space-time and conventional quantization techniques rely on a preexisting geometry such as Minkowski space. As soon as the geometry is promoted to a dynamical variable, its quantum fluctuations at the Planck scale $l_{\mathrm{P}}\simeq 10^{-33}\,\mathrm{cm}$ ruin the consistency of standard quantum field theory based on perturbative renormalization. 

Quantum gravity (see the books \cite{Kieffer} and \cite{Approaches}) is an attempt at constructing a quantum theory (not necessarily based on fields) whose long distance limit should be general relativity coupled to matter fields, governed by the action
\begin{equation}
S[g_{\mu\nu},A_{\mu},\phi,\psi]=\frac{1}{16\pi G}\int d^4x\big(R-2\Lambda \big)+\int d^{4}x \sqrt{-g}\,L_{\mbox{\tiny matter}}(g_{\mu\nu},A_{\mu},\phi,\psi)
\label{actiongravity}
\end{equation}
with $g_{\mu\nu}$ the metric, $A_{\mu}$ a gauge connection including the electroweak and strong sectors, $\psi$ fermionic matter fields and $\phi$ the Higgs boson.  In the sequel, we restrict our attention to pure gravity. We only briefly mention extensions including a non trivial cosmological constant $\Lambda$ or the matter fields.

Some approaches to a quantum theory of gravity interpret the failure of perturbative quantum field theory as an indication of  new physics at the Planck scale, just as the non renormalizability of the Fermi theory signals new degrees of freedom encoded in the electroweak theory. These approaches may include new degrees of freedom on existing space-time, as is the case in string theory, or involve radically new  ideas about the nature of space-time, as exemplified by non commutative geometry. The other approaches are more conventional. Some of them rely on renormalization group ideas by seeking for a non trivial UV fixed point, realizing Weinberg's asymptotic safety scenario. Renormalization group techniques are also essential in Ho\v rava-Lifshitz gravity, based on an improved power counting obtained at the price of a breakdown of Lorentz invariance, the latter being only recovered in the low energy limit. One may also resort to discretizations, like dynamical triangulations that encode all the geometry of space-time in triangulations with simplexes of fixed shape. In this case, the quantum theory is constructed as a sum over these triangulations, Alternatively, in a canonical approach, one can triangulate the space manifold and consider the holonomies of the Ashtekar connection as the fundamental variables. This is the route followed by loop quantum gravity and its space-time counterpart, spin foam models. Since these last two approaches to quantum gravity lie at the root of group field, let us give describe them shortly.

Dynamical triangulations and spin foam models rely on a path integral formulation of quantum gravity, which aims at defining the path integral over metrics 
\begin{equation}
\Psi(g\big|_{\partial {\cal M} })=\sum_{\mbox{\tiny topologies}\atop \partial {\cal M}\,\mbox{\tiny fixed}}
\int_{g|_{\partial {\cal M}}\,\mbox{\tiny fixed}}[Dg]\exp\frac{\mathrm{i}S[g]}{\hbar}\label{pathintegralquantumgravity}.
\end{equation}
It involves an integration over metrics modulo diffeomorphisms on a space-time manifold ${\cal M}$ which reduce to a fixed metric on the boundary. As a functional of the boundary metric, it can be considered as the wave function of the quantum gravitational field. Note that fluctuations of the topology of space-time are also allowed, even if these may be omitted in a first approximation.

In the context of dynamical triangulations (see the books \cite{quantumgeometry} and the review \cite{triangulations}, the path integral over metrics on a space-time of dimension $D$ is replaced by a sum over all possible $D$-dimensional triangulations with fixed boundary,
\begin{equation}
\sum_{\mbox{\tiny topologies}\atop \partial {\cal M}\,\mbox{\tiny fixed}}
\int_{g|_{\partial {\cal M}}\,\mbox{\tiny fixed}}[Dg]\exp\frac{\mathrm{i}S[g]}{\hbar}\quad\rightarrow\quad
\sum_{\mbox{\tiny triangulations}\, T\atop \partial {T}\,\mbox{\tiny fixed}}
\frac{1}{C_{T}}\exp\frac{\mathrm{i}S_{\mbox{\tiny Regge}}[T]}{\hbar}\label{pathintegralquantumgravity}
\end{equation}
The metric aspect of the geometry is recovered by assigning a fixed geometry to the simplexes of the triangulations, say regular simplexes of edge length $a$. $S_{\mbox{\tiny Regge}}[T]$ is the Regge action of the triangulation, which is a discretized version of the Einstein-Hilbert action and $C_{T}$ a combinatorial factor accounting for the discrete automorphisms of the triangulation. Two remarks are in order. First, the sum over triangulations, to be understood as a sum over ways of gluing of $D$-simplexes along their boundaries, naturally implements a sum over topologies. Second, the sum over triangulations can also be obtained as  the perturbative Feynman graph expansion of a tensor model, whose dynamical variable is a tensor $M_{i_{1},\dots,i_{D}}$, extending to higher dimensions the relation between matrix models and two dimensional gravity.

On the other hand, in loop quantum gravity and spin foam models (see the books by Rovelli \cite{Rovelli} and Thiemann \cite{Thiemann}, the lectures by Rovelli \cite{CarloZakopane} and by Giesel and Sahlmann at this school \cite{Sahlmann}, the lectures by Dona and Speziale \cite{Simonelectures} and the recent review by Perez \cite{Perez}) is rooted in the canonical quantization programme. We start by splitting the space-time manifold of dimension $D$ as ${\cal M}\sim {\Bbb R}\times \Sigma$ and consider the time evolution of triangulations of the space manifold $\Sigma$. In its simplest formulation, the loop quantum gravity Hilbert space is based on a triangulations of the space manifold by $(D\!-\!1)$-simplexes. The incidence relations of the triangulation are encoded in a graph $\Gamma$ and the dynamical variables are SU(2) elements associated to links of $\Gamma$ representing the holonomies of the Ashtekar connection. The dynamics is formulated in terms of spin foams which are 2-complexes made of vertices, edges and faces that describe histories of these boundary graphs. Spinfoams are dual to triangulations of space-time and their interaction vertices correspond to $D$-simplexes. The definition of a spin foam model is completed by assigning a weight to any spinfoam, which is a function of spins and intertwiners associated to its faces and edges, so that transition amplitudes can be computed by summing over all intermediate states as required by the superposition principle.

Group field theory arises from the marriage between tensor models and spin foam models.  It is a quantum field theory whose Feynman graph expansion reproduces spin foam amplitudes
\begin{align}
\int [{\cal D}\Phi]\,\exp S[\Phi]=\sum_{\mbox{\tiny triangulations}\, T\atop \Leftrightarrow\,\mbox{\tiny Feynman graphs}}
\frac{1}{C_{T}}{\cal A}_{T}
\end{align}
The group field theory Feynman graphs are in one to one correspondence with 2-complexes dual to triangulations. $C_{T}$ is the symmetry factor of the graph corresponding to $T$ and ${\cal A}_{T}$ the spin foam amplitude of the triangulation. The tensors in the models generating dynamical triangulations are replaced by  functions over $D$ copies of a group $G=SU(2),SO(3),SO(4),SL(2,\mathrm{C})$, $M_{i_{1},\dots,i_{D}}\rightarrow\Phi(g_{1},\dots,g_{D})$. The action can be written symbolically as $S[\Phi]\sim\Phi^{2}+\Phi^{D+1}$. The field $\Phi$ represents a $(D\!-\!1)$-simplex, the interaction a $D$-simplex made of $D\!+\!1$ $(D\!-\!1)$-simplexes. The application of Wick's theorem implements the gluing of $D$-simplexes along their boundary $(D\!-\!1)$-simplexes. Finally, the Feynman graph amplitudes, expressed as an integral over group elements, reproduce the spin foam amplitudes.

The first group field theory has been proposed by Boulatov \cite{Boulatov} for three dimensional quantum gravity, which is nothing but the topological $BF$ theory, and then generalized to $BF$ theory in four dimensions by Ooguri \cite{Ooguri}. The first four dimensional quantum gravity model has been formulated in group field theory by De Pietri, Freidel, Krasnov and Rovelli \cite{DePietriFreidelKrasnovRovelli}. Recent years have seen an outburst of new results, relying on two breakthroughs: the $\star$-product formulation of Baratin and Oriti \cite{OritiBaratin} and the colored models introduced by Gurau \cite{colored}, allowing Gurau to construct a $1/N$ expansion \cite{complete} and Ben Geloun and Rivasseau to define the first model renormalizable to all orders \cite{renormalizable}. 

The reader should be warned that these notes are lectures notes as opposed to a review. No attempt is made at giving a detailed account of the field. Our goal is to present the subject in its simplest possible form, starting from basic quantum field theory and matrix models, as can be found in the first chapters of \cite{ZinnJustin}. Almost all the material covered can be found in the literature, at the exception of the group field theory formulation of the lorentzian EPRL/FK model. This spin foam model is currently the best candidate for a quantum theory of gravity in dimension four and we refer the reader to the lectures by Rovelli at this school for an overview \cite{CarloZakopane}. The reader is also advised to consult more advanced and/or specialized reviews on group field theory. First of all, the review by Freidel \cite{Freidel} had a profound influence on  our presentation and presents interesting new ideas that can be the starting point for research in the field.
The recent review by Oriti \cite{Oritireview} presents general ideas about group field theory as well as an account of the $\star$-product formulation. State of the art of colored tensor models (as of mid 2011) is reviewed by Gurau and Ryan \cite{coloredreview}. The renormalization program in group field theory is outlined by Rivasseau in \cite{Vincent} and \cite{tensortrack} (see also \cite{update} for a up to date account). Finally, it is instructive to have a look at the review of Baratin and Oriti \cite{questions}, discussing some fundamental issues in group field theory.

This lecture is organized as follows. In the first section, we present basic aspects of quantum field theory and matrix models. The second section is devoted to general aspects of tensor models and group field theory and in the last section we examine properties of the group field formulation of $BF$ theory and of the EPRL/FK model. We conclude with a few possible research topics, like the construction of a continuum limit based on the double scaling limit or the relation to loop quantum gravity through Schwinger-Dyson equations. A few facts regarding the unitary irreducible representations of the Lorentz group and coherent states are collected in the appendix.

\section{QFT and Feynman graphs}

\subsection{Feynman graph expansions}

In its most general acceptance, quantum field theory can be defined as the quantum theory of systems with continuous degrees of freedom. Its basic object of interest is the functional integral expectation value of observables
\begin{equation}
\langle{\cal O}\rangle=\frac{\int [D\Phi]\,{\cal O}[\Phi]\exp\frac{\mathrm{i}S[\Phi]}{\hbar}}{\int [D\Phi]\,\exp\frac{\mathrm{i}S[\Phi]}{\hbar}}\label{Greens}
\end{equation}
The integration is over a suitable space of fields, which are functions $\Phi$ defined on a manifold. The observable ${\cal O}$ belongs to a certain class of functionals of the fields which may be taken as products of fields evaluated at different points ${\cal O}[\Phi]=\Phi(x_{1})\cdots\Phi(x_{n})$ to define $n$-point Green's functions $G(x_{1},\dots,x_{n})$. 

The dynamics of the theory is encoded in the action functional, which we often split $S[\Phi]=S_{0}[\Phi]+S_{\mbox{\tiny int}}[\Phi]$ as the sum of  a free action
\begin{equation}
S_{0}[\Phi]=\frac{1}{2}\int dxdy\,\Phi(x){\cal K}^{-1}(x,y)\Phi(y)
\end{equation}
and an interaction term
\begin{equation}
S_{\mbox{\tiny int}}[\Phi]=\sum_{n\geq 2}\frac{\lambda_{n}}{n!}\int dx_{1}\cdots dx_{n}{\cal V}_{n}(x_{1},\dots,x_{n})\Phi(x_{1})\cdots \Phi(x_{n})
\end{equation}
The kernel ${\cal K}^{-1}$ is the matrix inverse of a more fundamental objet ${\cal K}$ that appears in the Feynman rules. When ${\cal K}$ is invertible, it is defined by $\int dz\,{\cal K}(x,z){\cal K}^{-1}(z,y)=\int dz\,{\cal K}^{-1}(x,z){\cal K}(z,y)=\delta(x,y)$.

For instance, for a scalar field on Minkowski space with a local polynomial interaction
\begin{equation}
S[\Phi]=\frac{1}{2}\int dxdy\,\delta(x-y)\Phi(x)(\square-m^{2})\Phi(y)-\sum_{n}\frac{\lambda}{n!}\int dx\,\Phi^{n}(x)\label{QFTMinkowski}
\end{equation}
with $\square=\eta^{\mu\nu}\partial_{\mu}\partial_{\nu}$ the d'Alembertian and $\eta_{\mu\nu}=\mathrm{diag}(1,-1,\cdots,-1)$ the Minkowski metric. It is sometimes convenient to work in the euclidian setting where  \eqref{Greens} is replaced by (setting $\hbar=1$)
\begin{equation}
\langle{\cal O}\rangle=\frac{\int [D\Phi]\,{\cal O}[\Phi]\exp-S[\Phi]}{\int [D\Phi]\,\exp-S[\Phi]}\label{euclidian}
\end{equation}
where $S[\Phi]$ is the euclidian action. For instance, for a scalar field with a local polynomial interaction 
\begin{equation}
S[\Phi]=\frac{1}{2}\int dxdy\,\Phi(x)(-\Delta+m^{2})\Phi(y)+\sum_{n\geq 2}\frac{\lambda_{n}}{n!}\int dx\,\Phi^{n}(x)\label{examplefield}
\end{equation}
where $\Delta$ is the Laplacian.

It is convenient to collect all Green's functions into their generating functional
\begin{equation}
{\cal Z}[J]=\langle\exp\mathrm{i}J\cdot\Phi\rangle
=\frac{\int [D\Phi][\Phi]\exp\mathrm{i}\left\{S[\Phi]+J\cdot\Phi\right\}}{\int [D\Phi]\,\exp\mathrm{i}S[\Phi]}\label{ZJ}
\end{equation}
with $J\cdot\Phi=\int dx\,j(x)\Phi(x)$. Green's functions are recovered by functional differentiation of ${\cal Z}[J]$,
\begin{equation}
G(x_{1},\dots,x_{n})=(-\mathrm{i})^{n}\frac{\delta^{n}{\cal Z}}{\delta J(x_{1})\cdots\delta J(x_{n})}\bigg|_{J=0}
\end{equation}
To compute ${\cal Z}[J]$, one resorts to perturbation theory by expanding the result around the free field theory. For free fields the integral is Gau\ss ian and the result is given by Wick's theorem
\begin{equation}
\int [D\Phi]\,\Phi(x_{1})\cdots \Phi(x_{2n})\exp\mathrm{i}S_{0}[\Phi]=
\sum_{\mbox{\tiny pairings of }\atop\left\{1,2,\dots,2n\right\}} \mathrm{i}^{n}\,{\cal K}(x_{i_{1}},x_{i_{2}})\cdots {\cal K}(x_{i_{2n-1}},x_{i_{2n}})\label{Wick}
\end{equation}
where a pairing is a partition of  $\left\{1,2,\dots,2n\right\}=\left\{i_{1},i_{2}\right\}\cup\cdots\cup\left\{i_{2n-1},i_{2n}\right\}$ into pairs and vanishes for  an odd number of fields. The free field functional integral is normalized in such a way that $\int [D\Phi]\,\exp\mathrm{i}S_{0}[\Phi]=1$.

The perturbative expansion  is obtained by expanding $\exp\mathrm{i}S_{\mbox{\tiny int}}[\Phi]$ in powers of $\Phi$ and applying Wick's theorem to all the monomials.  Let us first consider consider the vacuum function (or partition function) expanded as 
\begin{equation}
{\cal Z}=\int[D\Phi]\exp\mathrm{i}S[\Phi]=\sum_{\gamma\atop
\mbox{\tiny vacuum Feynman graph}}\frac{1}{C_{\gamma}}\,{{\cal A}_{\gamma}}
\end{equation}
A vacuum Feynman graph is just a graph (i.e. a set of vertices $V$ joined by edges $E$) such that each vertex has valence at least $2$. The combinatorial factor $C_{\gamma}$ is the cardinal of the group of transformations of the half-edges that leave the graph unchanged. The amplitude is computed according to the following Feynman rules:
\begin{itemize}
\item assign variables $x_{i}$ to the half-edges of $\gamma$
\item associate every vertex of valence $n$  with $-\mathrm{i}\lambda_{n}{\cal V}_{n}(x_{i_{1}},\dots,x_{i_{n}})$
\item associate every edge with $\mathrm{i}\lambda{\cal K}(x_{i},x_{j})$
\item integrate over $x_{i}$ the product of the vertex and edge contributions
\begin{equation}
{\cal A}_{\gamma}=\int \prod_{\mbox{\tiny half-edges}}dx_{i}\prod_{\mbox{\tiny vertices}}\mathrm{i}\lambda_{n}{\cal V}_{n}(x_{i_{1}},\dots,x_{i_{n}})\prod_{\mbox{\tiny edges}}\mathrm{i}\lambda{\cal K}(x_{i},x_{j})
\end{equation}
Note that in a theory invariant under translations, one integrates on all but one of the vertices.
\end{itemize}
The expansion of the generating functions of Greens functions is obtained similarly
\begin{equation}
{\cal Z}[J]=\int[D\Phi]\exp\mathrm{i}\big\{S[\Phi]+J\cdot\Phi\big\}=\sum_{\gamma\atop
\mbox{\tiny Feynman graph}}\frac{1}{C_{\gamma}}{{\cal A}_{\gamma}}
\end{equation}
The Feynman graphs in this expansion involve also univalent vertices. The same Feynman rules are applied except for univalent vertices that contribute as $\mathrm{i}J(x)$. Because of the normalization by ${\cal Z}[J]$, the Feynman graphs involves in the expansion of ${\cal Z}$ are such that any connected component contains at least one univalent vertex. Then, the n-point Green's functions are obtained by differentiating with respect to $J(x)$. In this case, the univalent vertices, known in the particle physics terminology as  external legs, carry fixed space-time arguments $x_{n}$. 

\begin{figure}
\begin{center}
\begin{minipage}{3cm}
\includegraphics[width=3cm]{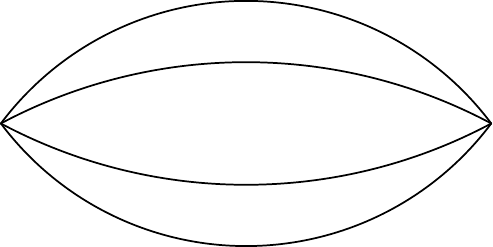}
\end{minipage}
\qquad\qquad
\begin{minipage}{4cm}
\includegraphics[width=4cm]{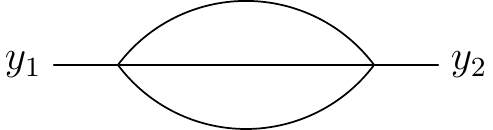}
\end{minipage}
\end{center}
\caption{Examples of Feynman graphs}
\label{graphex}
\end{figure}

To illustrate the Feynman rules, let us evaluate the two examples  drawn on figure \ref{graphex}. For the first  graph
\begin{equation}\frac{(-\mathrm{i}\lambda_{4})^{2}}{2\times4!}
\int dx_{1}\cdots dx_{8} \,\,V_{4}(x_{1},\dots,x_{4}){\cal K}(x_{1},x_{5}){\cal K}(x_{2},x_{6}){\cal K}(x_{3},x_{7}){\cal K}(x_{4},x_{8})V_{4}(x_{5},\dots,x_{8})
\end{equation}
while the second one evaluates to
\begin{multline}
\frac{(-\mathrm{i}\lambda_{4})^{2}}{3!}
\int dx_{1}\cdots dx_{8} \,\,
{\cal K}(y_{1},x_{1})
V_{4}(x_{1},\dots,x_{4})
{\cal K}(x_{2},x_{6}){\cal K}(x_{3},x_{7})
V_{4}(x_{5},\dots,x_{8})
{\cal K}(x_{8},y_{2})
\end{multline}

Let us conclude this section by a few remarks that will be useful in the sequel.

\begin{itemize}
\item
At the combinatorial level, it is simpler to work with $\log{\cal Z}$ and $\log{\cal Z}[J]$ whose expansion only involve connected graphs. Furthermore, when dealing with renormalization, it is convenient to introduce the Legendre $\Gamma[\Phi]$ transform of $\log{\cal Z}[J]$, with the classical field  $\Phi(x)=-{\mbox{ i}}\frac{\delta \log{\cal Z}[J]}{\delta J(x)}$, whose lowest order term is nothing but the action $S[\Phi]$. In many modern treatments of quantum field theory, one also uses the background field  effective action defined in the euclidian setting as
\begin{equation}
S_{\mbox{\tiny background}}[\Phi]=-\log\int[D\chi]\exp-S[\Phi+\chi]
\end{equation}
where $\Phi$ is a background field and $\chi$ a fluctuating field.

\item

We have been working with a real field that involve a symmetric kernel ${\cal K}(x,y)={\cal K}(y,x)$. In the case of a complex field, the quadratic  part of the action reads $\int dxdy\,{\cal K}^{-1}(x,y)\Phi(x)\Phi^{\ast}(y)$. The kernel only has hermitian symmetry ${\cal K}(x,y)={\cal K}^{\ast}(y,x)$ so that the resulting graphs have oriented edges. For fields having a more complicated structure, like matrix models or non commutative fields, it may also be useful to write the interaction using vertices with less symmetry than all permutations. In the example of matrix models, we formulate the theory in terms of vertices only invariant under circular permutations.  

\item

We have emphasized the role of the covariance ${\cal K}(x,y)$ (or propagator in the particle physics language) as opposed to the kernel ${\cal K}^{-1}(x,y)$ involved the free field action. This is because all we need to define the perturbative expansion is a Gau\ss ian measure with covariance ${\cal K}(x,y)$. This is defined even if ${\cal K}$ is not invertible since possible zero modes of ${\cal K}$ are set to zero. This can be seen by regularizing ${\cal K}(x,y)\rightarrow {\cal K}(x,y)+\epsilon\delta(x,y)$. As $\epsilon\rightarrow 0$, the zero modes of ${\cal K}$ are subject to a Gau\ss ian measure of width $\epsilon$ that set them to $0$.

\item

There are many equivalent formulations of the same quantum field theory that differ by a change of variable  in the functional integral,
$\Phi\rightarrow\Phi'$ where we express $\Phi$ in terms of a new field  $\Phi'$. The functional integrals are equal
\begin{equation}
\int[D\Phi]\,{\cal O}[\Phi]\exp\mathrm{i}S[\Phi]=\int[D\Phi]\,{\cal O'}[\Phi]\exp\mathrm{i}S'[\Phi]
\end{equation}
with ${\cal O}'[\Phi']={\cal O}[\Phi(\Phi')]$ and $
S'[\Phi']=S[\Phi(\Phi')]-\mathrm{i}\log\det\frac{\delta\Phi}{\delta\Phi'}$, this last term arising form the Jacobian of the transformation.

\end{itemize}

\subsection{Schwinger-Dyson equations}

\label{SDE}

\begin{figure}
\label{SDEfive}
\begin{equation}
\begin{minipage}{2.2cm}
\includegraphics[width=2.2cm]{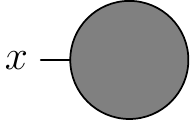}\end{minipage}\qquad=\qquad
\begin{minipage}{3cm}\includegraphics[width=3cm]{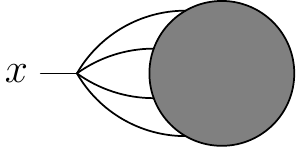}\end{minipage}\quad+\quad
\begin{minipage}{2cm}\includegraphics[width=2cm]{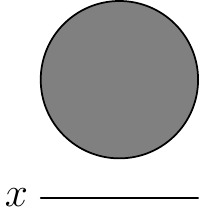}\end{minipage}
\end{equation}
\caption{Schwinger-Dyson equation for a $\Phi^{5}$ interaction}\label{SDEpic}
\end{figure}

Schwinger-Dyson equations are quantum analogues of the classical equations of motion. They follow form the invariance of the path integral representation of ${\cal Z}[J]$ under the change of variable $\Phi(x)\rightarrow\Phi(x)+\epsilon(x)$, with $\epsilon(x)$ an arbitrary infinitesimal function. Performing explicitly the infinitesimal change of variable and using $\Phi(x)=\mbox{\small  i}\frac{\delta \exp\mathrm{i}J\cdot\Phi}{\delta J(x)}$, it yields a functional differential equation for ${\cal Z}[J]$

\begin{align}
\frac{\delta {\cal Z}[J]}{\delta J(x)}=&
\bigg(\int dx_{1}{\cal K}(x,x_{1})J(x_{1})\bigg){\cal Z }[J]
\cr
&+\sum_{n\geq 2}{\textstyle \frac{\lambda_{n}}{(n-1)!}}
\int  dx_{1}\cdots dx_{n}{\cal K}(x,x_{1}){\cal V}_{n}(x_{1}.\dots.x_{n})\frac{\delta^{n}{\cal Z}[J]}{\delta J(x_{1})\cdots\delta J(x_{n})}\label{SchwingerDyson}
\end{align}
Graphically,  \eqref{SchwingerDyson} can be interpreted as follows: Choose a univalent vertex and follow the edge it is attached to. This edge may either end on another univalent vertex or on a vertex of higher degree. For instance, for a $\Phi^{5}$ interaction, it can be illustrated on figure \ref{SDEpic}.

When translated into an equation for the classical field  $\Phi(x)={\mbox{-\tiny i}}\frac{\delta \log{\cal Z}[J] }{\delta J(x)}$, they reduce to lowest order in $\hbar$ to the classical equations of motion. Indeed, if we restore $\hbar$ and define
\begin{equation}
\exp{ \frac{\mathrm{i}}{\hbar}}{\cal W}[j]=\int[D\Phi]\exp{\frac{\mathrm{i}}{\hbar}}\big\{S[\Phi]+J\cdot\Phi\big\}
\end{equation}
Then, we get in the limit $\hbar\rightarrow0$
\begin{equation}
\Phi(x)=-\sum_{n\geq 2}{ \frac{\lambda_{n}}{(n-1)!}}
\int  dx_{1}\cdots dx_{n}{\cal K}(x,x_{1}){\cal V}_{n}(x_{1}.\dots.x_{n})\Phi(x_{1})\cdots\Phi(x_{n})
\end{equation}

The most salient aspect of the Schwinger-Dyson equation is that they provide an equivalent formulation of quantum field theory.  Indeed, at the formal perturbative  level we are working at for the time being, the path integral representation for ${\cal Z}[J]$ in  \eqref{ZJ} is the unique solution to  \eqref{SchwingerDyson}, up to a multiplicative constant to be determined by the normalization. This is natural since a recursive resolution of  \eqref{SchwingerDyson} generates all the Feynman graph as a power series in the coupling constants $\lambda_{n}$. We refer to \cite{ZinnJustin} for a slick functional proof.

\subsection{Ultraviolet divergences and renormalization}

Let us consider a scalar field in $D$-dimensional Minkowski space with a local polynomial interaction, as given by  \eqref{examplefield}. It is convenient to work in momentum space by Fourier transforming all fields and Green's functions. A $n$-point connected Green's function $G_{n}(p_{1},\dots,p_{n})$ depends on external momenta and can be expressed as a sum over all connected graphs with $n$ external legs. Each edge of such a graph is given an arbitrary  orientation and comes equipped with a momentum $k_{e}$. Vertices  enforce a momentum conservation law while external legs are equipped with external momenta $p_{i}$. A global momentum conservation law can be factored as
\begin{equation}G_{n}(p_{1},\dots,p_{n})=
\delta(p_{1}+\cdots+p_{n}) \widetilde{G}_{n}(p_{1},\dots,p_{n})
\end{equation} 
$ \widetilde{G}_{n}(p_{1},\dots,p_{n})$ is computed by assigning a momentum space propagator to each edge 
\begin{equation}
\frac{\mathrm{i}}{k_{e}^{2}-m^{2}+\mathrm{i}\epsilon}
\end{equation}
and integrating over all independent momenta after taking into account momentum conservation at the vertices. The momentum space propagator is the Fourier transform of the kernel of the Klein-Gordon equation $(\square_{y}-m^{2})K(x,y)=\delta(x,y)$, computed using Feynman's prescription as
\begin{equation}
{\cal K}(x,y)=\lim_{\epsilon\rightarrow 0^{+}}\int\frac{dk}{(2\pi)^{D}}\,\frac{\mathrm{e}^{\mathrm{i}k(x-y)}}{k^{2}-m^{2}+\mathrm{i}\epsilon}
\end{equation}
Consider a connected graph with $e$ edges, $n$ external legs and $v_{n'}$ vertices of valence $n'$. Since there is one momentum per edge and one conservation law per vertex, there are $l=e-\sum_{n'}v_{n'}+1$ independent momenta. Therefore, the degree of divergence $\omega$, defined as the leading power of the integral as we rescale all momenta by $s$, with $s\rightarrow\infty$, is $\omega=lD-2e+\sum_{n'}\delta v_{n'}$. Since each edge is either connected to an external leg or to a vertex, we also have $2e=n+\sum_{n'}n'v_{n'}$. Altogether, the degree of divergence reads
\[
\omega=\sum_{n'}\Big(\frac{n'D}{2}-D-n'\Big)v_{n'}+n+D-\frac{nD}{2}
\] 
It is obvious that a graph with $\omega\geq 0$ is divergent. It is less obvious that a graph with $\omega<0$ for all its (1 particle irreducible, i.e. that cannot be disconnected by cutting a single line) subgraphs is convergent  (see the book \cite{ItzyksonZuber} for a proof). Theories that involve only interactions with $n'\leq\frac{D}{\frac{D}{2}-1}$ in dimension $D$ are especially nice: The only divergent Green's functions are those for which  $n\leq\frac{D}{\frac{D}{2}-1}$. The divergences correspond to the vertices already present in the action and can be removed by suitably choosing the coupling constants and rescaling the field. In the system of units we have adopted, the action is dimensionless so that the field has mass dimension $\frac{D}{2}-1$. Therefore, a theory is renormalizable if and only if its coupling have positive mass dimension. This is not  the case for gravity, whose coupling constant $G$ has mass dimension $2-D$. 

Ultraviolet divergences occur because the virtual particles propagating in the loops of a graph can have arbitrarily high momenta, even if the real particles on the external legs have fixed low momenta. To deal with this problem, one restricts the internal momenta to lie below a cut-off $\Lambda_{0}$. In a renormalizable theory, the parameters $\lambda_{n}^{\mbox{\tiny bare}}(\Lambda_{0},\mu,\lambda_{n}{\mbox{\tiny ren}})$ can be chosen in such a way that, once they have been fixed at a low energy scale $\mu$ to take the value $\lambda^{\mbox{\tiny ren}}$, all Green's functions remain finite, order by order in $\lambda^{\mbox{\tiny ren}}_{n}$. For a $\lambda\Phi^{n}$ interaction, $\lambda_{n}^{\mbox{\tiny bare}}$ is the sum of the divergent part of all graphs with $n$ external legs, once their possible subdivergences have been taken into account. The precise recursion relation is encoded in the Bogoliubov-Parasiuk-Hepp-Zimmermann formula (see the book \cite{ItzyksonZuber}).

More in general, renormalization stipulates that at a lower scale $\Lambda$ one should use an effective action $S_{\Lambda}$ obtained by integrating all the modes between $\Lambda$ and $\Lambda_{0}$

\[
S_{\Lambda}[\Phi]=-\log\int[D\chi]_{\Lambda,\Lambda_{0}}
\exp-S_{\Lambda_{0}}[\Phi+\chi]
\]   
where the  background field $\Phi$ has all its momenta below $\Lambda$ while the fluctuating field has momenta between $\Lambda$ and $\Lambda_{0}$. Performing the path integral, we obtain all types of interaction, including non renormalizable ones, even if we started with only renormalizable ones at scale $\lambda_{0}$. The  strength of an interaction with mass dimension $\delta<0$ is of order $\lambda(\Lambda)=\Lambda^{\delta}u(\log\Lambda)$, with $u(\log\Lambda)$ a dimensionless function that vary slowly with the energy scale. Accordingly,  effective field theories involve non renormalizable couplings obeying the previous scaling law, provided the cut-off remains finite. Moreover, these couplings are not all independent: Being obtained by integration over fast modes of a single action, they all belong to a single renormalization group trajectory. From this point of view, it the non renormalizable Einstein-Hibert action is perfectly acceptable as an low energy effective action. However, it necessarily involves an ultraviolet that cannot be taken to infinity.

\subsection{The propagator in Schwinger's proper time as one dimensional gravity with matter} 

\label{Schwingersec}
For our purposes, it is also fruitful to view quantum field theory on Minkowski space with action  \eqref{QFTMinkowski} as a second quantization of relativistic particles interacting locally.  The classical dynamics of a relativistic particle with space-time trajectory $x^{\mu}(s)$ is governed by the action $S[x]=-m\int dx\sqrt{-\dot x^{2}}$ with $\dot{x}^{\mu}=\frac{dx^{\mu}}{ds}$ and $\dot x^{2}=\eta_{\mu\nu}\dot{x}^{\mu}\dot{x}^{\nu}$. At the classical level, this action is equivalent to
\begin{equation}
S[x,e]=\frac{1}{2}\int ds\,\Big( e^{-1}\dot x^{2}-e\,m^{2}\Big)
\end{equation}
with $e(s)$ an auxiliary function. This action is invariant under reparametrizations of the worldline $s\rightarrow s'$ provided we transform $\dot{x}^{\mu}\rightarrow\dot{x}^{\mu}\frac{ds}{ds'}$ and $e\rightarrow e\frac{ds'}{ds}$. The dynamics of a relativistic point particle can be considered as a theory of gravity in one space-time dimension coupled to matter, where the auxiliary function $e$ is the einbein and $x^{\mu}$ a quantum field with $D$ components. Note that in this interpretation, the mass $m$ is a cosmological constant.  A mass term for the one dimensional field $x^{\mu}$ can be included as an harmonic oscillator potential $\frac{1}{2}\omega^{2}\eta_{\mu\nu}x^{\mu}x^{\nu}$.
 
To evaluate the path integral $\int[De][Dx]\,\exp\mathrm{i}S[e,x]$, one has to deal with reparametrization invariance in order not to overcount equivalent configurations. It is convenient to fix this ambiguity by imposing that $e(s)=\alpha>0$ is a constant that can be understood as the proper time. Then, the path integral reads
\begin{equation}
\int[De][Dx]\,\exp\mathrm{i}S[e,x]={\cal N}\int_{0}^{\infty} \!\!d\alpha\,\int[Dx]\,
\exp\mathrm{i}\Big\{\frac{1}{2}\int ds\,\big(\alpha^{-1}\dot x^{2}-\alpha\,m^{2}\big)\Big\}
\end{equation}
with ${\cal N}$ a normalizing factor. The path integral over $x^{\mu}$ is the standard quantum mechanical path integral for a free particle
\begin{equation}
\int[Dx]
\exp\mathrm{i}\Big\{\frac{1}{2}\int ds\,\alpha^{-1}\dot x^{2}\Big\}=\frac{{\cal N}'}{(2\pi\alpha)^{\frac{D}{2}}}\exp\mathrm{i}\frac{(x-y)^{2}}{2\alpha}
\end{equation}
with $x$ and $y$ the endpoints of the worldline. After a rescaling of $\alpha$ and an appropriate choice of the normalizing factors, we find
\begin{equation}
\int[De][Dx]\,\exp\mathrm{i}S[e,x]= \lim_{\epsilon\rightarrow 0^{+}}\frac{1}{(2\pi)^{D}}\int_{0}^{\infty}\!\!d\alpha\int dk\, \exp\mathrm{i}\Big\{k(x-y)+\alpha(k^{2}-m^{2}+\mathrm{i}\epsilon)\Big\}
\end{equation}
which is the Fourier transform of the Klein-Gordon propagator.  Note that the Feynman prescription $+\mathrm{i}\epsilon$ with $\epsilon>0$ ensures the convergence of the integral.

Thus, it is possible to understand the Feynman graph amplitudes in $D$-dimensional Minkowski space as one dimensional quantum gravity amplitudes with $D$ matter fields. The analogy goes on to the hamiltonian framework and we recover the Klein-Gordon equation as the constraint to be fulfilled by the wave function of the one dimensional system, $(\square-m^{2})\psi(x)=0$. A two dimensional version of this construction also exists: The point particle is replaced by a string and the corresponding quantum field theory that generates all string theory amplitude is known under the name of string field theory. In that respect, the task of finding a quantum field theory whose Feynman graphs describe four dimensional quantum gravity amplitudes seems hopeless. We will adopt an easier approach based on matrix models and their generalizations.

\subsection{Matrix models and $2d$ gravity}

Two dimensional gravity is a very interesting arena for testing some ideas that may be helpful in constructing a four dimensional theory of quantum gravity, in spite of the topological nature of the Einstein-Hilbert action in dimension two. For our purposes, the interesting aspect of this theory lies in the possibility of defining a continuum limit, including a sum over topologies, as the Feynman graph expansion of matrix models. We refer the reader to the review by Di Francesco, Ginsparg and Zinn-Justin \cite{largeN} for a thorough survey of matrix models applied to two dimensional gravity.

\begin{figure}
\centerline{\begin{minipage}{2.5cm}\includegraphics[width=2.5cm]{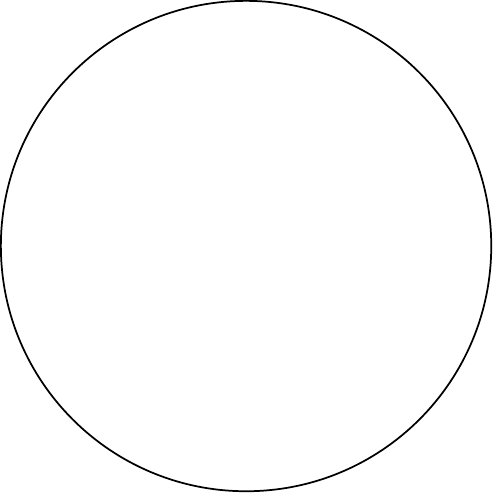}\end{minipage}
\qquad\qquad
\begin{minipage}{3cm}\includegraphics[width=3cm]{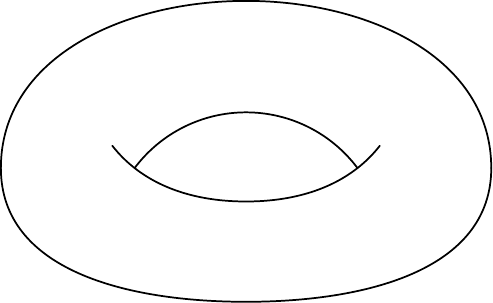}\end{minipage}
\qquad\qquad
\begin{minipage}{5cm}\includegraphics[width=5cm]{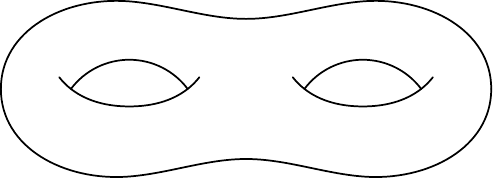}\end{minipage}}
\caption{Sphere, torus and genus two surface}
\label{surface}
\end{figure}

\begin{figure}
\centerline{\includegraphics[width=3cm]{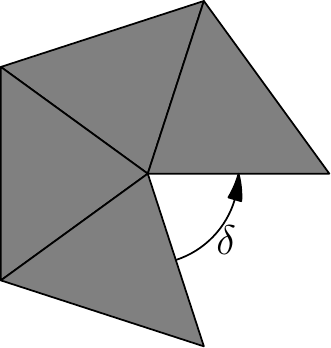}}
\caption{Deficit angle in dimension two}
\label{deficit}
\end{figure}

Let $\Sigma_{h}$ be a closed surface with $h$ handles. For instance, $\Sigma_{0}$ is a sphere and $\Sigma_{1}$ a torus, as illustrated on figure \ref{surface}. Up to homeomorphism, there is a single surface for each genus $h$. If we equip $\Sigma_{h}$ with a metric, then the Einstein-Hilbert action is nothing but the Euler characteristics $\chi(\Sigma_{h})=2-2h$
\begin{equation}
\frac{1}{2\pi}\int_{\Sigma_{h}}\sqrt{|g|} R=\chi(\Sigma_{h})
\end{equation}
We discretize the surface by assuming that it has been obtained by gluing together euclidian triangles along their sides to define a triangulation $T$ made of vertices $v$, edges $e$ and triangles $t$. Then the discrete analogue of the Einstein-Hilbert action is the Regge action
\begin{equation}
S_{\mbox{\tiny Regge}}=\frac{1}{2\pi}\sum_{v}\delta_{v}\qquad\mathrm{with}\qquad\delta_{v}=2\pi-\sum_{t\ni v}\alpha_{v,t}
\end{equation}
$\alpha_{v,t}$ is the angle at vertex $v$ in the triangle $t$ and $\delta_{v}$ the deficit angle at $v$ (see figure \ref{deficit}). $\delta_{v}$ is a discrete analogue of the curvature: If the triangles lie flat in the plane we have $\delta_{v}$=0, otherwise $\delta_{v}>0$ for positively curved spaces and $\delta_{v}<0$ for negatively curved ones.  Taking into account $\sum_{v\in t}\alpha_{v,t}=\pi$, we recover the Euler characteristics $S_{\mbox{\tiny Regge}}=t(T)-e(T)+v(T)=\chi(T)$, with $f(T)$, $e(T)$ and $v(T)$ the number of triangles, edges and vertices in $T$ and $\chi(T)=\chi(\Sigma_{h})$ if $T$ is a triangulation of a genus $h$ surface.

\begin{figure}
\centerline{\includegraphics[width=3cm]{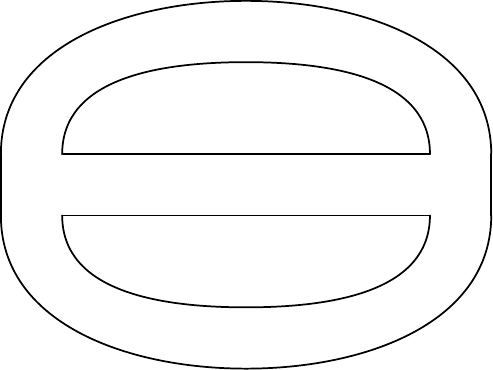}}
\caption{Triangulation of a sphere with 2 triangles}
\label{ribbonpic}
\end{figure}

The sum over triangulations weighted by  $\exp-\kappa S_{\mbox{\tiny Regge}}$ is generated by the following Feynman graph expansion of
\begin{equation}
{\cal F}(\lambda,N)=\log\int[DM]\,\exp\bigg\{-\frac{N}{2}\mbox{Tr}\,M^{2}+\frac{\lambda N}{3}\mbox{Tr}\,M^{3}\bigg\}=
\sum_{T\atop
\mbox{\tiny triangulations}}\,\frac{\lambda^{t(T)}N^{\chi(T)}}{C_{T}}\label{matrixmodel}
\end{equation}
where the integral is over hermitian $N\times N$ matrices with the measure $\prod_{i<j}d\mbox{Im}(M_{ij})\prod_{i\leq j}d\mbox{Re}(M_{ij})$, with a suitable normalizing factor. For instance, the graph depicted in figure \ref{ribbonpic} yields $\frac{\lambda^{2}N^{2}}{2}$.

Let us explain how this follows from a Feynman graph expansion. Wick's theorem for matrices relies on the following normalized Gau\ss ian integral
\begin{equation}
\int[DM]\,\exp-\big\{\textstyle{\frac{N}{2}}\mbox{Tr}\,M^{2}\big\}\,M_{ij}M_{kl}=\frac{1}{N}\,\delta_{il}\delta_{jk}\label{Wickmatrices}
\end{equation}
as well as generalization to products of an even number of matrix elements involving sums of pairing as in  \eqref{Wick}.  Note that one can use $M_{ji}=M^{\ast}_{ij}$ to compute pairings that involve complex conjugate matrix elements. 

\begin{figure}
\centerline{\includegraphics[width=3cm]{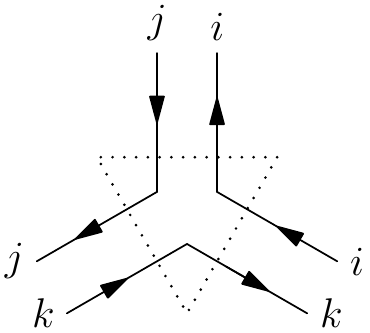}}
\caption{Matrix model interaction dual to a triangle}
\label{phi3}
\end{figure}

The basic variable is the matrix $M_{ij}$ associated with the edges of the triangulation. The trivalent vertex $\mbox{Tr}\,M^{3}=\sum_{i,j,k}M_{ij}M_{jk}M_{ki}$ represents the three sides of a triangle (see figure \ref{phi3}) and the propagators implements the gluing of the triangles along their sides. Since matrices carry two indices, the perturbative  expansion of the matrix model  \eqref{matrixmodel} involves  Feynman graphs whose edges are made of two lines. These graphs are ribbon graphs $\gamma$ which possess, in addition to edges $e(\gamma)$ and vertices $v(\gamma)$, faces $f(\gamma)$ made of closed lines of matrix indices. Ribbon graphs are dual to triangulations of orientable surfaces through the correspondence $v(\gamma)\leftrightarrow t(T)$, $e(\gamma)\leftrightarrow e(T)$ and $f(\gamma)\leftrightarrow v(T)$. The combinatorial factor $C_{T}$ is the symmetry factor of the dual graph and the logarithm restricts the summation to connected graphs. We have been working with complex hermitian matrices $M_{ji}=M_{ij}^{\ast}$, which correspond to orientable surfaces,  but several other choices are possible. For instance, if we work with real symmetric matrices, we have to sum over triangulations of possibly non orientable surfaces. This is reflected in the fact that for a real symmetric matrix, 
the propagator involves an extra term with a twist in the matrix indices. Note what we called triangulation is in fact a generalization of what is  called triangulation in mathematics. The latter are defined as simplicial complexes (see \ref{vocabulary}), while the triangulations we are talking about may fail to be simplicial complexes when the graphs have self-loops (edges with both ends attached to the same vertex) of multiple edges (pairs of vertices joined by several edges in parallel). 

To define the continuum limit, we collect terms having the same genus, so that
\begin{equation}
{\cal F}(\lambda,N)=\sum_{h}N^{2-2h}{\cal F}_{h}(\lambda)\label{largeN}
\end{equation}
The sum defining ${\cal F}_{h}(\lambda)$ can be shown to be convergent for $\lambda$ below a critical value $\lambda_{c}$ and exhibit a scaling behavior of the type
\begin{equation}
{\cal F}_{h}(\lambda)\mathop{\sim}\limits_{\lambda\rightarrow\lambda_{c}}(\lambda_{c}-\lambda)^{(1-\frac{\gamma}{2})(2-2h)}
\end{equation}
with $\gamma$ a fixed critical exponent ($\gamma=-\frac{1}{2}$ for pure gravity, as considered here). Therefore, in the double scaling limit $N\rightarrow\infty$ and $\lambda\rightarrow \lambda_{c}$ with $N(\lambda_{c}-\lambda)^{1-\frac{\gamma}{2}}=\mathrm{e}^{-\kappa}$ fixed, the matrix model expansion can be considered as a definition of the path integral of quantum gravity  \eqref{pathintegralquantumgravity} for closed manifolds.

To deal with surfaces with boundaries, it is convenient to expand the connected expectation values of products of $b$ traces  
\begin{equation}
\big\langle\mbox{Tr}\,(z_{1}-M)^{-1}\cdots\mbox{Tr}\,(z_{b}-M)^{-1}\big\rangle_{\mbox{\tiny c}}=
\sum_{T\,\mbox{\tiny triangulations}\atop b(T)=b}\frac{1}{C_{T}}\,\frac{\lambda^{t(T)}N^{\chi(T)}}{z_{1}^{n_{1}(T)+1}\cdots z_{b}^{n_{b}(T)+1}}
\end{equation}
where the sum  is over triangulations with $b(T)=b$ connected components and $\chi(T)=t(T)-e(T)+v(T)=2-2h-b$ its Euler characteristics. 
Connected correlations are  defined using the cumulants, for instance 
\begin{equation}
\big\langle\mbox{Tr}\,(z_{1}-M)^{-1}\mbox{Tr}\,(z_{2}-M)^{-1}\big\rangle_{\mbox{\tiny c}}=
\big\langle\mbox{Tr}\,(z_{1}-M)^{-1}\mbox{Tr}\,(z_{2}-M)^{-1}\big\rangle-\big\langle\mbox{Tr}\,(z_{1}-M)^{-1}\rangle\langle\mbox{Tr}\,(z_{2}-M)^{-1}\big\rangle\label{matrixmodelboundary}
\end{equation}
They obey Schwinger-Dyson equations  which translate into the loop equations of 2d quantum gravity.

Finally, let us sketch how matter can be included in the matrix model formulation of quantum gravity. We choose the $q$-state Potts model as a simple form of matter. On a given triangulation, it is defined by assigning a spin $\sigma_{t}\in\left\{1,2,\dots,q\right\}$ to each triangle. Two spins interact if and only if they belong to nearby triangles and the partition function is 
\begin{equation}
{\cal Z}_{\mbox{\tiny Potts}}(T)=\sum_{\{\sigma_{t}\}}\exp-\beta H(\{\sigma_{t}\})\qquad\mathrm{with}
\qquad H=-J\sum_{e\,\mbox{\tiny edges  of }T}\delta_{\sigma_{t(e)},\sigma_{t'(e)}}
\end{equation}
with $t(e)$ and $t'(e)$ the two triangles sharing the edge $e$ in $T$. Gravity coupled to the Potts model is generated by the following  multi-matrix model
\begin{equation}
\log\int[DM]\,\exp\bigg\{-{\textstyle \frac{N}{2}}\!\!\!\!\sum_{1\leq\alpha,\beta\leq q}\!\!\!\!Q_{\alpha,\beta}
\mbox{Tr}\,M_{\alpha}M_{\beta}
+{\textstyle \frac{\lambda N}{3}}\sum_{1\leq\alpha\leq q}\mbox{Tr}\,M_{\alpha}^{3}\bigg\}=
\!\!\!\!\sum_{T\atop
\mbox{\tiny triangulations}}\!\!\!\!\frac{\lambda^{t(T)}N^{\chi(t)}{\cal Z}_{\mbox{\tiny Potts}}(T)}{C_{T}}\label{matrixmodelpotts}
\end{equation}
where $Q_{\alpha,\beta}$ is the quadratic form 
\begin{equation}
Q_{\alpha,\beta}=-\frac{1}{(\mathrm{e}^{-\beta J}-1)^{2}(\mathrm{e}^{-\beta J}+q-1)}\big(1-\delta_{\alpha,\beta}\big)
+\frac{\mathrm{e}^{-\beta J}+q-2}{(\mathrm{e}^{-\beta J}-1)^{2}(\mathrm{e}^{-\beta J}+q-1)}\delta_{\alpha,\beta}
\end{equation}  
Thus, the spin degrees of freedom are encoded in a extra index in the matrix $M_{ij}\rightarrow M_{ij,\alpha}$ which obeys it own dynamics governed by the matrix $Q_{\alpha,\beta}$.
 
\section{From simplicial gravity to group field theory}

\subsection{Simplexes and triangulations}

\label{vocabulary}

\begin{figure}
\centerline{\begin{minipage}{2.5cm}\includegraphics[width=2.5cm]{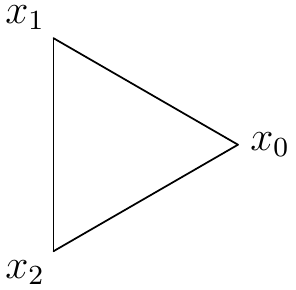}\end{minipage}
\hskip1cm
\begin{minipage}{3.3cm}\includegraphics[width=3.3cm]{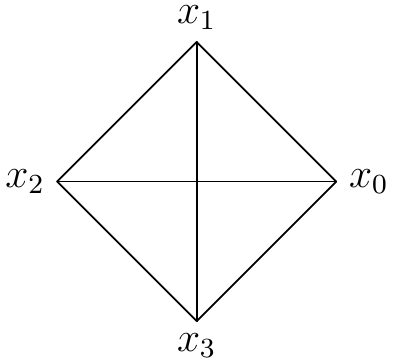}\end{minipage}
\hskip1cm
\begin{minipage}{3.3cm}\includegraphics[width=3.3cm]{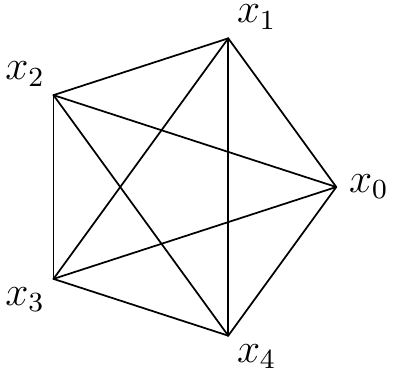}\end{minipage}}
\caption{Simplexes of dimension two, three and four}
\label{simplexes}
\end{figure}

In order to present in a self-content form tensor models and group field theories, let us introduce some mathematical terminology following \cite{quantumgeometry}. A $n$-simplex is the convex hull of $n+1$ points in ${\Bbb R}^{d}$ with $d\geq n$
\begin{equation}
\sigma_{n}=\langle x_{1}\dots x_{n}\rangle=\bigg\{\sum_{i}\lambda_{i}x_{i}\,\big|\,\lambda_{i}\in[0,1]\quad\mbox{with}\quad\sum_{i}\lambda_{i}=1\bigg\}
\end{equation}
We assume that these points dot not belong to a $n-1$ dimensional subspace of ${\Bbb R}^{d}$, otherwise the simplex would be degenerate. Note that at this stage there is no metric assigned to $\sigma_{n}$ and any realization of $\sigma_{n}$ may be taken. A standard choice is to take the subspace of ${\Bbb R}^{n+1}$ made of points with all coordinates positive and summing to 1. For instance,  a 0-simplex is a vertex, a 1-simplex is a segment, a  2-simplex is a triangle, a 3-simplex is a tetrahedron and a 4-simplex is sometimes referred as a pentachoron, see figure \ref{simplexes}.  A $m$ dimensional face of a simplex $\langle x_{1}\dots x_{n}\rangle$ is the convex hull of a subset of $m$ points $\langle x_{i_{1}}\dots x_{i_{m}}\rangle$.  The boundary $\partial\sigma$ of a $n$-simplex $\sigma$ is the set of its $n-1$ dimensional faces,
\begin{equation}
\partial \langle x_{0}\cdots x_{n}\rangle=
\big\{\langle x_{1}\cdots x_{n}\rangle,\langle x_{0}x_{2}\cdots x_{n}\rangle,\dots,
\langle x_{0}\cdots x_{i-1} x_{i+1}\cdots x_{n}\rangle,\dots,\langle x_{0}\cdots x_{n-1}\rangle\big\}
\end{equation}

A simplicial complex is a finite collection of simplexes $K=\left\{ \sigma\right\}$ such that: $ i)$ If $\sigma'$ is a face of $\sigma\in K$ then $\sigma'\in K$. $ii)$ If $\sigma,\sigma'\in K$ with $\sigma\cap\sigma'\neq\emptyset$ then $\sigma\cap\sigma'$ is a face of $\sigma$ and of $\sigma'$. Its dimension is the dimension of its top dimensional simplex. Two simplicial complexes are isomorphic if there is a bijection between them preserving the incidence relations. An abstract triangulation (i.e. a triangulation not associated to any a priori given space) is a isomorphism class of simplicial complexes. An abstract triangulation is a triangulation of a topological space ${\cal X}$ is there is an homeomorphism from the simplicial complex $K$ to ${\cal X}$.

An abstract triangulation is associated to a pseudo-manifold of dimension $n$ if:  $i)$ For any $\sigma\in K$, there is $\sigma'\in K$ of dimension $n$ such that $\sigma\subset\sigma'$. $ii)$ Any $\sigma\in K$ of dimension $n-1$ is contained in one or two simplexes of dimension $n$. In the first case $\sigma$ is a boundary simplex while in the second case it is an inner one.  $iii)$ Any two simplexes of dimension $n$ are connected by a sequence of simplexes of dimension $n$ that share a  simplex of dimension $n-1$. The first axiom excludes lower dimensional pieces, the second one states that there is no branching and the third one is a kind of connectedness.

To understand how a pseudo-manifold may fail to be a manifold, let us introduce some more terminology. Two simplicial complexes are combinatorially equivalent if they can be included in finer simplicial complexes that are isomorphic. For any vertex $v$, we define the star of $v$ in $K$ as the simplicial complex made of all the faces of the simplexes  in $K$ that contain $v$. The link of $v$ in $K$ is obtained from its star by removing all simplexes that contain $v$. Then a pseudo-manifold of dimension $n$ defines a (combinatorial) manifold if the link of any vertex is combinatorially  equivalent  to the boundary of a of a $n$-simplex.

The simplexes we have defined up to now are not oriented and $\langle x_{\tau(0)}\cdots x_{\tau(n)}\rangle=\langle x_{0}\cdots x_{n}\rangle$ for any permutation $\tau$. To orient the simplexes, we introduce for each ordered set of vertices $ x_{0},\cdots ,x_{n}$ two simplexes $\langle x_{0}\cdots x_{n}\rangle$ and $\langle x_{0}\cdots x_{n}\rangle^{\ast}$ that differ by their orientations
\begin{equation}
\langle x_{\tau(0)}\cdots x_{\tau(n)}\rangle=\left\{
\begin{array}{ll}
\langle x_{0}\cdots x_{n}\rangle& \mbox{if }\tau\,\mbox{is even}\cr
\langle x_{0}\cdots x_{n}\rangle^{\ast}& \mbox{if }\tau\,\mbox{is odd}
\end{array}
\right.
\end{equation}
This notion is useful to define oriented triangulations. We also define the oriented boundary $\partial\sigma$ of an oriented $n$-simplex $\sigma$ is the set of its oriented $n-1$ dimensional faces,
\begin{equation}
\partial \langle x_{0}\cdots x_{n}\rangle=
\big\{\langle x_{2}\cdots x_{n}\rangle,\langle x_{0}x_{2}\cdots x_{n}\rangle^{\ast},\dots,
\langle x_{0}\cdots x_{i-1} x_{i+1}\cdots x_{n}\rangle^{{\ast}^{i-1}},\dots,\langle x_{0}\cdots x_{n-1}\rangle^{{\ast}^{n-1}}\big\}\label{orientedboundary}
\end{equation}
where ${\ast}^{n}=1$ if $n$ is even and ${\ast}^{n}=\ast$ if $n$ is odd.

\subsection{Simplicial gravity}

In general, the most natural way of attempting at a non-perturbative definition  of the quantum gravity path integral  \eqref{pathintegralquantumgravity}  consists in discretizing the geometry. This is also the approach used in elementary quantum mechanics and lattice gauge theory. Therefore, we choose a triangulation $T$ of  a $D$ dimensional  space-time ${\cal M}$ made of $n$-simplexes $\sigma_{n}\in{\cal S}_{n}$ for $n\in\{ o,\dots,D\}$. Working in the euclidian setting for simplicity, metric degrees of freedom are captured by assigning lengths to the edges of the $D$-simplexes. This specifies completely the embedding of the simplex in ${\Bbb R}^{D}$ up to rotation, translation and reflection, as can be checked by the counting of degrees of freedom: $\frac{D(D+1)}{2}=D(D+1)-D^{2}-\frac{D(D-1)}{2}$. Thus, the $n$-volume of any $n$-simplex is known. In particular, the cosmological constant term in the gravity action  \eqref{actiongravity} is proportional to the sum over $D$-simplexes of their volumes, 
\begin{equation}
\frac{\Lambda}{G}\int_{{\cal M}}d^{D}x\,\sqrt{|g|}\rightarrow \lambda\sum_{\sigma_{D}\in{\cal S}_{D}}
V_{\sigma_{D}}
\end{equation}
Because Newton's constant $G$ has mass dimension $2\!-\!D$, a discretization of the Einstein-Hilbert action involves volumes of $(D\!-\!2)$-simplexes. This leads to the $D$-dimensional version of Regge's action
\begin{equation}
\frac{\Lambda}{G}\int_{{\cal M}}d^{D}x\,\sqrt{|g|} R\rightarrow \kappa\sum_{\sigma_{D-2}\in{\cal S}_{D-2}}
V_{\sigma_{D\!-\!2}}\delta_{\sigma_{D-2}}
\end{equation} 
with curvature measured by the deficit angle $\delta_{\sigma_{D-2}}$ defined as follows. In any $D$-dimensional simplex containing the $(D\!-\!2)$-dimensional simplex $\sigma_{D\!-\!2}$, consider the  two distinguished $(D\!-\!1)$-dimensional simplexes $\sigma_{D-1}$ and $\sigma_{D-1}'$ that contain $\sigma_{D-2}$. If we further assume all simplexes to be oriented, the embedding of $\sigma_{D}\in{\Bbb R}^{D}$ allows us to associate $\sigma_{D-1}$ and $\sigma_{D-1}'$ with outward pointing normals $N$ and $N'$. Then we define the angle $\alpha_{\sigma_{D},\sigma_{D-2}}$  of $\sigma_{D}$ at $\sigma_{D-2}$ as the angle between $N$ and $N'$. Then, the  deficit angle is defined as
\begin{equation}
\delta_{\sigma_{D-2}}=2\pi-\!\!\!\sum_{\sigma_{D}\supset \sigma_{D-2}}\alpha_{\sigma_{D},\sigma_{D-2}}
\end{equation}
To define discretized path integrals, there are two alternative choices. Following Ponzano and Regge, we may fix a triangulation and sum over the lengths associated to its edges. Alternatively, we may fix the size of the simplexes and sum over triangulations. In this case, the choice of the triangulation encodes all the degrees of freedom of the theory. This is the approach known as dynamical triangulations. From now one, we focus on the latter approach and defer a brief overview of the Ponzano-Regge model to the section \ref{BF} devoted to spin foam models .

If we restrict our attention to regular euclidian simplexes of size $a$, the deficit angle only counts how many $D$ simplexes contain a given $D$-2 simplex $\sigma_{D-2}$. Then, the discretized Einstein-Hilbert action with cosmological constant reads
\begin{equation}
S(T)=\,\frac{a^{D-2}}{G'}\,n_{D-2}(T)+\bigg(\frac{\Lambda'a^{D}}{G'}+\frac{a^{D-2}}{G''}\bigg)\, n_{D}(T)
\end{equation} 
where $n_{k}(T)$ is the number of $k$-dimensional simplexes in the triangulation and $\Lambda'$, $G'$ and $G''$ are proportional to $\Lambda$ and $G$.  An arbitrary metric is approximated by a sufficiently fine triangulation. Because the length $a$ is fixed, the only degrees of freedom lie in the choice of the triangulations. Therefore, the path integral  \eqref{pathintegralquantumgravity} becomes a sum over triangulations with fixed boundary, 
\begin{equation}
\sum_{\mbox{\tiny topologies}\atop \partial {\cal M}\,\mbox{\tiny fixed}}
\int_{g|_{\partial {\cal M}}\,\mbox{\tiny fixed}}[Dg]\exp\mathrm{i}S[g]
\quad\rightarrow\quad\sum_{T\,\mbox{\tiny triangulations}}\frac{1}{C_{T}}\exp\mathrm{i}S(T)\label{pathintegralsimplicial}
\end{equation}
with $C_{T}$ a combinatorial factor, usually chosen to be the cardinal of the automorphism group of $T$. In this sum, we do not necessarily restrict ourselves to triangulations with a fixed topology. In this way, it implements a sum over topologies. When $D=2$, we recover the  matrix model expansion \eqref{matrixmodel}. This remains true in $D>2$, provided one uses tensor models.

\subsection{Tensor models as generalized matrix models }

We aim at generalizing the matrix models to higher dimensions as tensor models. Recall that in matrix model formulation of quantum gravity in $D=2$ dimensions, the matrix $M_{ij}$ represents a $(D\!-\!1)$-simplex (edge), its indices $i$ and $j$ two $(D\!-\!2)$-simplexes (points) and the interaction $\mbox{Tr}\,M^{3}$ the boundary of a $D$-simplex (triangle). Detailed proofs of the following results can be found in an article by De Pietri and Petronio \cite{dipietri}.

The generalization to $D>2$ dimensions goes as follows:
\begin{figure}
\centerline{\includegraphics[width=5cm]{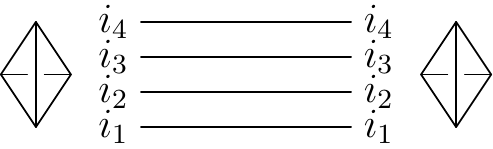}}
\caption{Propagation of a tetrahedron}
\label{propam}
\end{figure}

\begin{figure}
\centerline{\includegraphics[width=4cm]{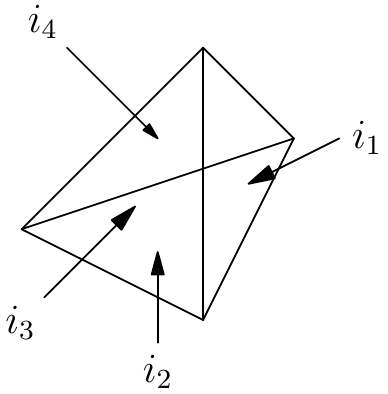}}
\caption{A tetrahedron and its four triangles}
\label{tetrahedrontensor}
\end{figure}

\begin{itemize}
\item
The basic field $M_{i_{1}\dots i_{D}}$ is a tensor with $D$ indices. It represents a $(D\!-\!1)$-simplex and its indices stand for the $D$ $(D\!-\!2)$-simplexes on the boundary of this $(D\!-\!1)$-simplex, as illustrated on figure \ref{tetrahedrontensor}. The indices belong to an index  set $I$ which is for the time being left arbitrary.  

\item
As for matrix models, the tensor $M_{i_{1}\dots i_{D}}$ may be real or complex and obey different transformation laws under permutations of its indices. To fix the notations, we momentarily choose the one analogous to hermitian matrices
\begin{equation}
M_{i_{\tau(1)}\cdots i_{\tau(D)}}=\left\{
\begin{array}{ll}
M_{i_{1}\cdots i_{D}}& \mbox{if }\tau\,\mbox{is even}\cr
(M_{i_{1}\cdots i_{D}})^{\ast}& \mbox{if }\tau\,\mbox{is odd}
\end{array}
\right.\label{reality}
\end{equation}
Many other choices are also possible and lead to summations over different class of triangulations. For instance, for a real symmetric tensor, we are no longer able to assign an orientation to the simplexes. In this case, the sum involves possibly non orientable triangulations.   If we do not impose any invariance under permutations, we are led to colored tensor models (see \cite{coloredreview}) which we will briefly discuss in section \ref{coloredsec} in the context of group field theory. 

\item
The quadratic term in the action enforces the gluing of two ($D$-1) simplexes with opposite orientations
\begin{equation}\frac{1}{2}|M|^{2}=
\frac{1}{2}\sum_{i_{1},\dots,i_{D}}
M_{i_{1}\cdots i_{D}}(M_{i_{1}\cdots i_{D}})^{\ast}
\end{equation} 
It represents the propagation of a $(D\!-\!1)$-simplex through an edge made of $D$ strands that stand for the $(D\!-\!2)$-simplexes.  For example, in four dimensions, we have four strands corresponding to the four faces of a tetrahedron, as illustrated in \ref{propam}.

\begin{figure}
\centerline{\begin{minipage}{4cm}\includegraphics[width=4cm]{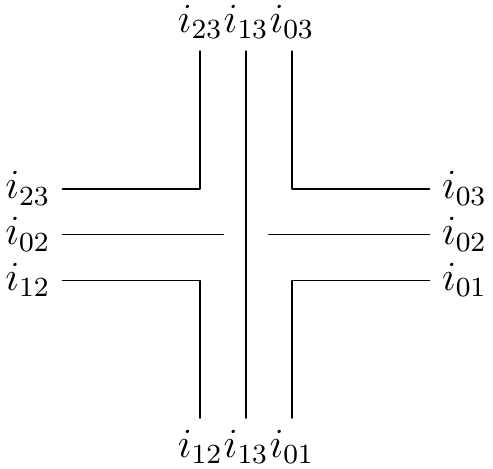}\end{minipage}
\hskip2cm
\begin{minipage}{5cm}\includegraphics[width=6cm]{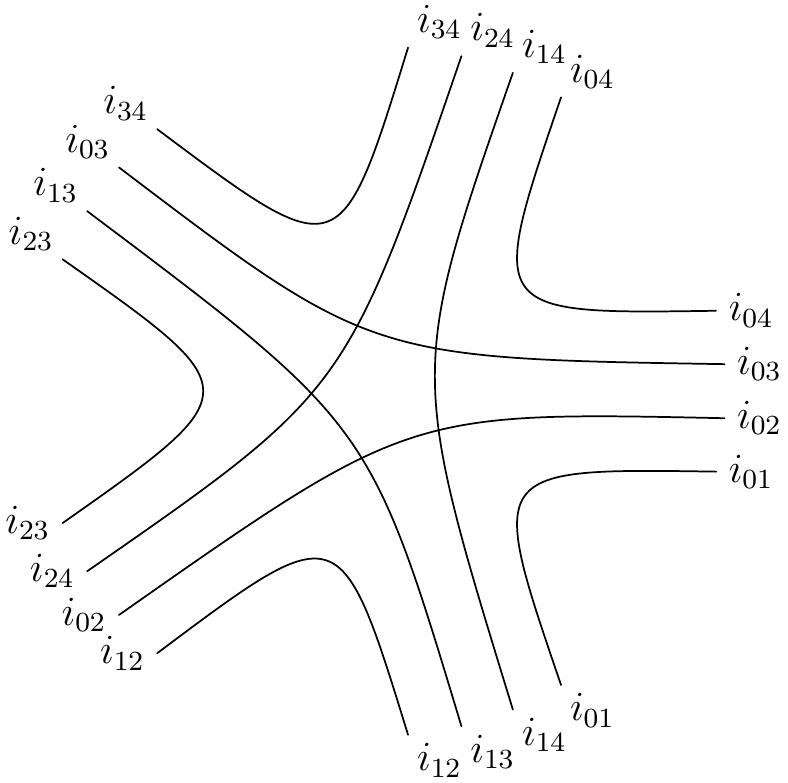}\end{minipage}}
\caption{Vertex in $D=3$ and $D=4$}\label{intmatpic}
\end{figure}

\item
The interaction term reproduces the $D\!+\!1$ $(D\!-\!1)$-simplexes that form the boundary of an oriented  $D$-simplex, given in the oriented boundary\eqref{orientedboundary}
\begin{equation}
V(M)=\frac{\lambda}{D+1}\sum_{i_{k,l}\,\mbox{\tiny with }\atop
1\leq k\neq l\leq D+1}\bigg\{
\prod_{0\leq k\leq D}(M_{i_{k,1},\dots,i_{k,k-1},i_{k,k+1},\dots i_{k,D}})^{\ast^{k}} \prod_{0\leq k<l\leq D}\delta_{i_{k,l},i_{l,k}}\bigg\}\label{tensorinteraction}
\end{equation}
$M^{\ast^{k}}$ is the complex conjugate of $M$ if $k$ is odd and $M$ if $k$ is even. There is a sum over the indices $i_{k,l}$ associated to the $(D\!-\!2)$-simplexes obtained by removing the vertices $k$ and $l$ from the $D$-simplex. These ($D\!-\!2)$-simplexes arise from the boundaries of each of the $(D\!-\!1)$-simplexes in the boundary of the $D$-simplex. Since each of these ($D\!-\!2)$-simplexes appears twice in the boundary, we identify $i_{k,l}$ and $i_{l,k}$. It is also worthwhile to note that the symmetry properties of the tensor ensure the reality of the action. Finally, the factor $\frac{1}{D+1}$ reflects the invariance under cyclic permutations of the labels.
\end{itemize}

To illustrate the general formula, let us treat in detail the cases $D=2,3,4$. We use the invariance of the field \eqref{reality} to rewrite the interaction in more conventional form, depicted in dimension three and four on picture \ref{intmatpic}.

\begin{itemize}

\item D=2: the boundary of a triangle 
\begin{equation}
\partial\langle x_{0}x_{1}x_{2}\rangle=\left\{\langle x_{1}x_{2}\rangle,\langle x_{0}x_{2}\rangle^{\ast},\langle x_{0}x_{1}\rangle\right\}
\end{equation}
so that the interaction is the matrix model interaction
\begin{align}
V(M)&=\frac{\lambda}{3}\sum_{i_{01},i_{02},i_{12}\atop i_{10},i_{20},i_{21}}
M_{i_{01},i_{02}}
(M_{i_{10},i_{21}})^{\ast}
M_{i_{20}i_{21}}\,
\delta_{i_{01},i_{10}}\delta_{i_{21},i_{12}}\delta_{i_{20},i_{02}}\cr
&=
\frac{\lambda}{3}\sum_{i_{01},i_{02},i_{12}}
M_{i_{12}i_{02}}
M_{i_{02}i_{12}}
M_{i_{12}i_{01}}
\end{align}

\item D=3: the boundary of a tetrahedron is 
\begin{equation}
\partial\langle x_{0}x_{1}x_{2}x_{3}\rangle=\left\{\langle x_{1}x_{2}x_{3}\rangle,\langle x_{0}x_{2}x_{3}\rangle^{\ast},\langle x_{0}x_{1}x_{3}\rangle,\langle x_{0}x_{1}x_{2}\rangle^{\ast}\right\}
\end{equation}
Each of these four terms correspond to a triangle, which is itself made of three edges.  This leads to the interaction, after the identifications provided by the Kronecker symbols
\begin{align}
V(M)&=\frac{\lambda}{4}\sum_{i_{01},i_{02},i_{03},i_{12},i_{13},i_{23}}
M_{i_{01}i_{02}i_{03}}(M_{i_{01}i_{12}i_{13}})^{\ast}M_{i_{02}i_{12}i_{23}}(M_{i_{03}i_{13}i_{23}})^{\ast}\cr
&=\frac{\lambda}{4}\sum_{i_{01},i_{02},i_{02}, i_{12},i_{13},i_{23}}
M_{i_{01}i_{02}i_{03}}
(M_{i_{03}i_{13}i_{23}})^{\ast}
M_{i_{23}i_{02}i_{12}}
(M_{i_{12}i_{02}i_{01}})^{\ast}
\end{align}

\item D=4: the boundary of a 4-simplex is 
\begin{equation}
\partial\langle x_{0}x_{1}x_{2}x_{3}x_{4}\rangle=\left\{\langle x_{1}x_{2}x_{3}x_{4}\rangle,\langle x_{0}x_{2}x_{3}x_{4}\rangle^{\ast},\langle x_{0}x_{1}x_{3}x_{4}\rangle,\langle x_{0}x_{1}x_{2}x_{4}\rangle^{\ast},\langle x_{0}x_{1}x_{2}x_{3}\rangle\right\}.
\end{equation} 
Each of these five terms correspond to a tetrahedron, which is itself made of four triangleS. Accordingly, the interaction reads, after the identifications provided by the Kronecker symbols
\begin{align}
V(M)&=\frac{\lambda}{5}\sum_{i_{ab}}
M_{i_{01}i_{02}i_{03}i_{04}}
(M_{i_{01}i_{12}i_{13}i_{14}})^{\ast}
M_{i_{02}i_{12}i_{23}i_{24}}
(M_{i_{03}i_{13}i_{23}i_{34}})^{\ast}
M_{i_{04}i_{14}i_{24}i_{34}}\cr
&=\frac{\lambda}{5}\sum_{i_{ab}}
M_{i_{01}i_{02}i_{03}i_{04}}
M_{i_{04}i_{14}i_{24}i_{34}}
M_{i_{34}i_{03}i_{13}i_{23}}
M_{i_{23}i_{24}i_{02}i_{12}}
M_{i_{12}i_{13}i_{14}i_{01}}
\end{align}

\end{itemize}

In complete analogy with matrix models, we compute the path integral for a tensor model as
\begin{align}
{\cal Z}(\lambda,N)=\int DM\,\exp\big\{-\frac{1}{2}|M|^{2}+V(M)\big\}=\sum_{{\cal G}\atop\mbox{\tiny stranded graph}}\,\frac{\lambda^{v({\cal G})}N^{f({\cal G})}}{C({\cal G})}
\end{align}
The expansion is over stranded graphs, i.e. graphs whose edge are made of $D$ strands. $N$ is the cardinal of the index set $I$ of the tensor, $C({\cal G})$ the symmetry factor of the graph and $v({\cal G})$ the number of vertices of ${\cal G}$. The number of faces $F({\cal G})$ is the number of cycles obtained by following the strands through the edges and the vertices. In $D=2$ it coincides with the notion of face of a ribbon graph. We integrate over the real and imaginary parts of the tensor
\begin{equation}
DM=\prod_{i_{1}<i_{2}<\cdots<i_{D}}d\mbox{Im}(M_{i_{1}\cdots i_{D}})\prod_{i_{1}\leq i_{2}\leq\cdots\leq i_{D}}d\mbox{Re}(M_{i_{1}\cdots i_{D}})
\end{equation}
with a suitable normalizing factor.

 This expansion relies on the tensor model generalization of Wick's theorem for matrices  \eqref{Wickmatrices} given by the Gau\ss ian integral
\begin{equation}
\int[DM]\,\exp-\big\{ {\textstyle \frac{1}{2}}|M|^{2}\big\}\,M_{i_{1}\cdots i_{D}}M_{j_{1}\dots j_{D}}=\frac{1}{D!}
\sum_{\tau\atop\mbox{\tiny odd permutation}}\delta_{i_{1},j_{\tau(1)}}\cdots \delta_{i_{D},j_{\tau(D)}}
\label{Wicktensor}
\end{equation}
supplemented by a sum over pairings if more tensors are involved. As for matrix models, pairings involving complex conjugate fields can be computed using the reality conditions  \eqref{reality}, for instance
\begin{equation}
\int[DM]\,\exp-\big\{ {\textstyle \frac{1}{2}}|M|^{2}\big\}\,M_{i_{1}\cdots i_{D}}M^{\ast}_{j_{1}\dots j_{D}}=\frac{1}{D!}
\sum_{\tau\atop\mbox{\tiny even permutation}}\delta_{i_{1},j_{\tau(1)}}\cdots \delta_{i_{D},j_{\tau(D)}}
\label{Wicktensor}
\end{equation}
This ensure that two $(D\!-\!1)$-simplexes have opposite orientations when identified, as requested by the orientability of the space we triangulate.

\begin{figure}
\begin{equation*}
\parbox{5cm}{\includegraphics[width=5cm]{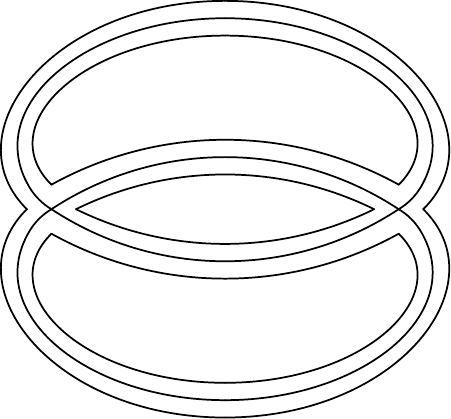}}\quad\Leftrightarrow\quad\parbox{5cm}{\includegraphics[width=5cm]{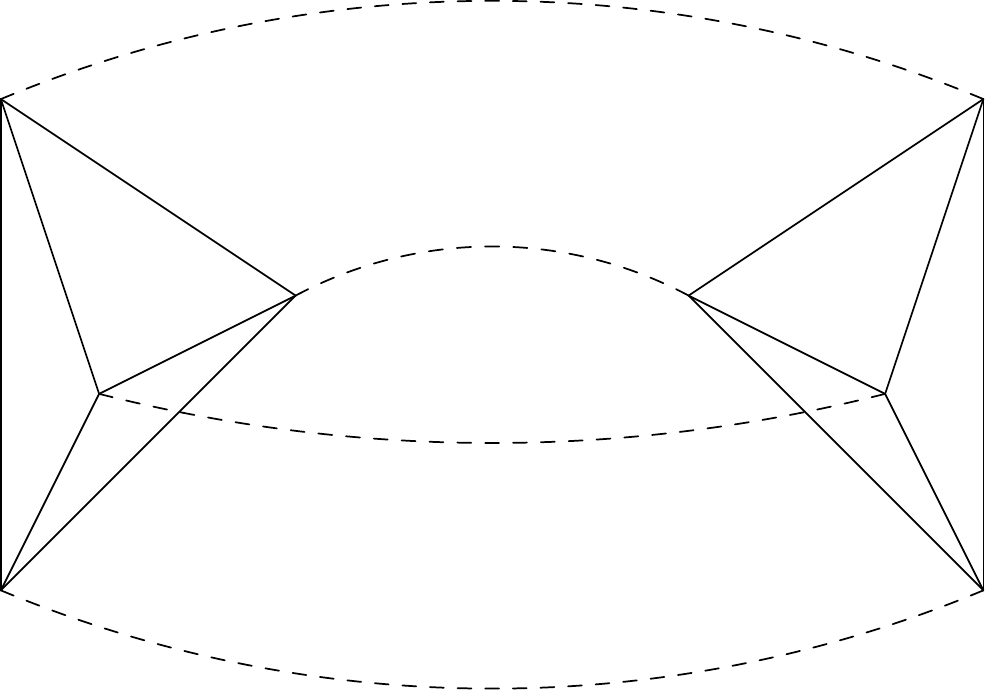}}
\end{equation*}
\caption{Triangulation of a 3d sphere with 2 tetrahedra}
\label{2tetratensor}
\end{figure}

Alternatively, every vertex of the graph represents of $D$ simplex and the edges with their strands yield a prescription for gluing these $D$-simplexes along their $(D\!-\!1)$-simplexes by an identifications of their $(D\!-\!2)$-simplexes. Loosely speaking, any such gluing can be viewed as a triangulation, so that we can also write
\begin{align}
{\cal Z}(\lambda,N)=\int DM\,\exp\big\{-|M|^{2}+V(M)\big\}=\sum_{T\atop\mbox{\tiny triangulation}}\,\frac{\lambda^{n_{D}(D)}N^{n_{D-2}(T)}}{C_{T}}
\end{align}
in complete analogy with the two dimensional case  \eqref{matrixmodel}. We recover the path integral of simplicial quantum gravity  \eqref{pathintegralsimplicial} by analytical continuation of ${\cal Z}(\lambda,N)$. For instance, the stranded graph in figure \ref{2tetratensor} evaluates to  $\lambda^{2}N^{6}$ which correspond to the six edges and two tetrahedra. It is the 3-dimensional analogue of the ribbon graph given in figure \ref{ribbonpic}.

However, there is a notable difference between matrix and tensor models. In the former case, the Feynman graph only yield triangulations of surfaces (apart from self-loops and and graph with multiple edges), which are orientable for hermitian matrices and non orientable for real symmetric matrix. In the case of tensor models, the situation is more involved and the summation also contains triangulations of pseudo-manifolds.  Moreover, many possible field contents are available: the field may be real or complex and may obey various invariance properties under permutations of its indices. This generates sums over different kind of triangulations. In the sequel, we leave open the field content of the theory and focus on the construction of a vertex suitable for spin foam models.

\subsection{Spin foam models of quantum gravity}

\label{spinfoamsec}

Spin foam models are discretized versions of the path integral of quantum gravity, devised in order to define the scalar product between spin network states in loop quantum gravity. The latter is an approach to quantum gravity rooted in the canonical formalism. Space-time is assumed to be decomposed as ${\cal M}={\Bbb R}\times \Sigma$ and the basic degrees of freedom are the holonomies of the Ashtekar connection $A$, an SU(2) connection on $\Sigma$.  Let $\Gamma$ be a graph defined by its links $L$ and nodes $N$ as well as source $s:\, L\rightarrow N$ and target $t:\, L\rightarrow N$ maps\footnote{We use the word links and nodes instead of edges and vertices because we reserve the latter for spin foams.}. This graph is obtained from a triangulation of space by tetrahedra: a node is associated to each tetrahedron and there is a link between two nodes if the corresponding tetrahedra share a triangle.

The gauge invariant graph Hilbert space is defined as
\begin{multline}
{\cal H}_{\Gamma}=L^{2}\big(\mathrm{SU(2)}^{L}/\mathrm{SU(2)}^{N}\big)=\\
\Big\{\psi_{\Gamma}(h_{l})\,\,\,\,\mbox{such that}\,\,\,\,\psi_{\Gamma}(g_{s_{l}}h_{l}g^{-1}_{t_{l}})=\psi_{\Gamma}(h_{l})\,\,\,\,\mbox{for all}\,\,\,\,g_{n}\in \mathrm{SU(2)}\Big\}
\end{multline}
with $s_{l}$ and $t_{l}$ the source and target nodes of the link $l$. This is reminiscent of lattice gauge theory, with gauge fields on the links subjected to gauge transformations associated with the nodes. Note that the graphs have to be oriented, but reversing the orientation of the edge $l$ yields an isomorphic Hilbert space, provided we trade $h_{l}$ for  $h_{l}^{-1}$. Moreover, the graph need not to be connected, if $\Gamma=\Gamma_{1}\cup\Gamma_{2}$, then ${\cal H}_{\Gamma}={\cal H}_{\Gamma_{1}}\otimes{\cal H}_{\Gamma_{1}}$.
 
The gravitational wave function  $\psi_{\Gamma}(h_{l})$  depends on the SU(2) link variables $h_{l}$ which represent the holonomies of the Ashtekar connection. On these wave functions, $A$ acts by multiplication and its canonically conjugate momentum, the densitized triad $E$,  by functional differentiation. Moreover,  physically relevant operators, like area and volume operators, are well defined on this Hilbert space and have a discrete spectrum in a basis build on spin networks, which are graphs whose links are labeled by SU(2) representations and nodes by intertwiners between the representations associated to the  incoming and outgoing links.

In the canonical formalism, the dynamics of general relativity is equivalent to three sets of constraints associated with symmetries. The first one is the Gau\ss \, law, associated with the SU(2) gauge symmetry of the connection $A$.  Since  states in ${\cal H}_{\Gamma}$ are gauge invariant, it is already implemented. The second one is the vector constraint that implements space diffeomorphism invariance. Roughly speaking, if we think of the graph as embedded in ${\Sigma}$, one may take the vector constraint into account by declaring that two graphs are equivalent if one can be obtained from the other by the action of a diffeomorphism of ${\Sigma}$. Thus the precise location of $\Gamma$ in $\Sigma$ does not matter and the only relevant information is of combinatorial nature. Again, it can be considered as implemented in our construction because we only used abstract graphs that do not refer to any embedding.

Lastly, the scalar constraint supplements the vector constraint to recover space-time diffeomorphism invariance. It is implemented through the definition of a new scalar product in ${\cal H}_{\Gamma}$ that modifies the canonical scalar product in $L^{2}(\mathrm{SU(2)}^{L}/\mathrm{SU(2)}^{N})$. This procedure is similar to the one encountered in the dynamics of a relativistic particle: enforcing the mass-shell condition, associated to reparametrization of the trajectories, forces us to modify the scalar product on the wave function by inserting $\delta(k^{2}-m^{2})$.

Inspired by Feynman's path integral, the  scalar product implementing the hamiltonian constraint is defined by inserting  sums over intermediate states
\begin{equation}
\langle\psi_{\Gamma'}|\psi_{\Gamma}\rangle=\sum_{\Gamma_{1},\dots,\Gamma_{k}} 
\langle\psi_{\Gamma'}|\psi_{\Gamma_{k}}\rangle\cdots \langle\psi_{\Gamma_{i+1}}|\psi_{\Gamma_{i}}\rangle\cdots\langle\psi_{\Gamma_{1}}|\psi_{\Gamma}\rangle
\end{equation}
Here, $\Gamma_{i}$ differs form $\Gamma_{i+1}$ by some elementary operations like splitting or joining nodes. This sequence of graphs defines a two dimensional object ${\cal C}$. Its vertices encode the elementary operations on the nodes to obtain $\Gamma_{i+1}$ from $\Gamma_{i}$. If the graphs are spin networks, edges and faces of ${\cal C}$ are decorated by interwiners and representations and define a spin foam. 

\begin{figure}
\centerline{\includegraphics[width=6cm]{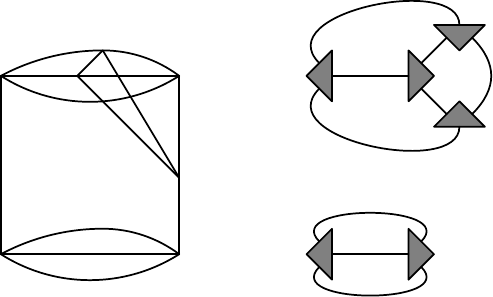}}
\caption{Simple evolution of a triangulation of the sphere}
\label{sfpic}
\end{figure}

Following closely Rovelli's lecture notes at this school \cite{CarloZakopane}, let us give a purely combinatorial definition of spin foam models. An oriented 2-complex is defined as a set of faces $F$, edges $E$ and vertices $V$ together with two incidence relations. The first incidence relation defines, for each $e\in E$, its  the source $s(e)\in V$ and  target $t(e)\in V$. The second one defines the boundary of any face $\partial f$ as a cyclically ordered set of edges that bound the face. We assume that each edge belongs to the boundary of at least one face. The edges that belong to the boundary of a single face are the boundary edges. They are called links  and we retain the term edge only for the edges that are not links. We also assume that the vertices incident to at least a link are incident to exactly one edge. These boundary vertices are the nodes and we call vertices only the inner vertices. This restrictions allows us to think of the boundary of a 2-complex as a graph $\partial{\cal C}=\Gamma$ and of ${\cal C}$ as an history of graphs whose nodes join and split, as illustrated in figure \ref{sfpic}. 

The amplitude associated to ${\cal C}$ is the state ${\cal H}_{\partial{\cal C}}$ defined as 
\begin{equation}
{\cal A}_{{\cal C}}(h_{l})=\int_{SU(2)} \prod_{v,f}dh_{v,f}\,\prod_{f}\delta(h_{f})\,\prod_{v}{\cal A}_{v}(h_{v,f},h_{l})\label{SFamplitude}
\end{equation}
with a group variable $h_{l}$ associated to every link on the boundary graph and an integration over group variables $h_{v,f}$ for each pair of a vertex $v$ and a face  $f$ incident to $v$. By isolating vertices, we decompose ${\cal C}=\cup_{v\in V}{\cal C}_{v}$ into elementary 2-complexes made of single vertices, whose amplitude ${\cal A}_{v}$ must be given a priori. Each of these elementary 2-complexes defines a boundary graph $\Gamma_{v}=\partial C_{v}$, whose link variables  are $h_{v,f}^{\pm}$ depending on the orientation. For every face of ${\cal C}$, $h_{f}$ is the product around $\partial f$ of the link variables $h_{l}^{\pm}$ and $h_{v,f}^{\pm}$ and the sign $\pm$ depend on the relative orientation of the links and the faces. The Dirac distribution $\delta(h_{f})$ ensures the proper gluing of all the variables around the face.

The explicit expression of the vertex amplitudes 
${\cal A}_{v}(h_{v,f},h_{l})$ can be found in section  \ref{BF} for BF theory and in section \ref{EPRLFK} for the EPRL/FK model. In these cases, the explicit expressions involve sum over spins  attached to the faces and intertwiners attached to the edges so that we recover the previous definition of spin foam models.  In the general case, the vertex amplitude is constrained by gauge invariance on the nodes.  If we denote by $g_{n_{v}}$ gauge transformations on the nodes of $\Gamma_{v}$, gauge invariance reads
\begin{equation}
 {\cal A}_{v}(g_{s_{l_{v}}}h_{l_{v}}g_{t_{l_{v}}}^{-1})={\cal A}_{v}(h_{l_{v}}).
 \end{equation}

These 2-complexes and spin foams are related to triangulations of space-time as follows. If $T$ is a triangulation of a space-time ${\cal M}$ of dimension $D$, we define a 2-complex ${\cal C}$ as the dual 2-skeleton of $T$. The vertices of ${\cal C}$ are the $D$-simplexes of $T$, edges $(D\!-\!1)$-simplexes and faces $(D\!-\!2)$-simplexes. Two vertices are joined by an edge if the corresponding 
$D$-simplexes  share the $(D\!-\!1)$-simplex associated to the edge, oriented in an arbitrary way. In the models we consider, any change of orientation simply induces a replacement of a group variable by its inverse. To obtain the incidence relation between faces and edges, consider a generic slice of the triangulation by a plane. A $(D\!-\!2)$-simplex appears as a point surrounded by a cyclically ordered sequence of $(D\!-\!1)$-simplexes. By construction, in dimension $D$ we recover only those 2-complexes for which every vertex has valence $D+1$ and every edge belongs to the boundary of $D$ faces. The orientation of the face is provided by the common orientation of the $D$-simplexes of the triangulation, if space-time is oriented. Although the orientation of the faces appears in the spin foam amplitudes  \eqref{SFamplitude}, it turns out that it does not play any role in the two models we consider $SU(2)$ BF theory and the Lorentzian EPRL/FK. This is not a generic feature of spin foam models but a peculiarity of the finite dimensional representations of SU(2)  and of the prinicpal series of irreducible  representations of $SL(2,{\Bbb C})$, any of them being equivalent its  complex conjugate.

The boundary graph $\partial{\cal C}$ is the 1-skeleton of the boundary  triangulation $\partial T$. Its nodes correspond to $(D\!-\!1)$-simplexes and its links to $(D\!-\!2)$-simplexes. The basic variables are group elements  $h_{l}$ attached to $(D\!-\!2)$-simplexes of the boundary triangulations. Then, the elementary amplitudes for a $D$ simplex are glued together to reproduce the amplitude associated to the 2-skeleton of the triangulation. Finally, the spin foam amplitude  \eqref{SFamplitude}, eventually supplemented with a weighted sum over 2-compexes with fixed boundary is the spin foam definition of the quantum gravity path integral  \eqref{pathintegralquantumgravity}
\begin{equation}
\Psi(g\big|_{\partial {\cal M} })=\sum_{\mbox{\tiny topologies}\atop \partial {\cal M}\,\mbox{\tiny fixed}}
\int_{g|_{\partial {\cal M}}\,\mbox{\tiny fixed}}[Dg]\exp\frac{\mathrm{i}S[g]}{\hbar}
\qquad\rightarrow\qquad
\psi_{\Gamma}(h_{l})=\sum_{\mbox{\tiny 2-complexes}\atop \partial{\cal C}=\Gamma}
w_{\cal C}\,{\cal A}_{\cal C}(h_{l})\label{spinfoamsum}
\end{equation}
with $w_{\cal C}$ a suitable  combinatorial weight that may also include powers of a coupling constant.


\subsection{General definition of a group field theory}

\label{GFT}

Let us now give a general definition of a group field theory by combining the combinatorics of tensor models with the group theoretical definition of quantum gravity amplitudes provided by spin foam models. The basic field of the theory is a function $\Phi(g_{1},\dots,g_{D})$ on $D$ copies of the group $G$. It represents a $(D\!-\!1)$-simplex and the group elements $g_{1},\dots,g_{D}$ are associated with its $D$ faces of dimension  $D-\!2\!$. Intuitively, the group elements define a connection whose curvature is concentrated on the $(D\!-\!2)$-simplexes, as is the case in Regge calculus. 

For definiteness, we assume that the field obeys the reality conditions
\begin{equation}
\Phi(g_{\tau(1)},\dots,g_{\tau(D)})=\left\{
\begin{array}{ll}
\Phi(g_{1},\dots, g_{D})& \mbox{if }\tau\,\mbox{is even}\cr
\Phi^{\ast}(g_{1},\dots, g_{D})& \mbox{if }\tau\,\mbox{is odd}
\end{array}
\right.\label{realityphi}
\end{equation}
but the field may as well be taken to be real or to obey different symmetry relations. An important alternative to these symmetry rules are provided by colored models, see section \ref{coloredsec}.

\begin{figure}
\centerline{\includegraphics[width=4cm]{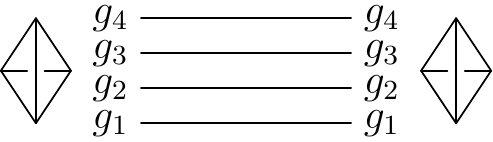}}
\caption{Propagation of a tetrahedron in a four dimensional group field theory}
\label{pgftpic}
\end{figure}

In complete analogy with tensor models, we choose the quadratic term to be the trivial one
\begin{equation}
|\Phi|^{2}=\int_{G^{D}}\prod dg_{i}\,\Phi(g_{1},\dots,g_{D})\Phi^{\ast}(g_{1},\dots,g_{D})\label{trivialpropagator}
\end{equation}
This ensures an identification of the group elements associated to the ($D\!-\!2$)-simplexes in the gluing of the $D$ simplexes (see figure \ref{pgftpic}). Alternative choices are discussed below.

The interaction vertex is required to reproduce the vertex amplitude ${\cal A}_{v}(h_{l})$ of the spin foam model associated with a $D$-simplex. The latter involves $\frac{D(D+1)}{2}$ group elements $h_{l}$ associated to the links of the complete graph which is the 1-skeleton of the dual of the boundary of the $D$-simplex.  Labeling the vertices of the $D$ simplex with letters $a,b\in\{1,\dots,D\}$, the variables $h_{l}$ are written as $h_{l}=h_{ab}$ with the convention that $h_{ba}=h_{ab}^{-1}$. This means that we orient the links of the boundary from $a$ to $b$ if $a<b$. We introduce $D(D+1)$ variables $g_{a,b}$ such that $h_{ab}=g^{}_{ba}g_{ab}^{-1}$. The variable $g_{a,b}$ is attached to the $(D\!-\!2)$-simplex which is the intersection of the $(D\!-\!1)$-simplexes associated to the nodes $a$ and $b$. Then, the interaction term is build combining the tensor model interaction \eqref{tensorinteraction} and the spin foam vertex amplitude
\begin{equation}
V(\Phi)=\frac{\lambda}{D+1}\int_{G^{D(D+1)}}\prod_{0\leq a\neq b\leq D} dg_{ab}\quad\bigg\{
{\cal A}_{v}\big(g^{}_{ba}g_{ab}^{-1}\big)
\prod_{0\leq a\leq D}\Phi^{\ast^{a}}\big(g_{a,b\neq a}\big)\bigg\}\label{interactionGFT}
\end{equation}
where $\Phi\big(g_{a,b\neq a}\big)$ is a shorthand for $\Phi\big(g_{a,1},\dots,g_{a,a-1},g_{a,a+1},\dots g_{a,D+1}\big)$. $\Phi^{\ast^{a}}=\Phi$ if $a$ is even and $\Phi^{\ast}$ if $a$ is odd.

Let us note that we have made a specific choice of orientation of the boundary graph, which amounts to the choice of an orientation of  the strands that meet at the vertex. This choice of orientation is irrelevant if the  vertex amplitude is invariant under the inversion of any of its argument $h_{l}$. We assume that this invariance holds in the following discussion. This is the case for the two models we consider in the next sections. If this is not the case, the amplitude depends on the orientations of the faces and the group field must be supplemented with an extra ${\Bbb Z}_{2}$ index $\Phi_{\pm}$ conserved along the strands. This extra index allows to distinguish the two orientations of the faces.

Working with real fields, this is nothing but  the interaction term introduced in \cite{Freidel}. Here, we are working with complex fields and the complex conjugation is necessary in order to have a real potential. The reality of the potential is necessary in order to make sense at a non perturbative level of a path integral of the type $\int [D\Phi]\exp\mathrm{i}\{|\Phi|^{2}+V(\Phi)\}$ since, in even space-time dimension, the interaction is an odd monomial.
The situation is similar to the case of Chern-Simons theory, the matrix models of $D=2$ quantum gravity or the theory of Airy functions: an integral like $\int_{\Bbb R} dx\, \mathrm{e}^{\mathrm{i}x^{3}}$ may be defined as a semi-convergent one while $\int_{\Bbb R} dx\, \mathrm{e}^{\pm x^{3}}$ is meaningless. 

By performing a circular permutation of the argument of the fields, which is an odd permutation for $D$ even and an even permutation for $D$ odd, we can rewrite the interaction for $D$ even as
\begin{equation}
V(\Phi)=\frac{\lambda}{D+1}\int_{G^{D(D+1)}}\prod_{1\leq a\neq b\leq D+1} dg_{a,b}\quad\bigg\{
{\cal A}_{v}\big(g^{}_{ba}g_{ab}^{-1}\big)
\prod_{0\leq a\leq D}\Phi^{\ast^{a}}\big(g_{a(a+1)},\dots,g_{aD},g_{a,0}\dots,g_{a(a-1)}\big)\bigg\}\label{interactionGFTeven}
\end{equation}
and for $D$ odd
\begin{equation}
V(\Phi)=\frac{\lambda}{D+1}\int_{G^{D(D+1)}}\prod_{1\leq a\neq b\leq D+1} dg_{a,b}\quad\bigg\{
{\cal A}_{v}\big(g^{}_{ba}g_{ab}^{-1}\big)
\prod_{0\leq a\leq D}\Phi\big(g_{a(a+1)},\dots,g_{aD},g_{a,0}\dots,g_{a(a-1)}\big)\bigg\}\label{interactionGFTodd}
\end{equation}

\begin{figure}
\centerline{\begin{minipage}{4cm}\includegraphics[width=4cm]{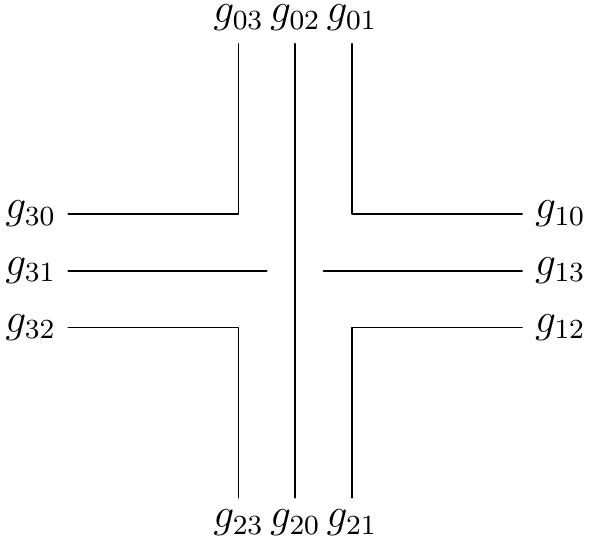}\end{minipage}
\hskip2cm
\begin{minipage}{4cm}\includegraphics[width=4cm]{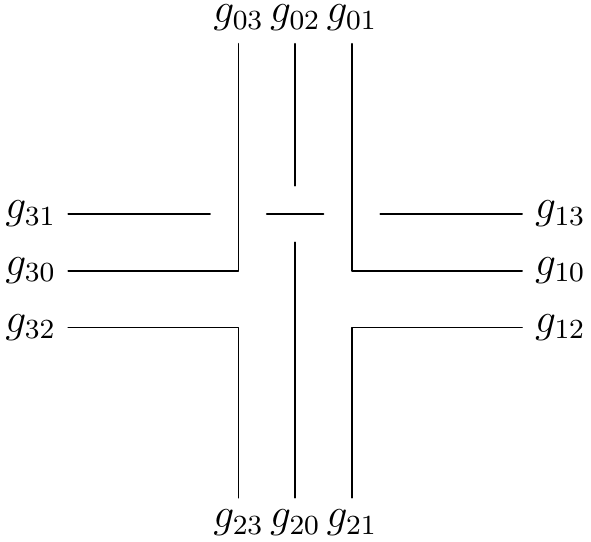}\end{minipage}}
\caption{Two equivalent forms of the vertex in $D=3$}
\label{V3pic}
\end{figure}

\begin{figure}
\centerline{\begin{minipage}{6cm}\includegraphics[width=6cm]{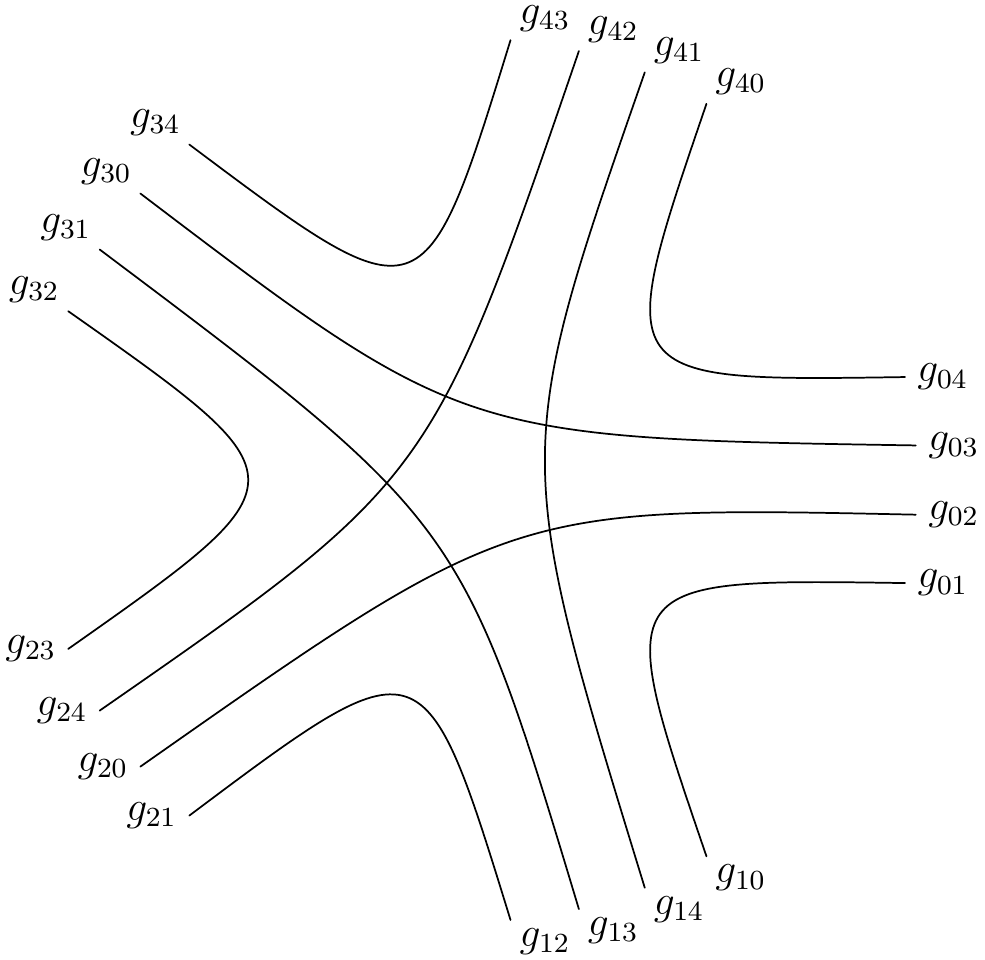}\end{minipage}}
\caption{Vertex iand $D=4$}
\label{V4pic}
\end{figure}

For definiteness, let us explicit the interaction in low dimension, using the symmetry of the fields under permutations to rewrite the interaction in various forms. These are illustrated on figure \ref{V3pic} and \ref{V4pic}.

\begin{itemize}
\item D=2
\begin{align}
&\quad\frac{\lambda}{3}\int _{G}\prod_{0\leq a<b\leq 2}dg_{ab}\quad{\cal A}_{v}(g^{}_{ba}g_{ab}^{-1})\quad
\Phi(g_{01},g_{02})
\Phi^{\ast}(g_{10},g_{12})
\Phi(g_{20},g_{21})\cr
&=\frac{\lambda}{3}\int _{G}\prod_{0\leq a<b\leq 2}dg_{ab}\quad{\cal A}_{v}(g^{}_{ba}g_{ab}^{-1})\quad
\Phi(g_{01},g_{02})
\Phi(g_{20},g_{21})
\Phi(g_{12},g_{10})
\end{align}
\item D=3
\begin{align}
&\quad\frac{\lambda}{4}\int _{G}\prod_{0\leq a<b\leq 3}dg_{ab}\quad{\cal A}_{v}(g^{}_{ba}g_{ab}^{-1})\quad
\Phi(g_{01},g_{02},g_{03})
\Phi^{\ast}(g_{10},g_{12},g_{13})
\Phi(g_{20},g_{21},g_{23})
\Phi^{\ast}(g_{30},g_{31},g_{32})\cr
&=\quad\frac{\lambda}{4}\int _{G}\prod_{0\leq a<b\leq 3}dg_{ab}\quad{\cal A}_{v}(g^{}_{ba}g_{ab}^{-1})\quad
\Phi(g_{01},g_{02},g_{03})
\Phi^{\ast}(g_{30},g_{31},g_{32})
\Phi(g_{23},g_{20},g_{21})
\Phi^{\ast}(g_{12},g_{13},g_{10})\cr
&=\quad\frac{\lambda}{4}\int _{G}\prod_{0\leq a<b\leq 3}dg_{ab}\quad{\cal A}_{v}(g^{}_{ba}g_{ab}^{-1})\quad
\Phi(g_{01},g_{02},g_{03})
\Phi(g_{30},g_{32},g_{31})
\Phi(g_{23},g_{20},g_{21})
\Phi(g_{13},g_{12},g_{10})
\end{align}
\item D=4
\begin{align}
&\frac{\lambda}{5}\int _{G}\prod_{0\leq a<b\leq 4}dg_{ab}\quad{\cal A}_{v}(g^{}_{ba}g_{ab}^{-1})\quad\cr
&\quad\Big\{
\Phi(g_{01},g_{02},g_{03},g_{04})
\Phi^{\ast}(g_{10},g_{12},g_{13},g_{14})
\Phi(g_{20},g_{21},g_{23},g_{24})
\Phi^{\ast}(g_{30},g_{31},g_{32},g_{34})
\Phi(g_{40},g_{41},g_{42},g_{43})\Big\}\cr
&=\frac{\lambda}{5}\int _{G}\prod_{0\leq a<b\leq 4}dg_{ab}\quad{\cal A}_{v}(g^{}_{ba}g_{ab}^{-1})\quad\cr
&\quad\Big\{
\Phi(g_{01},g_{02},g_{03},g_{04})
\Phi(g_{40},g_{41},g_{42},g_{43})
\Phi(g_{34},g_{30},g_{31},g_{32})
\Phi(g_{23},g_{24},g_{20},g_{21})
\Phi(g_{12},g_{13},g_{14},g_{10})\Big\}\nonumber\\
\end{align}
\end{itemize}

As was already pointed out in \cite{Freidel}, the interaction is invariant under a gauge symmetry and a global symmetry. The global symmetry is a consequence of the fact that the vertex amplitude only depends on the products $g^{}_{ba}g_{ab}^{-1}$. Thus, the potential obeys
\begin{equation}
V(T_{g}\Phi)=V(\Phi)\qquad\mathrm{with}\qquad T_{g}\Phi(g_{1},\dots,g_{D})=\Phi(g_{1}g,\dots,g_{D}g)\label{globalGFT}
\end{equation}
Its analogue in electrodynamics is translation invariance under $A(x)\rightarrow A(x-a)$.

The gauge invariance is a consequence of the gauge invariance at the nodes  of the spin foam vertex amplitude, ${\cal A}(g^{}_{a}h^{}_{ab}g_{b}^{-1})={\cal A}_{v}(h_{ab})$ for all $g_{a}\in G$. In the group field theory formalism, it translates into
\begin{equation}
V(\Phi+\Upsilon)=V(\Phi)\qquad\mbox{with}\qquad \int_{G}dg\,\Upsilon(gg_{1},\dots,gg_{D})=0 \label{gaugeGFT}
\end{equation}
which is easily checked using the invariance of the Haar measure under $g\rightarrow hg$ and $g\rightarrow g^{-1}$. It is similar to the gauge invariance of electrodynamics $A\rightarrow A+\chi$ with $d\chi=0$. On a simply connected space-time, the Poincar\'e lemma states that $\chi=d\Lambda$, so that we recover the ordinary formulation of gauge invariance.

The trivial group field theory propagator \eqref{trivialpropagator} is invariant under the global symmetry  \eqref{globalGFT}, $|T_{g}\Phi|^{2}=|\Phi|^{2}$ but not under gauge transformations \eqref{gaugeGFT}. However, we may always split the field as $\Phi=\Phi_{\mbox{\tiny gauge}}+\Phi_{\mbox{\tiny inv}}$ with
\begin{equation}
 \int_{G}dg\,\Phi_{\mbox{\tiny gauge}}(gg_{1},\dots,gg_{D})=0\quad\mbox{and}\quad
\int_{G}dg\,\Phi_{\mbox{\tiny inv}}(gg_{1},\dots,gg_{D})=\Phi_{\mbox{\tiny inv}}(g_{1},\dots,g_{D})\label{decomposition}
\end{equation}
Equivalently, the invariant part of the field obeys
\begin{equation}
\Phi_{\mbox{\tiny inv}}(gg_{1},\dots,gg_{D})=\Phi_{\mbox{\tiny inv}}(g_{1},\dots,g_{D})\label{invariant}
\end{equation}
Gauge degrees of freedom decouple in the sense that $|\Phi|^{2}=|\Phi_{\mbox{\tiny gauge}}|^{2}+|\Phi_{\mbox{\tiny inv}}|^{2}$ and $V(\Phi)=V(\Phi_{\mbox{\tiny gauge}})$, so that the group field theory path integral splits as
\begin{multline}
\int[D\Phi]\exp\{-|\Phi|^{2}+V(\Phi)+J\cdot\Phi\}=\int[D\Phi_{\mbox{\tiny gauge}}]\exp\{-|\Phi_{\mbox{\tiny gauge}}|^{2}+
J_{\mbox{\tiny gauge}}\cdot\Phi_{\mbox{\tiny gauge}}\}\times\cr\int[D\Phi_{\mbox{\tiny inv}}]\exp\{-|\Phi_{\mbox{\tiny inv}}|^{2}+V(\Phi_{\mbox{\tiny inv}})
+J_{\mbox{\tiny inv}}\cdot\Phi_{\mbox{\tiny inv}}\}
\end{multline}
where  a source term $J(g_{1},\dots,g_{D})$, obeying the same reality conditions \ref{realityphi} as $\Phi$ , has been introduced with
\begin{equation}
J\cdot\Phi=\int_{G}\prod dg_{i}\,J(g_{1},\dots,g_{D})\Phi(g_{1},\dots,g_{D})
\end{equation}
and decomposed as $J=J_{\mbox{\tiny gauge}}+J_{\mbox{\tiny inv}}$ as in  \eqref{decomposition}. Thus, the effect of the gauge degrees of freedom manifest themselves only at vanishing coupling, i.e. for Feynman graph made of single edges, and for a non invariant source. It is therefore convenient to factor them out and consider a path integral over gauge invariant terms. 

It is often useful to allow more general quadratic terms of the form
\begin{equation}
\Phi\cdot{\cal K}^{-1}\cdot\!\Phi^{\ast}=\int_{G}\prod dg_{i}dg'_{i} \,{\cal K}^{-1}(g_{1},\dots,g_{D};g'_{1},\dots,g'_{D})\Phi(g_{1},\dots,g_{D})\Phi^{\ast}(g'_{1},\dots,g'_{D})
\end{equation}
We require ${\cal K}^{-1}$  to be real and symmetric, 
\begin{equation}
{\cal K}^{-1}(g'_{1},\dots,g'_{D};g_{1},\dots,g_{D})={\cal K}^{-1}(g_{1},\dots,g_{D};g'_{1},\dots,g'_{D})
\end{equation} 
The global invariance  \eqref{globalGFT} remains true if 
\begin{equation}
{\cal K}^{-1}(g_{1}g,\dots,g_{D}g;g'_{1}g,\dots,g'_{D}g)={\cal K}^{-1}(g_{1},\dots,g_{D};g'_{1},\dots,g'_{D})
\end{equation}
 and gauge degrees of freedom decouple provided ${\cal K}^{-1}$ commutes with the projection onto invariant states,
\begin{equation}
\int_{G}dg\,{\cal K}^{-1}(gg_{1},\dots,gg_{D};g'_{1},\dots,g'_{D})=\int_{G}dg\,{\cal K}^{-1}(g_{1},\dots,g_{D};gg'_{1},\dots,gg'_{D})
\end{equation}
The group field theory formulation of Wick's theorem is analogous to its tensor model formulation,
\begin{equation}
\int[D\Phi]\exp\{-{\textstyle\frac{1}{2}}\Phi\cdot{\cal K}^{-1}\cdot\!\Phi^{\ast}\}\,
\Phi(g_{1},\dots,g_{D})\Phi(g'_{1},\dots,g'_{D})=
\sum_{\tau\atop\mbox{\tiny odd permutation}}{\cal K}(g_{1},\dots,g_{D};g'_{\tau(1)},\dots,g'_{\tau(D)})\label{WickGFT}
\end{equation}
where ${\cal K}$ and ${\cal K}^{-1}$ are related by
\begin{equation}
\int_{G}\prod dg'_{i}
\,{\cal K}^{-1}(g_{1},\dots,g_{D};g'_{1},\dots,g'_{D}){\cal K}(g'_{1},\dots,g'_{D};g''_{1},\dots,g''_{D})=\prod_{i}\delta\big(g_{i}(g''_{i})^{-1}\big)
\end{equation}
As for tensor models, If more than two fields are involved, one has to sum over pairings. Moreover, pairings involving the complex conjugate field are computed using the reality conditions to enforce the orientability.

As already mentioned in the introductory section on quantum field theory, ${\cal K}$ need not to be invertible to provide a well defined Feynman graph expansion. All we need is to have a Gau\ss ian measure $\int[D\Phi]_{\cal K}$ with a well defined covariance ${\cal K}$ to perform the wick contractions, like for instance
\begin{equation}
\int[D\Phi]_{\cal K}\,\Phi(g_{1},\dots,g_{D})\Phi(g'_{1},\dots,g'_{D})=\sum_{\tau\atop\mbox{\tiny odd permutation}}{\cal K}(g_{1},\dots,g_{D};g'_{\tau(1)},\dots,g'_{\tau(D)})\label{gaussianmeasure}
\end{equation}
If ${\cal K}$ is invertible, it reduces to the standard Gau\ss ian measure $[D\Phi]_{\cal K}=[D\Phi]\exp\{-{\textstyle\frac{1}{2}}\Phi\cdot{\cal K}^{-1}\cdot\!\Phi^{\ast}\}$,

This allows to choose as a covariance the projector onto gauge invariant states
\begin{equation}
{\cal K}(g_{1},\dots,g_{D};g'_{1},\dots,g'_{D})=\int_{G}dh\,\prod_{i}\delta\big(g_{i}(g'_{i})^{-1}h\big)\label{invariantpropagator}
\end{equation}
This propagator usually leads to divergent Feynman graph amplitudes, because of the distributional nature of the Dirac distribution $\delta$. It is convenient to regulate the theory by replacing the Dirac distribution by the heat kernel on the group $G$ (which we assume to be compact for simplicity)
\begin{equation} 
{\cal K}_{\alpha}(g_{1},\dots,g_{D};g'_{1},\dots,g'_{D})=\int_{G}dh\,\prod_{i}\bigg\{\sum_{\rho} d_{\rho}\mathrm{e}^{-\alpha C_{\rho}}
\mbox{Tr}\,_{V_{\rho}}\big[g_{i}(g'_{i})^{-1}h\big]\bigg\}\label{heat}
\end{equation} 
where $\mbox{Tr}\,_{V_{\rho}}$ is the trace in the irreducible representation $\rho$, $d_{\rho}$ its dimension and $C_{\rho}$. For $G=\mathrm{SU(2)}$, $d_{j}=2j+1$ and $C_{j}=j(j+1)$.

The group field theory Feynman are stranded graphs made of edges consisting of $D$ strands. These strands represent the $(D\!-\!2)$-simplexes that carry curvature in Regge calculus. They define $(D\!-\!1)$-simplexes propagating along the edges of the stranded graph. These $(D\!-\!1)$-simplexes interact on the boundary of a $D$-simplex. Altogether, a stranded graph yields a prescription on how to glue together $D$-simplexes, as in tensor models, with the further information provided by the vertex amplitude. This leads to the following Feynman graph expansion
\begin{equation}
\int[D\Phi]_{\cal K}\exp V(\Phi)=\sum_{{\cal G}\,
\mbox{\tiny stranded graph}\atop\mbox{\tiny without external legs}}\frac{{\cal A}_{{\cal G}}}{C_{{\cal G}}}
\end{equation}
Because the graph does not carry external legs, it only involves closed strands. We construct  a 2-complex by gluing a disk to each of these closed strands. This way we obtain particular 2-complexes such that each vertex has $D+1$ incident edges and each edge has $D$ incident faces. These are  the 2-complexes corresponding to triangulations, i.e. in our context spaces obtained by gluing $D$-simplexes. The group field theory amplitude involves an integration over two variables $g^{+}_{v,f}$ and $g^{-}_{v,f}$ for each strand $f$ (strands represent $(D\!-\!2)$-simplexes and are in bijection with faces of the 2-complex) and vertex $v$ whose amplitude is nothing but the spin foam amplitude. It is equivalent to an integration over the variables $h_{v,f}$ with the constraint $\delta(\prod_{v\in\partial f}h_{v,f})$, for a normalized measure on a compact group. With the heat kernel propagator ${\cal K}_{\alpha}$, only gauge invariant states are involved in the functional integral. Then, the perturbative expansion of the group field theory partition function reads
\begin{equation}
\int[D\Phi]_{\cal K}\exp V(\Phi)=\sum_{
\mbox{\tiny Feynman graph }{}\Leftrightarrow\mbox{\tiny 2-complexes}}w_{\cal C}\,{\cal A}_{\cal C}\label{GFTSF}
\end{equation}
Thus, the group field theory perturbative expansion reproduces exactly a sum over the spin foam amplitudes  \eqref{SFamplitude}, with a combinatorial weight equal to the symmetry factor of the graph. Working with the heat kernel propagator we obtain regularized spin foam amplitudes.  

This is a discretized version of the quantum gravity path integral for  the partition function and corresponds to  a manifold without boundary. To deal with the quantum gravity wave function, one has to insert observables associated with some boundary state.

\subsection{Boundary states}

\label{boundarysec}
The case of 2-complexes with fixed  boundary $\Gamma$ is more involved since it does not immediately correspond to group field theory Green's functions defined as
\begin{equation}
\int[D_{{\cal K}}^{}\,\Phi]\exp V(\Phi)\,\prod_{n\in N}\Phi(g_{n,1},\dots,g_{n,D})=\sum_{{\cal G}\,
\mbox{\tiny stranded graph}\atop\mbox{\tiny with }N\,\mbox{\tiny external legs}}\frac{{\cal A}_{{\cal G}}(g_{n,a})}{C_{{\cal G}}}
\end{equation}
However, for any graph $\Gamma$ encoding the incidence relations of a triangulation of space, we associate the observable \cite{SFobservables}
\begin{equation}
{\cal O}_{\Gamma}[ \Phi](h_{l})=\prod_{l\,\mbox{\tiny link of }\Gamma}\delta(g_{s(l),l}h_{l}g^{-1}_{t(l),l})
\prod_{n\in N}\Phi(g_{n,l_{1}},\dots,g_{n,l_{D}})\label{observable}
\end{equation}
Note that the group variables $g_{n,l}$ correspond to the half-links of the graph $\Gamma$, or equivalently, to a pair formed by a node and a link attached to it.

\begin{figure}
\begin{equation}
\begin{minipage}{3cm}\includegraphics[width=4cm]{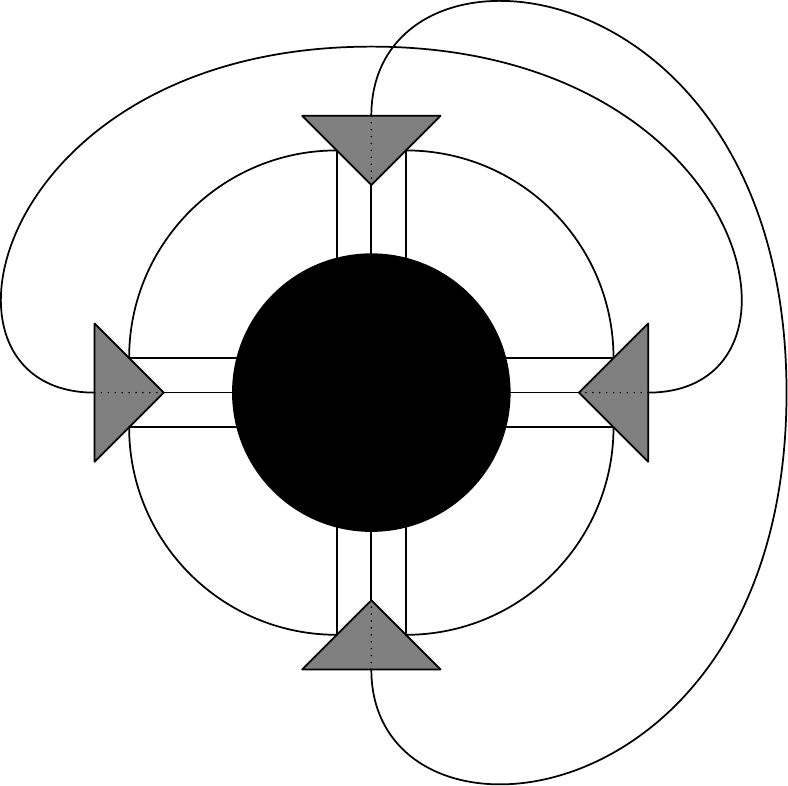}\end{minipage}
\end{equation}
\caption{A boundary graph (dashed) triangulating a sphere with 4 triangles}\label{boundarypic}
\end{figure}

Its normalized connected expectation value of is sum over all spin foams with fixed number of  boundary  $(D\!-\!1)$-simplexes. 
\begin{equation} 
\langle{\cal O}_{\Gamma}[\Phi](h_{l})\rangle_{c}=\sum_{T\, \mbox{\tiny possibly disconnected triangulations with }\atop n\,\mbox{\tiny boundary ({\it D }-1)-simplexes }}\frac{{\cal A}_{T/\Gamma}(h_{l})}{C_{T}}\label{observableexp}
\end{equation}
The subscript $\langle\cdots\rangle_{c}$ refers to the connected expectation value. For instance, if $\Gamma$ is connected $\langle{\cal O}_{\Gamma}\rangle_{c}=\langle{\cal O}_{\Gamma}\rangle$ and if $\Gamma=\Gamma_{1}\cup\Gamma_{2}$ has two connected components,  $\langle{\cal O}_{\Gamma}\rangle_{c}=\langle{\cal O}_{\Gamma_{1}}{\cal O}_{\Gamma_{2}}\rangle-\langle{\cal O}_{\Gamma_{1}}\rangle\langle{\cal O}_{\Gamma_{2}}\rangle$. This is similar to the matrix model loop functions \eqref{matrixmodelboundary}. The sum runs over possibly disconnected triangulations in order to encompass processes like the one illustrated by figure \ref{sfpic}. The insertion of the boundary graph imposes new relations between the boundary $(D\!-\!1)$-simplexes which leads to the new triangulation $T/\Gamma$. Taking the connected expectation value imposes that the latter be connected.

\begin{figure}
\begin{equation}
\begin{minipage}{3cm}\includegraphics[width=3cm]{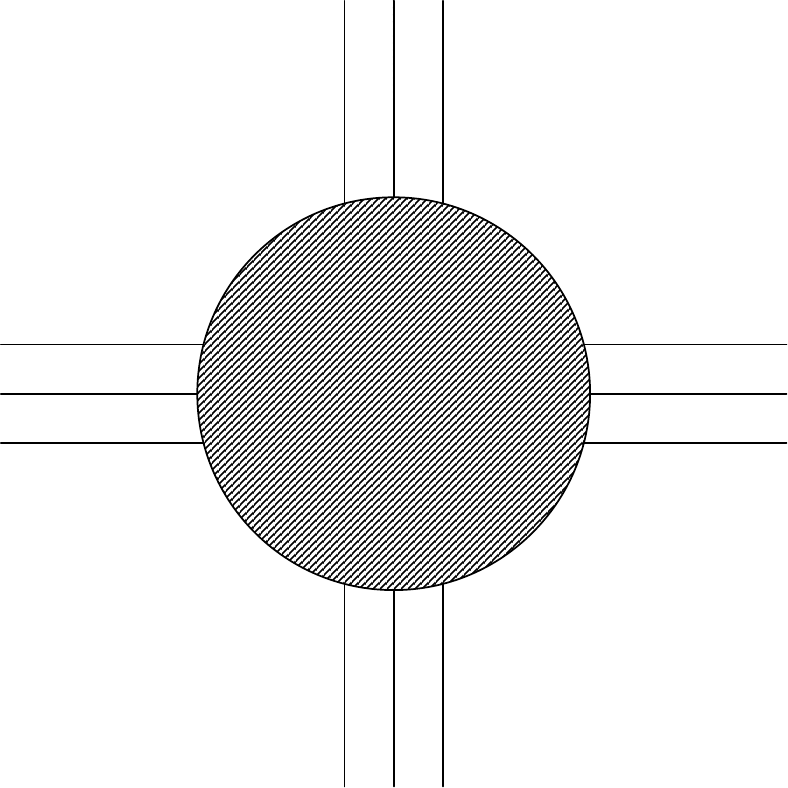}\end{minipage}
=
\begin{minipage}{3cm}\includegraphics[width=3cm]{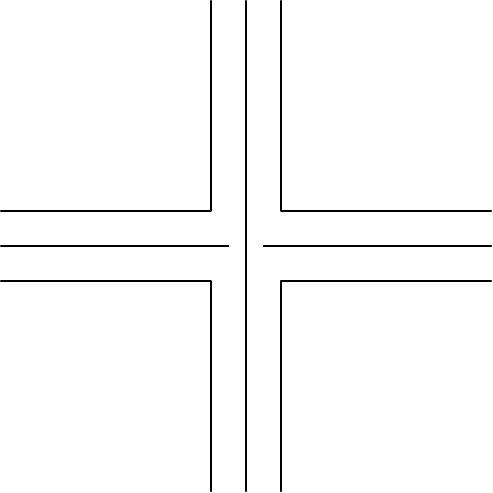}\end{minipage}
+\cdots+
\begin{minipage}{3cm}\includegraphics[width=3cm]{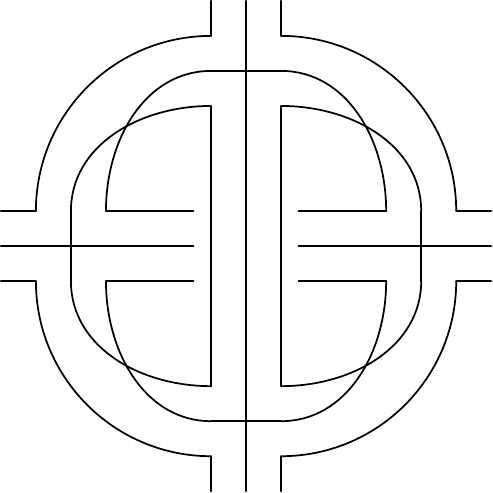}\end{minipage}
+\cdots
\quad
\nonumber
\end{equation}
\caption{Contributions to the four point function}
\label{4pointboundary}.
\end{figure}

\begin{figure}
\begin{equation}
\begin{minipage}{3cm}\includegraphics[width=3cm]{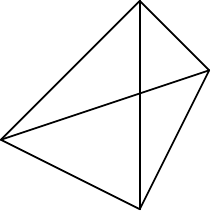}\end{minipage}
\qquad\qquad
\begin{minipage}{3cm}\includegraphics[width=3cm]{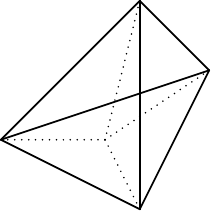}\end{minipage}
\nonumber
\end{equation}
\caption{The 1-4 Pachner move}
\label{Pachner}
\end{figure}

The expectation value of the boundary graph depicted in figure \ref{boundarypic} involves the four point function, as illustrated in figure \ref{4pointboundary}. 
The two terms in this equation correspond to a single tetrahedron and to a tetrahedron split into four by adding an extra point (1-4 Pachner move, see figure  \ref{Pachner}).

Let us explain what we mean by the quotient $T/\Gamma$. Each group field theory stranded graph defines a  procedure to glue together $D$-simplexes. The external legs of the graph define the boundary $(D\!-\!1)$-simplexes that share $(D\!-\!2)$-simplexes given by the open strands. These open strands define a boundary graph whose nodes are the external legs which are linked if they are related by an external strand.
Then, the observable associated to the graph $\Gamma$ enforces an identification of the boundary $(D\!-\!1)$-simplexes following the incidence relations encoded in $\Gamma$. These may impose further relations with respect to the ones already present in $\partial T$. The resulting triangulation defines $T/\Gamma$. Recall that we have identified a triangulation with a stranded graph so that the sum in \eqref{observableexp} is simply a sum over stranded graphs with $n$ external legs. 

This also occurs in the case of matrix models \cite{quantumgeometry}. In this case, the observables are product of traces of powers of the matrix and correspond to boundary graphs which are disjoint unions of circles,
\begin{align}
{\cal O}[M]=\prod_{k}\mbox{Tr}\,[M^{n_{k}}]
\end{align} 
The perturbative expansion reproduces the Euler characteristics of a triangulation with boundary $\chi=f-e+v=2-2h-b$, with $b$ the number of boundaries. 

Let us illustrate this phenomenon in the simple case of the expectation value of $\mbox{Tr}\,[M^{3}]$, which represents a circle made of three arcs. At lowest order in $\lambda$, it reads
\begin{equation}
\langle\mbox{Tr}\,[M^{3}]\rangle=\langle\mbox{Tr}\,[M^{3}]\,\frac{\lambda}{3}\mbox{Tr}\,[M^{3}]\rangle_{0}+O(\lambda^{3})
\end{equation}
with $\langle\cdots\rangle_{0}$ the expectation value with respect to the Gau\ss ian measure on matrices \eqref{Wickmatrices}.  There are two different ways to perform the Wick contraction: one of them yields a factor of $N$ and corresponds to a triangle bounding the circle while the second one evaluates to $N^{-1}$ and is obtained from the first one by further identifying all three points on the triangle.

Note that because the field  $\Phi(g_{n,1},\dots,g_{n,D})$ is only invariant under even permutations of the group elements, the ordering of the half-edges at a node matters up to an even permutation. In the case $D=3$, even permutations are cyclic permutations and the boundary graph is a ribbon graph which is the 1-skeleton dual to a triangulation of a surface.

\section{Group field formulation of $BF$ theory and quantum gravity}

\subsection{Discretization of BF theories}

\label{BF}

On a space-time manifold ${\cal M}$ of dimension $D$, $BF$ theory is a topological field theory involving a non abelian gauge field $A$ with curvature  $F=dA+ A\wedge A$  and a $(D\!-\!2)$-form $B$ with values in the Lie algebra of the gauge group. Its action is $S[A,B]=\int_{\cal M}\mbox{Tr}\,(B\wedge F)$, where the trace stands is taken is the adjoint representation. Its relevance to quantum gravity is twofold. First, three dimensional quantum gravity can be written as a $BF$ theory. In the euclidian setting, the gauge group is $SU(2)$, the connection is the spin connection and $B$ is the dreibein $e$. Second, quantum gravity in dimension four can be seen as a constraint $BF$ theory, as briefly explained in section \ref{EPRLFK}.

At a formal path integral level, $B$ is nothing but a Lagrange multiplier enforcing the flatness of the connection,
\begin{equation}
{\cal Z}_{{\cal M}}=\int [{\cal D}A][{\cal D}B]\,\mathrm{e}^{\mathrm{i}\int_{\cal M}\mathrm{Tr}(B\wedge F)}=
\int [{\cal D}A]\delta(F)
\end{equation}
On a manifold with a boundary, the path integral yields a wave function $\psi_{\partial{\cal M}}(A)$ which depends on the values of the gauge field $A$ on the boundary. It is not to surprising that $\psi_{\partial{\cal M}}$ only depends on $A$: $B$ is canonically conjugate to $A$ and it is known from quantum mechanics that boundary conditions in the path integral are imposed only for configuration variables, not momentum ones.

To actually define the path integral we use a discretization involving spin foams \cite{Baez}. Let us triangulate the manifold and let ${\cal C}$ be the 2-skeleton of the dual of the triangulation. Since $B$ is a $(D-2)$-form, it is naturally associated to the faces of ${\cal C}$. In this setting, the connection is represented as group variables $h_{e}$ associated to the edges of ${\cal C}$ and the flatness condition states that the ordered product of the group elements around each face equals the identity. Therefore, the partition function reads
\begin{equation}
{\cal Z}_{{\cal C}}=\int_{\mathrm{SU(2)}} dh_{e}\,\prod_{f\in F}
\delta\Big(\mathop{\prod}\limits^{\rightarrow}_{e\in\partial f}h_{e}^{\epsilon_{e,f}}\Big)\label{discretizedBF}
\end{equation}  
where the incidence matrix $\epsilon_{e,f}=\pm1$ depending on the relative orientation of the face $f$ and the edge $e$. The ordered product is taken around the boundary of $f$. To write the amplitude we have chosen an orientation of  the edges and faces of ${\cal C}$ but the amplitude does not depend on these choices.  Let us note that this expression is still formal since it involves products of Dirac distributions. It may be regularized by replacing the latter by heat-kernels, as in   \eqref{heat}. It is interesting to note that in dimension two, the heat kernel regularized amplitude in nothing but the discretization of Yang-Mills theory, with the Wilson action replaced by the heat kernel one. In three dimensions,  group variables are holonomies of the spin connection and the flatness condition is equivalent to Einstein's equation for pure gravity. The standard formutation of Regge's theory in terms of $6j$ symbols is recovered by expanding the Dirac distributions on characters 
\begin{equation}
\delta(g)=\sum_{j}d_{j}\mbox{Tr}\,_{V_{j}}(g)\label{delta}
\end{equation}
with $V_{j}$ the spin $j$ representation of SU(2) of dimension $d_{j}=2j+1$. Then, an integration over group variables yields the $6j$ symbols.

Finally, note that there is an analogous discretization on manifolds with boundary, with extra variables on the links of the boundary graph $\Gamma=\partial{\cal C}$. In this case, we insert group elements associated to the links on the boundary of the faces in \eqref{discretizedBF}.


\subsection{Group field theory formulation}

According to the general construction describded in section \ref{GFT}, the group field theory that generates $BF$ amplitudes follows from the general form of the vertex amplitude, which is the amplitude for a single $(D\!+\!1)$-simplex. Let us label $a,b\in\{1,\dots,D\}$ the points of this $(D\!+\!1)$-simplex. We also label by $a$ the $(D\!-\!1)$-simplex obtained by removing $a$ and by $ab$ with $a<b$ the $(D\!-\!2)$-simplex obtained by removing $a$ and $b$. Thus, the boundary graph is made is equipped with link variables $h_{ab}$. The corresponding vertex amplitude reads
\begin{equation}
{\cal A}_{v}(h_{ab})=\int_{G}\prod_{a}dg_{a}\,\prod_{0\leq a<b\leq D}\delta\big(h^{}_{ab}g^{}_{a}g_{b}^{-1}\big)\label{vertexBF}
\end{equation}
We further introduce $D(D+1)$ variables $g_{ab}$ with $a\neq b$ such that  $h_{ab}=g^{}_{ba}g_{ab}^{-1}$, so that the vertex amplitude reads
\begin{equation}
{\cal A}_{v}(g_{ab})=\int_{G}\prod_{a}dg_{a}\,\prod_{a\neq b}\delta\big(g^{}_{ba}g_{ab}^{-1}g^{}_{a}g_{b}^{-1}\big)\label{BFvertex}
\end{equation}

Then, the general discussion of section \ref{GFT} applies and the interaction potential reads
\begin{equation}
V(\Phi)=\frac{\lambda}{D+1}
\int_{G}\prod_{a}dg_{a}\,\prod_{a\neq b}\delta\big(g^{}_{ba}g_{ab}^{-1}g^{}_{a}g_{b}^{-1}\big)
\prod_{c}\Phi^{\ast^{c-1}}(g_{c,d\neq c})\label{BFvertex}
\end{equation}
Note that because of $\delta(g)=\delta(g^{-1})$, this interaction does not depend on the various orientation choice we have made, encoded in the variables $h_{ab}=g^{}_{ba}g_{ab}^{-1}$, with $a<b$. The quadratic term may be chosen to be the trivial one, the projector onto invariant states or its regularized version. If we choose the projector onto the invariant states, given by the propagator  \eqref{invariantpropagator}, the variables $g_{a}$ in the vertex are redundant since they also project onto invariant states. Dropping these variables from the vertex yields the most concise form of the interaction that is used in actual computations. However, these variables also ensure manifest gauge invariance at the level of the interaction. 

To illustrate our discussion, let us work out the three and four dimensional cases. First, in dimension three, the potential reads
\begin{multline}
V(\Phi)=\frac{\lambda}{4}
\int_{G}\prod_{1\leq a\leq 4}dg_{a}\,\prod_{1\leq a\neq b\leq 4}dg_{ab}\,\\
\Big\{\delta\big(g^{}_{12}g_{21}^{-1}g^{}_{1}g_{2}^{-1}\big)
\delta\big(g^{}_{13}g_{31}^{-1}g^{}_{1}g_{3}^{-1}\big)
\delta\big(g^{}_{14}g_{41}^{-1}g^{}_{1}g_{4}^{-1}\big)
\delta\big(g^{}_{23}g_{32}^{-1}g^{}_{2}g_{3}^{-1}\big)
\delta\big(g^{}_{24}g_{42}^{-1}g^{}_{2}g_{4}^{-1}\big)
\delta\big(gv_{34}g_{42}^{-1}g^{}_{3}g_{4}^{-1}\big)\\
\Phi(g_{12},g_{13},g_{14})\Phi^{\ast}(g_{21},g_{23},g_{24})
\Phi(g_{31},g_{32},g_{34})
\Phi^{\ast}(g_{41},g_{42},g_{43})\Big\}
\end{multline}
If we use the projector onto the invariant states as a propagator, we can drop the variables $g_{a}$. Eliminating the variables $g_{ab}$ with $b<a$, we obtain the conventional form of the interaction, as initially proposed by Boulatov \cite{Boulatov}.
\begin{equation}
V(\Phi)=\frac{\lambda}{4}
\int_{G}\prod_{0\leq a<b\leq 3}dg_{ab}\,
\Phi(g_{01},g_{02},g_{03})\Phi^{\ast}(g_{03},g_{13},g_{23})
\Phi(g_{23},g_{02},g_{12})
\Phi^{\ast}(g_{12},g_{13},g_{01})
\end{equation}
In dimension four, the very same analysis leads to
\begin{multline}
V(\Phi)=\frac{\lambda}{5}
\int_{G}\prod_{0\leq a<b\leq 4}dg_{ab}\,
\Phi(g_{01},g_{02},g_{03},g_{04})
\Phi(g_{40},g_{14},g_{24},g_{34})
\Phi(g_{34},g_{03},g_{13},g_{23})\cr
\Phi(g_{23},g_{24},g_{02},g_{12})
\Phi(g_{12},g_{13},g_{14},g_{01})
\end{multline}
This is the interaction introduced by Ooguri \cite{Ooguri}.

The perturbative expansion of the $BF$ group field theory generates the spin foam amplitudes given in \eqref{GFTSF}. At a formal level, the integral over the variables $g_{a,b}$ can be performed, leaving an integral over the variables $g_{a}$ with a product of a Dirac distribution along every closed strand. In each $D$-simplex, there are $D+1$  variables $g_{a}$ attached to its $(D-1)$ simplexes, so that in the whole 2-complex we relabel them as $g_{a}\rightarrow g_{e,v}$. Then, the discretized $BF$ amplitude  \eqref{discretizedBF} is recovered from  \eqref{SFamplitude} after the change of variables $h_{e}=g^{}_{s(e),e}g_{t(e),e}^{-1}$.

Choosing the projector onto invariant states as a propagator, it is also possible to make contact with the standard expression in terms of spins and intertwiners. To  simplify the notations, let us restrict our analysis to the case $G=SU(2)$. We expand the field using the Peter-Weyl theorem as
\begin{equation}
\Phi(g_{1},\dots,g_{D})=\sum_{j_{1},\dots,j_{D}\atop m_{1},\dots,m_{D},m'_{1},\dots,m'_{D}}
\Phi^{j_{1},\dots,j_{D}}_{ m_{1},\dots,m_{D},m'_{1},\dots,m'_{D}}{\cal D}_{m_{1},m'_{1}}^{j_{1}}(g_{1})\cdots{\cal  D}_{m_{D},m'_{D}}^{j_{D}}(g_{D})
\end{equation}
with ${\cal D}_{m,n}^{j}(g)$ the Wigner matrices. The projector onto the invariant states enforces the right invariance of the field, $\Phi(g_{1},\dots,g_{D})=\int dg\,\Phi(g_{1}g,\dots,g_{D}g)$. Using the integration formula
\begin{equation}
\int dg\,{\cal D}_{m'_{1},m''_{1}}^{j_{1}}(g_{1})\cdots{\cal  D}_{m'_{D},m''_{D}}^{j_{D}}(g_{D})
=\sum_{i\,\,\mbox{\tiny intertwiner between}\atop j_{1},\dots,j_{D}}
i_{m_{1}',\dots,m'_{D}}i^{\ast}_{m''_{1},\dot,m''_{D}}
\end{equation}
with $i_{m_{1},\dots,m_{D}}$ an orthonormal basis of invariant tensors in $V_{j_{1}}\otimes\dots\otimes V_{j_{D}}$, the invariance of the field translates into an expansion
\begin{equation}
\Phi(g_{1},\dots,g_{D})=\sum_{j_{1},\dots,j_{D},m_{1},\dots,m_{D}\atop\mbox{\,\tiny i intertwiner between } j_{1},\dots,j_{D}}
M_{m_{1},\dots,m_{D}}^{j_{1},\dots,j_{D};\,i}\,
{\cal D}_{m_{1},m'_{1}}^{j_{1}}(g_{1})\cdots{\cal  D}_{m_{D},m'_{D}}^{j_{D}}(g_{D})\,i^{\ast}_{m'_{1},\dots,m'_{D}}
\label{expansionM}
\end{equation}
The basic variables are now the complex numbers
\begin{equation}
M_{m_{1},\dots,m_{D}}^{j_{1},\dots,j_{D};\,i}=\sum_{m'_{1},\dots,m'_{D}}
\Phi^{j_{1},\dots,j_{D}}_{ m_{1},\dots,m_{D},m'_{1},\dots,m'_{D}}i_{m'_{1},\dots,m'_{D}}\label{intertwiner}
\end{equation}
These numbers obey additional constraints arising form the reality conditions  \eqref{realityphi}. Within this framework, the propagator is trivial and the interaction can be expressed in terms of Wigner's coefficients. This formalism leads to rather complicated calculations but has two important advantages.

\begin{figure}
\centerline{\includegraphics[width=5cm]{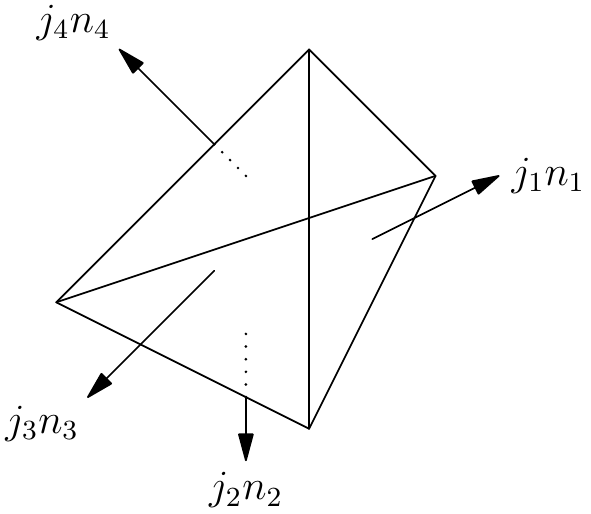}}
\caption{Coherent states and normals to the tetrahedron}
\label{normals}
\end{figure}

First, it has a clear geometrical interpretation in relation to the quantum triangle and tetrahedron \cite{quantumtetrahedron} . In dimension three, there is an intertwiner between $j_{1}$, $j_{2}$ and $j_{3}$ if and only if they obey the triangular inequality $|j_{1}-j_{2}|\leq j_{3}\leq j_{1}+j_{2}$ and when the interwiner exists, it is unique. Therefore spins  faithfully represent the lengths of the sides of a triangle. A similar interpretation holds in dimension four, where the four spins  $j_{1},\dots,j_{4}$ represent the areas of the faces of a quantum tetrahedron and the intertwiner $i$, which may be chosen to be a fifth spin obtained by splitting the four valent graph into two three valent ones, is the area of a parallelogram inside the tetrahedron .  Thus, the five spin indices describe a quantum tetrahedron while the magnetic indices $m_{1},\dots,m_{4}$ just serve to create the 2-complex and count its faces.  

Coherent states provide a more geometrical description of the quantum tetrahedron as follows. Intertwiners label a basis of invariant tensors in the tensor product ${\cal H}^{j_{1}}\otimes\cdots\otimes {\cal H}_{j_{4}}$. An arbitrary element of this tensor product can be expanded over the coherent states (see appendix \ref{coherent}) $|j_{1},n_{1}\rangle\otimes\cdots\otimes|j_{4},n_{4}\rangle$. This state is invariant if and only if the coherent states and the spins obey the closure condition
\begin{equation}
j_{1}n_{1}+j_{2}n_{2}+j_{3}n_{3}+j_{4}n_{4}=0
\end{equation}
Then, each unit vector can be interpreted as a normal to a face of the tetrahedron, with $j_{i}$ the area of the face $i$ in Planck units. In fact, four vectors obeying the closure condition are the normals of a tetrahedron, defined up to rotation and inversion. This can be checked by a simple count of degrees of freedom $4\times 3-3-3= 6$. The first term stands for the four vectors $j_{i}n_{i}$, to which we subtract $3$ degrees of freedom for the closure constraint and three other degrees for freedom because of rotational invariance. This yields six degrees of freedom corresponding to the lengths of the edges of the tetrahedron, which specify the geometry up to rotation and inversion.


Second, it can be generalized to to the quantum group $U_{q}(\mathrm{SU(2)})$, with $q$ a root of unity. Quantum groups allows us to treat $BF$ theory with a cosmological constant in dimensions three and four.   All explicit formul\ae {} for intertwiners generalize to the $q$-deformed case, provided integers are replaced by $q$-integers $[n]_{q}=\frac{q^{n}-q^{-n}}{q-q^{-1}}$.  The corresponding amplitude can then be reproduced as a tensor model using the sum over spins, as presented by Boulatov in dimension three \cite{Boulatov} and Ooguri in dimension four \cite{Ooguri}. 

Apart form the group and spin formulations, there is a third formulation based on the group Fourier transform \cite{OritiBaratin}. First, we define the analogues of plane waves as
\begin{equation}
e_{g}(x)=\exp\mathrm{i}\mbox{Tr}\,(xg)
\end{equation}
with $x\in\frak{su}(2)\simeq{\Bbb R}^{3}$ a Lie algebra element and $g\in\mathrm{SU}(2)$, considered as $2\times2$ matrices. Plane waves are multiplied using the non commutative (but associative) product
\begin{equation}
e_{g}\ast e_{g'}(x)=e_{gg'}(x)\label{planewaveproduct}
\end{equation}
It is convenient to introduce the group Fourier transform
\begin{equation}
\hat{f}(x)=\int_{\mathrm{SU}(2)} dg\, f(g)e_{g}(x)
\end{equation}
and extend the plane wave product  \eqref{planewaveproduct} by linearity to all functions on ${\Bbb R}^{3}$. Furthermore, 
\begin{equation}
\delta_{x}(y)=\int_{\mathrm{SU}(2)} dg\, e_{g^{-1}}(x)e_{g}(y)
\end{equation}
play a role analogous to Dirac distributions for the deformed product
\begin{equation}
\int_{{\Bbb R}^{3}} dy\,\delta_{x}\ast \hat{f}(y)=\int_{{\Bbb R}^{3}} dy\,\hat{f}\ast\delta_{x}(y)=
\hat{f}(x)
\end{equation}
Then, it is possible to rewrite entirely the group field theory in terms of the Fourier transformed field
\begin{equation}
\hat{\Phi}(x_{1},\dots,x_{D})=\int_{\mathrm{SU(2)}^{D}}dg_{1}\cdots dg_{D}\,
\Phi(g_{1},\dots,g_{D})\,e_{g_{1}}(x_{1})\cdots e_{g_{D}}(x_{D})
\end{equation}
The invariance property of the field  \eqref{invariant} translates into
\begin{equation}
\hat{\Phi}(x_{1},\dots,x_{D})=\hat{C}\ast\hat{\Phi}(x_{1},\dots,x_{D})
\quad\mbox{with}\quad\hat{C}(x_{1},\dots,x_{D})=\delta_{0}(x_{1}+\cdots+x_{D})
\end{equation}
Thus, in dimension three (resp. in dimension four), the invariance property is related to a non commutative closure condition on the normals to the edges of a triangle (resp. to the faces of a tetrahedron). This formulation is especially handy when discussing symmetries \cite{Diff} and likely to shed a new light on the coupling to matter since matter fields are described by a non commutative field theory involving precisely the deformed product  \eqref{planewaveproduct}, as shown in \cite{matterstar}.

\subsection{Analysis of the divergences}

\begin{figure}
\centerline{\includegraphics[width=6cm]{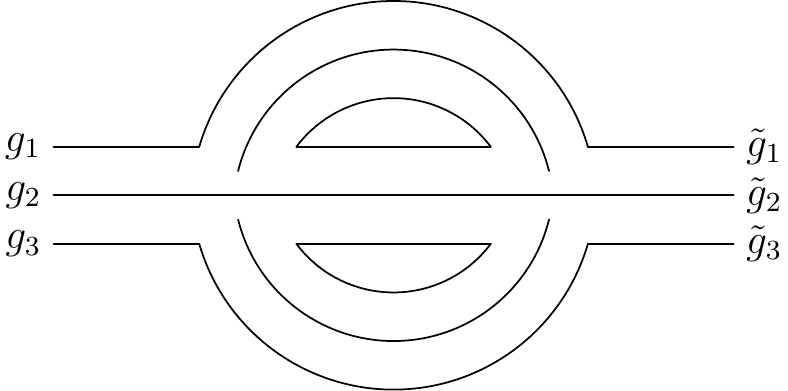}}
\caption{First correction to the 2-point function}
\label{2point}
\end{figure}

The group field theory amplitude are in general divergent, as is illustrated in the following section. For notational simplicity, let us focus on the three dimensional case, higher dimensional cases can be treated similarly. We begin by computing the first correction to the two point function, given by the graph depicted in figure \ref{2point}.

\begin{figure}
\centerline{\includegraphics[width=10cm]{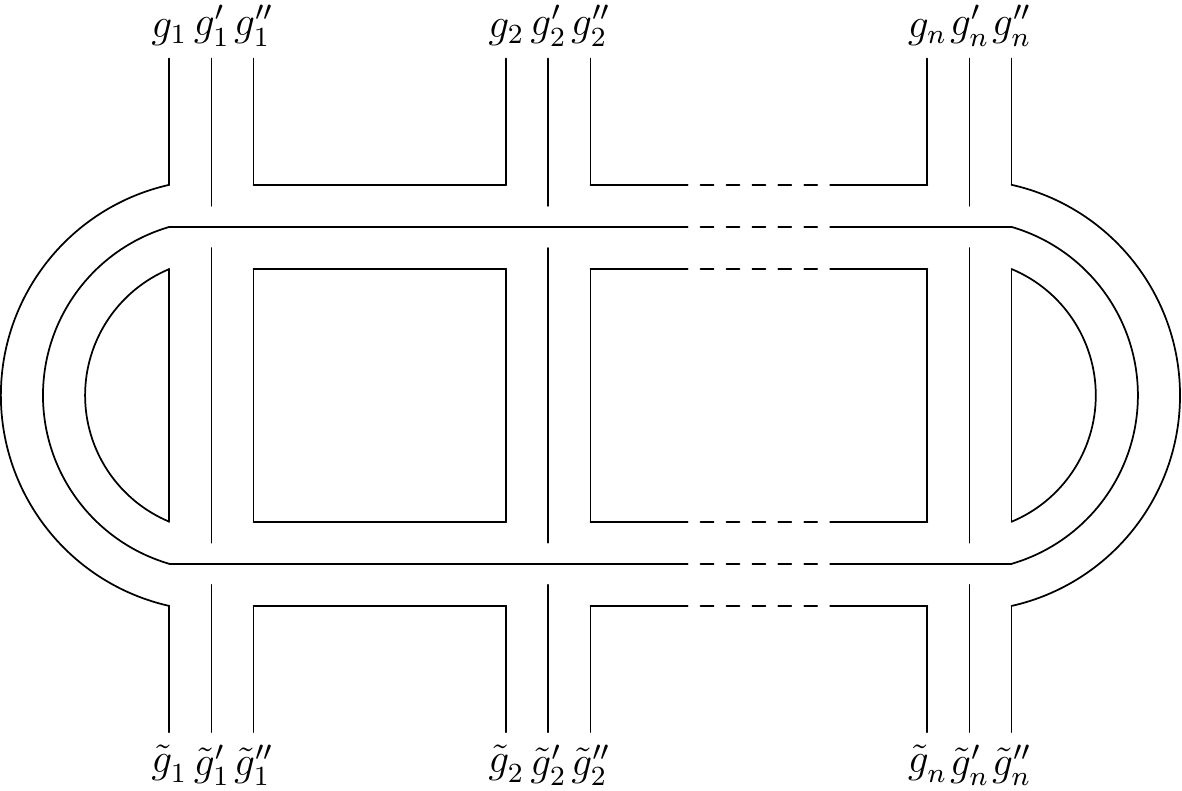}}
\caption{A divergent contribution to the $n$-point function }
\label{npoint}
\end{figure}

The propagators on the external legs are projectors that do not contribute to the amplitude because they can be absorbed by the projectors on the internal lines. Then the amplitude reads
\begin{equation}
\lambda^{2}\int_{\mathrm{SU(2)}}dh_{1}dh_{2}dh_{3}\,\delta(g_{1}\tilde{g}_{1}^{-1}h_{1})\delta(g_{2}\tilde{g}_{2}^{-1}h_{2})\delta(g_{3}\tilde{g}_{3}^{-1})\delta(h_{1}h_{2}^{-1}h_{3})\delta(h_{2}h_{3}^{-1})\delta(h_{3}h_{1}^{-1})
\end{equation}
Two of the Dirac distributions enforce the condition $h_{1}=h_{2}=h_{3}$ while the remaining one is redundant, so that the amplitude is
\begin{equation}
\delta(1)\int_{\mathrm{SU(2)}}dh\,\delta(g_{1}\tilde{g}_{1}^{-1}h)\delta(g_{2}\tilde{g}_{2}^{-1}h)\delta(g_{3}\tilde{g}_{3}^{-1}h)
\end{equation}
The prefactor is divergent and can be regularized either using the heat kernel instead of the the Dirac distribution or by restricting the sum to spin below a cut-off $N$. In this last case, $\delta(1)$ is replaced by
\begin{equation}
\lambda^{2}\delta_{N}(1)=\sum_{j\leq N}(2j+1)^{2}\mathop{\sim}\limits_{N\rightarrow\infty} {\textstyle \frac{4}{3}}N^{3}
\end{equation}
Therefore, in the large $N$ limit the amplitude is equivalent to
\begin{equation}
{\textstyle \frac{4}{3}}\lambda^{2}N^{3}\int_{\mathrm{SU(2)}}dh\,\delta(g_{1}\tilde{g}_{1}^{-1}h)\delta(g_{2}\tilde{g}_{2}^{-1}h)\delta(g_{3}\tilde{g}_{3}^{-1}h)
\end{equation}
This is encouraging since it is nothing but the propagator multiplied by a diverging constant so that it can be easily renormalized. By the same token, the graph corresponding to the 1-4 Pachner move given in picture \ref{4pointboundary} evaluates to the vertex multiplied by a diverging factor $\delta(1)$. These two example seem to indicate that the $BF$ group field theory is renormalizable, since the divergences are proportional to the terms already present in the action. However, the correction to the $2n$ point function given in figure \ref{npoint} is divergent for all $n$ and requires its own counterterm. This spoils renormalization in the naive sense, but we shall see in section \ref{coloredsec} that renormalizable group field theories exist, provided a suitable propagator is used. 

Applying the Feynman rules, the amplitude can be expressed as
\begin{multline}
\lambda^{2n}\int_{\mathrm{SU(2)}}\prod_{0\leq i\leq n+1}dh_{i}\quad
\delta\big[h_{0}h_{1}^{-1}\big]
\delta\big[g_{1}\tilde{g}_{1}^{-1}h_{0}\big]
\delta\big[h_{n}h_{n+1}^{-1}\big]
\delta\big[g_{n}''(\tilde{g}''_{n})^{-1}h_{n+1}\big]
\delta\big[h_{0}h''_{1}\cdots h''_{n}h_{n+1}^{-1}(h'_{n})^{-1}\cdots (h_{1}')^{-1}\big]\\
\prod_{1\leq i\leq n-1}
\bigg\{\delta\big[h_{i}h''_{i}h_{i+1}^{-1}(h'_{i})^{-1}\big]
\delta\big[g_{i}''(g''_{i+1})^{-1}h_{i}'\big]\delta\big[\tilde{g}_{i}(\tilde{g}_{i+1})^{-1}h_{i}''\big]\bigg\}
\prod_{1\leq i\leq n}\delta\big[g'_{i}(\tilde{g}'_{i})^{-1}h_{i}\big]
\end{multline}
We use $\delta(h_{0}h_{1}^{-1})$ and $\delta(h_{n}h_{n+1}^{-1})$ to set $h_{0}=h_{1}$ and $h_{n+1}=h_{n}$. Then $h_{0}h''_{1}\cdots h''_{n}h_{n+1}^{-1}(h'_{n})^{-1}\cdots (h_{1}')^{-1}=1$, using the $n-1$ relations $h_{i}h''_{i}h_{i+1}^{-1}(h'_{i})^{-1}=1$. Consequently the graph diverges as
\begin{multline}
\lambda^{2n}\delta(1)\int_{\mathrm{SU(2)}}\prod_{1\leq i\leq n}\!\!\!dh_{i}\,\,
\delta\big[g_{1}\tilde{g}_{1}^{-1}h_{1}\big]
\delta\big[g_{n}''(\tilde{g}''_{n})^{-1}h_{n}\big]\cr
\prod_{1\leq i\leq n-1}
\big\{\delta\big[h_{i}h''_{i}h_{i+1}^{-1}(h'_{i})^{-1}\big]
\delta\big[g_{i}''(g''_{i+1})^{-1}h_{i}'\big]\delta\big[\tilde{g}_{i}(\tilde{g}_{i+1})^{-1}h_{i}''\big]\big\}
\!\!\!\prod_{1\leq i\leq n}\delta\big[g'_{i}(\tilde{g}'_{i})^{-1}h_{i}\big]
\end{multline}
Moreover, it can be easily seen that this graph has no divergent subgraph. Accordingly, it should come with its own counterterm, which is not initially present in the Lagrangian for $n>2$. This means that the Boulatov model in $D=3$ is not renormalizable in the naive sense of the term. A similar conclusion has been obtained by analyzing higher order corrections  to the two point function which require a Laplacian like propagator  \cite{JosephValentin}. The non renormalizabilty of the group field formulation of $BF$ theory reveals its incompleteness and can be interpreted in several ways.

First, one may argue that the divergences are solely due to the topological nature of the theory. Such a theory has no local degrees of freedom and should be invariant under refinements of the triangulation, such as Pachner moves. A simple example of a Pachner move is the $1\rightarrow 4$ move where a tetrahedron is split in four tetrahedra by adding an extra point. Any Pachner move introduces new edges in the graph, each equipped with a group element $h_{e}$. Its also creates new closed strands with Dirac distributions which enforce  $h_{e}=1$. However, these Dirac distributions are not all independent and give rise to a  power of the divergence $\delta(1)$. Thus, divergences may be traced back to topological invariance and we may expect a theory of quantum gravity in four dimensions, which has local degrees of freedom, to be free of many of these divergences. Unfortunately, a preliminary analysis shows that this is not the case.

It is also possible to follow the standard interpretation of non renormalizable theories in physics as effective theories with a physical cut-off. Such a cut-off is provided by the cosmological constant $\Lambda$ which amounts to the replacement of the group $G$ by a quantum group at roots of unity. In that case only a finite number of representations (for SU(2) those with $j\leq \Lambda^{-1/2}$, if $\Lambda>0$ is expressed using Planck units) come into play so that there is no divergence. Moreover, $BF$ theory with a cosmological constant can be shown to be invariant under refinement of the triangulation, thus leading to a well defined topological field theory. Four dimensional quantum gravity models with a non zero cosmological constant are also finite though they exhibit local degrees of freedom, see \cite{Muxin} and \cite{Winston}. Last but not least, a non vanishing cosmological constant is also required on experimental grounds by recent cosmological observations of supernov\ae.  However, the upper bound on the cosmological constant is roughly 120 orders of magnitude smaller than the natural scale of quantum gravity. It is then necessary to be able to master the limit $\Lambda\rightarrow0$, or equivalently of large cut-off on the spins, in the quantum gravity regime.

These divergences can also be seen as the need for a modification of the propagator, which we have taken to be the trivial one or the projector onto gauge invariant states. This propagator enforces a the gluing of two $D-1$ simplexes (triangles in $D=3$ and tetrahedra in $D=4$) that belong to neighboring $D$ simplexes. The situation is similar to that of an ordinary quantum field theory with a propagator $C(p,q)=\delta(p+q)$ which solely enforces momentum conservation instead of 
\begin{equation}
C(p,q)=\frac{\mathrm{i}\delta(p+q)}{p^{2}-m^{2}+\mathrm{i}\epsilon}
\end{equation}
as is usually the case in quantum field theory. In this case, the internal edge of the graph represent virtual particles and the numerator $p^{2}-m^{2}+\mathrm{i}\epsilon$ suppresses virtual particles far away from the mass $p^{2}-m^{2}=0$, while external legs carry real particles on the mass shell. In the group field theory context, edges of the graph can be considered as propagators for virtual $D-1$ simplexes and it is sensible to modify the propagator in such a way that $D-1$ simplexes that differ significantly form the real ones on the external legs are suppressed. In its euclidian version, the quantum field theory propagator
\begin{equation}
C(p,q)=\frac{\delta(p+q)}{p^{2}+m^{2}}=\delta(p+q)\int_{0}^{\infty}d\alpha\,\mathrm{e}^{-\alpha(p^{2}+m^{2})}
\end{equation}
allows for a proper identification of the various scales of the theory, as required by Wilsods formulation of renormalization theory. In this framework, effects of unobserved fluctuations of very high energy, or equivalently, very small scales,  are encoded in a low energy or long distance effective action. 
In the context of group field theory  such a propagator should lead to a series of effective actions obtained one from another by successive integration over unobserved geometries, closer and closer to the observed one described by the external legs of the graphs.  

The two simple graphs we have worked out in detail show that the divergences are due to redundancies in the Dirac distributions enforcing the flatness condition. Each closed strand yields such a Dirac distribution and the redundancy appear whenever the discs bounding the closed strand form a closed surface called a bubble \cite{PerezRovelli2}. This intuitive idea has been made precise in \cite{MatteoValentin2} where divergences are associated with the second Betti number of a twisted cohomology introduced in \cite{BarrettNaish}.  Roughly speaking, this goes a follows. The basic variables are SU(2) elements $h_{e}$ associated to the edges. Then the amplitude on a fixed 2-complex ${\cal C}$ given in  \eqref{discretizedBF} is regularized using the heat kernel and  rewritten as
\begin{equation}
{\cal Z}({\cal C})=\int _{\mathrm{SU(2)}}\prod_{e}dh_{e}\,\exp-S_{\alpha}[h_{e}]
\end{equation}
with $\alpha$ the heat kernel regularizing parameter and $S_{\alpha}$ an action functional. Then. in the limit $\alpha\rightarrow 0$, we have ${\cal Z}\simeq \alpha^{-\omega({\cal C})}$ with $\omega({\cal C})$ the degree of divergence.  $\omega({\cal C})$  is computed using the stationary phase approximation where ${\cal Z}({\cal C})$ is approximated as an integral over classical solutions and quadratic fluctuations around these solutions. For $BF$ theory classical solutions are flat connections which define the twisted cohomology. Recall that a cohomology is defined by a sequence of vector spaces and linear maps
\begin{equation} 
\cdots\mathop{\longrightarrow}\limits^{d^{n\!-\!1}}V_{n-1}\mathop{\longrightarrow}\limits^{d^{n}}V_{n}\mathop{\longrightarrow}\limits^{d^{n\!+\!1}}V_{n+1}\mathop{\longrightarrow}\limits^{d^{n\!+\!2}}\cdots
\end{equation}
such that $d^{i}d^{i\!+\!1}=0$. Then, the Betti numbers are defined as $b_{n}=\mathrm{dim}(\mathrm{Ker}\,d^{n}/\mathrm{Im}\,d^{n\!-\!1})$ and the degree of divergence is $\omega({\cal C})=b_{2}$. For an abelian group or if ${\cal C}$ is simply connected (i.e. every closed curve can deformed into a point so that all classical solutions are pure gauge)  we use the ordinary ordinary cell cohomology instead of twisted cohomology \cite{MatteoValentin1}. 

Let us mention that there are other power counting results among which bubble counting for the so called type I graphs \cite{FreidelGurauOriti} and bounds obtained by the Cauchy-Schwartz inequality \cite{CauchySchwartz}, which are compared to the approach presented here in \cite{MatteoValentin3}. Note that the approach based on the Cauchy-Schwartz also provides some non perturbative results since it not only deals with the behavior of single graphs but also with the summation of the perturbative series. This series is in general divergent since the origin only lies on the boundary of the analycity domain. Provided some regularity requirements are met (see the appendix on the Nevanlinna-Sokal theorem in \cite{CauchySchwartz}), a formal power series $F(\lambda)=\sum_{n}a_{n}\lambda^{n}$ defines an analytic function using the Borel resummation procedure. We assume that $G(\lambda)=\sum_{n}\frac{a_{n}}{n!}\lambda^{n}$ is convergent and define $F(\lambda)=\int_{0}^{\infty}dt\,\mathrm{e}^{-t}G(t\lambda)$. A Borel summable modification of the group field theory of three dimensional $BF$ theory has been proposed in \cite{FreidelLouapre} by introducing an other term in the interaction, the so called "pillow" 
\begin{equation}
V_{\mbox{\tiny pillow}}(\Phi)=\int_{\mathrm{SU(2)}} dg_{1}\cdots dg_{6}\;\Phi(g_{1},g_{2},g_{3})\Phi(g_{3},g_{4},g_{5})\Phi(g_{5},g_{4},g_{6})\Phi(g_{6},g_{2},g_{1})
\end{equation}
which represents a tetrahedron obtained by gluing two tetrahedra along two common faces.

All the general results we have mentioned in this section only concern divergences of the partition function. In the group field theory formalism, this means that we only have the degree of divergence of the graph without external legs. In order to deal with the renormalization of group field theory, it is necessary to handle graphs with external legs. The latter correspond to two-complexes with boundary and it is likely that the technique based on the stationary phase approximation and twisted cohomology is versatile enough to encompass these cases.

\subsection{Colored group field theories}

\label{coloredsec}

The need to control the divergences occurring in the expansion of the group field theory led Gurau to introduce colored group field theories \cite{colored}.
These models provide an interesting alternative to the formulations of group field theories we have presented so far.  In dimension $D$, we start with $D+1$ complex (bosonic or fermionic) fields $\Phi^{a}(g_{1},\dots,g_{D})$ which carry an extra index $a\in\left\{0,1,\dots,D\right\}$ called the color. Although the field is complex, we do not usually impose any symmetry properties on its arguments and we consider $\Phi^{a}$ and its complex conjugate $\overline{\Phi}{}^{a}$ as independent fields.

The propagator (or covariance) connects the fields $\Phi^{a}$ with its complex conjugate $\overline{\Phi}^{a}$ so that the Gau\ss ian measure analogous to  \eqref{gaussianmeasure} 
\begin{equation}
\int[D\Phi]_{{\cal K}}\,\overline{\Phi}\,{}^{a}(g_{1},\dots,g_{D})\Phi^{a'}(g'_{1},\dots,g'_{D})=\delta^{a,a'}{\cal K}(g_{1},\dots,g_{D};g'_{1},\dots,g'_{D})
\end{equation}
and
\begin{equation}
\int[D\Phi]_{{\cal K}}\,\overline{\Phi}\,{}^{a}(g_{1},\dots,g_{D})\overline{\Phi}\,{}^{a'}(g'_{1},\dots,g'_{D})=
\int[D\Phi]_{{\cal K}}\,{\Phi}^{a}(g_{1},\dots,g_{D})\Phi^{a'}(g'_{1},\dots,g'_{D})=0
\end{equation}
It is essential to note that the propagators preserve the colors and that there is no summation over permutations.

The interaction does not mix the $\Phi^{a}$ with its complex conjugate and reads\footnote{It differs from the ordinary formulation of colored tensor models by a relabeling of the colors $0\leftrightarrow D$, $1\leftrightarrow D\!-\!1$, $2\leftrightarrow D\!-\!2$, $\dots$}
\begin{equation}
V(\Phi^{a},\overline{\Phi}\,{}^{a})=\int _{G}\prod_{0\leq a\neq b\leq D}dg_{ab}\quad{\cal A}_{v}(g^{}_{ba}g_{ab}^{-1})
\prod_{0\leq a\leq D}\Phi^{a}(g_{a(a+1)},\dots,g_{aD},g_{a1},\dots,g_{a(a-1)})\quad+\quad \mathrm{c.c.}
\end{equation}
Note that we have added the complex conjugate involving the field $\overline{\Phi}\,{}^{a}$ in such a way that the action is real. As usual, the vertex amplitude only depends on the $\frac{D(D+1)}{2}$ products $g^{}_{ba}g_{ab}^{-1}$.

Let us point out some characteristic features of the graphs involved in the perturbative expansion of colored tensor models. First, the fields are not constrained by any permutation of its $D$ arguments. This means that there is no permutation of the $D$ strands so that the stranded graph is entirely determined by its ribbon underlying graph. Second, the colors labels the edges and the structure of the strands can be fully recovered from the coloring of the edges. Third, vertices are of valence $D+1$ and come in two types: a black one for the $\Phi^{a}$ interaction and a white one for the $\overline{\Phi}\,{}^{a}$ interaction with edges connecting only white vertices to black ones. Such kind of graphs are called bipartite graphs.  Finally, a color in $\left\{0,1,\dots,D\right\}$ is assigned to every edge in such a way that the colors of the edges incident to every vertex are different, which is called a proper coloring of the edges with $D+1$ colors. Moreover, colors of the edges are in the cyclic order $0,1,\cdots ,D$ for white vertices and $D,D-1,\cdots ,0$   for black ones. Such a graph can be reconstructed form an ordinary (i.e. without a cyclic order of the edges at the vertices) as follows. First, it is known in graph theory that every bipartite graph with vertices of valence $D+1$ admits a proper coloring of the edges with $D+1$ colors (see for instance \cite{graph}). Then, we choose a suitable cyclic ordering at each vertex.

In the case of $BF$ theory, the propagator can be chosen as
\begin{equation}
{\cal K}(g_{1},\dots,g_{D};g'_{1},\dots,g'_{D})=
\int_{G}dh\,\delta(g_{1}h(g'_{1})^{-1})\cdots\delta(g_{D}(g'_{D})^{-1}h)
\end{equation}
and the vertex amplitude 
\begin{equation}
{\cal A}_{v}(g^{}_{ab}g_{ba})=\frac{\lambda}{D+1}\,\prod_{0\leq a<b\leq D}\delta(g^{}_{ba}g_{ab}^{-1})
\end{equation}
Performing the integration over half of the variables, the interaction reads, in dimensions $D=$ 2, 3, 4
\begin{itemize}
\item D=2
\begin{equation}
\frac{\lambda}{3}\int _{G}\prod_{0\leq a<b\leq 2}dg_{ab}\quad
\Phi^{0}(g_{01},g_{02})
\Phi^{2}(g_{02},g_{12})
\Phi^{1}(g_{12},g_{02})
\quad+\quad \mathrm{c.c.}
\end{equation}
\item D=3
\begin{equation}
\frac{\lambda}{4}\int _{G}\prod_{0\leq a<b\leq 3}dg_{ab}\quad
\Phi^{0}(g_{01},g_{02},g_{03})
\Phi^{3}(g_{03},g_{13},g_{23})
\Phi^{2}(g_{23},g_{02},g_{12})
\Phi^{1}(g_{12},g_{13},g_{01})
\quad+\quad \mathrm{c.c.}
\end{equation}
\item D=4
\begin{multline}
\frac{\lambda}{5}\int _{G}\prod_{0\leq a<b\leq 4}dg_{ab}\quad\Big\{
\Phi^{0}(g_{01},g_{02},g_{03},g_{04})
\Phi^{4}(g_{04},g_{14},g_{24},g_{34})\times\cr
\Phi^{3}(g_{34},g_{03},g_{13},g_{23})
\Phi^{2}(g_{23},g_{24},g_{20},g_{12})
\times\Phi^{1}(g_{12},g_{13},g_{14},g_{01})\Big\}
\quad+\quad \mathrm{c.c.}
\end{multline}
\end{itemize}

For $p\in\left\{0,1,\cdots,D\right\}$, $p$-bubbles of a given graph are defined as follows. $0$-bubbles are vertices and 1-bubbles edges. For $p>1$, $p$-bubbles with colors $i_{1},\dots,i_{p}$ are connected components of made of edges of colors $i_{1},\dots,i_{p}$. Let us denote ${\cal B}^{p}$ the space of $p$-bubbles. There are boundary operators $\partial^{p}:\,{\cal B}^{p}\rightarrow{\cal B}^{p-1}$ taking a $p$-bubble to a $(p\!-\!1)$-bubble
\begin{equation}
\partial^{p} b_{i_{1},\cdots,i_{p}}=\sum_{q=0}^{p} (-1)^{q} \Big(\!\!\!\!\!\sum_{\rho\atop\mbox{\tiny connected component}}\!\!\!\!\!b^{(\rho)}_{i_{1},\dots,i_{q-1},i_{q+1},\dots,i_{q}}\Big)
\end{equation}  
These operators obey $\partial^{p-1}\partial^{p}=0$  and define an homology 
\begin{equation}
\dots\mathop{\longleftarrow}\limits^{{}\;\partial^{p-1}}{\cal B}^{p-1}\mathop{\longleftarrow}\limits^{\;\partial^{p}}{\cal B}^{p}\mathop{\longleftarrow}\limits^{\;\partial^{p+1}}{\cal B}^{p+1}\mathop{\longleftarrow}\limits^{\;\partial^{p+2}}\dots
\end{equation}
This homology is instrumental in understanding the relation between the group field theory graph and the underlying simplicial complex. Furthermore, it provides an exact power counting theorem in the abelian case \cite{linhom}.

Let us now discuss a few salient features of colored tensor models.

An important breakthrough allowed by colored tensor model lies in the construction of a $\frac{1}{N}$ expansion similar to the matrix models case 
given in  \eqref{largeN} which opens the possibility of defining a continuum limit through a double scaling limit. The first result in this direction is the identification of the dominant terms as triangulations of the sphere $S^{D}$ \cite{complete}. We only outline this result and urge the reader to consult \cite{complete}. For any stranded graph ${\cal G}$ of a colored tensor model, we define its jackets as the ribbon graphs with the same vertices and edges as ${\cal G}$ and faces of colors $\left\{\sigma^{q}(0)\sigma^{q+1}(0)\,|\, q\in{\Bbb Z}_{D+1}\right\}$ with $\sigma$ a cycle of length $D+1$, i.e. a permutation of the colors that does not leave any subset of colors globally invariant. There are $D!$ such permutations but the jacket does not depend on the orientation of the cycle, so that ${\cal G}$ has $\frac{D!}{2}$ jackets. Then, we define the convergence degree of ${\cal G}$ as the sum of the genera of all its jackets, $\omega({\cal G})=\sum_{{\cal J}}g_{\cal J}$.  It is convenient to rescale the coupling constant as
\begin{equation}
\lambda\quad\rightarrow\quad\frac{\lambda}{\delta(1)^{\frac{(D-1)(D-2)}{4}}}
\end{equation}
where $\delta(1)$ has been regularized, either by a cut-off on the spins or using the heat kernel. It plays a role analogous to the size $N$ of the matrices in the large $N$ limit of matrix models. By a series of combinatorial reductions performed on the edges of the graph, one obtains the jacket bound
\begin{equation}
\big|{\cal A}_{\cal G}\big|\quad\leq\quad(\lambda\overline{\lambda})^{\frac{V}{2}}\,\delta(1)^{D-1-\frac{D-2}{D!}\omega({\cal G})}
\end{equation}
for the $BF$ amplitude of a graph with $V$ vertices. By construction, we always have $\omega({\cal G})\geq 0$ and it may be shown that $\omega({\cal G})=0$ implies that ${\cal G}$ corresponds to a triangulation of the sphere $S^{D}$, the converse holding in $D=2$.

Colored models also enjoy a extended version of the group field theory symmetries, where fields with different colors have different symmetries transformations. This is used in \cite{Diff}, where analogues of diffeomeorphisms as local translations of the vertices of the tetrahedra are constructed. Such an invariance may shed a new light on the divergences as the latter are related to spikes, i.e. points free to go to infinity inside the triangulation. An example of a spike is given by the point added in the Pachner move given in figure \ref{Pachner}. Let us illustrate the idea of extended symmetries  on the different example of global translations mentioned in section \ref{GFT}. In a general group field theory, the vertex amplitude only depends on the $\frac{D(D+1}{2}$ variables $g^{}_{ba}g_{ab}^{-1}$, which lead us to the invariance under the global translation   \eqref{globalGFT} involving a single group element. In a colored tensor model, this transformation may be extended to
\begin{multline}
T_{g_{1},\dots,g_{D}}\Phi^{a}\big(g_{a(a+1)},\dots,g_{aD},g_{a,0},\dots,g_{a(a-1)}\big)=\cr
\Phi^{a}\big(g_{a(a+1)}g_{a+1},\dots,g_{aD}g_{D},g_{a,0}g_{0},\dots,g_{a(a-1)}g_{a-1}\big)
\end{multline}
The color indices are essential in order to define different transformations on the $(D\!-\!1)$-simplexes.

Finally, the first renomalizable four dimensional group field theory has been constructed by Ben Geloun and Rivasseau \cite{renormalizable} using colored models.  This construction involves an abelian version of the model with group $\mathrm{U(1)}$. Since it is four dimensional it involves stranded graphs with four strands and five colors. Thus, we start with five fields $\phi^{a}(g_{1},\dots,g_{4})$ on $U(1)^{4}$. The colors $a\in\left\{1,\dots,5\right\}$ have a trivial kinetic term while the color $0$ has a kinetic term of the time 
\begin{equation}
\overline{\Phi}(g_{1},\dots,g_{4})\big(\sum_{i=1}^{4}\Delta_{g_{i}}+m^{2}\big){\Phi}(g_{1},\dots,g_{4})
\end{equation}
Then, we integrate  over all but the  color $0$ with the standard colored interaction we get an effective action and keep only the leading order terms in $N$, where $N$ is a momentum cut-off. The resulting theory for the color $0$ is proven to be renormalizable using techniques form multiscale analysis \cite{Vincentbook}. Note that this model involves terms of degree 6 in the field as well as a non-local one of the type $\big(\int \phi\big)^{2}$. This construction has been adapted to dimension three by Ben Geloun and Samary  \cite{Ousmane}. It also leads to a detailed computation of the $\beta$ function and a discussion of asymptotic freedom \cite{Beta}. A systematic presentation of renormalizable models is given by Ben Geloun  and Livine \cite{EteraJoseph}.

\subsection{The Lorentzian EPRL/FK model}

Before we come to grips with its group field theory formulation, we give a very brief overview of the EPRL/FK spin foam model of Lorentzian quantum gravity in dimension four. This model has been defined by Engle, Pereira, Rovelli and Livine \cite{EPRL} and Freidel and Krasnov \cite{FK}, building on earlier work of Livine and Speziale \cite{coherent}. We adopt here the formulation presented in \cite{simple} and refer to the lectures by Rovelli \cite{CarloZakopane} for a thorough overview, including recent applications. 

Recall that the action of pure gravity can be written as
\begin{equation}
S[e,\omega]=\frac{1}{32\pi G}\int_{\cal M} \mbox{Tr}\,\, B\wedge R\label{actiongravity}
\end{equation}
where $R=d\omega+\omega\wedge\omega$ is the curvature of the Lorentzian spin connection $\omega$ with values in the Lie algebra of $\mathrm{SL}(2,{\Bbb C})$ and $B$ is a 2-form with values in the Lie algebra of the Lorentz group
\begin{equation}
B=\star(e\wedge e)+\frac{1}{\gamma}\,e\wedge e\label{constraintB}
\end{equation}
with $e$ the vierbein and $\gamma$ the Immirzi parameter. All the field carry internal Lorentz indices raised or lowered  with the tensors $\eta^{IJ}$ and $\eta_{IJ}$. Trace and star are shorthands for $\mbox{Tr}\,\,B\wedge R=B^{IJ}\wedge R_{IJ}$ and $(\star B)_{IJ}=\frac{1}{2}\epsilon_{IJKL}B^{KL}$. In the Einstein-Cartan formulation of gravity, the second term involving the Immirzi parameter is absent. This term has no effect in the classical formulation of pure gravity but is instrumental in making contact with loop quantum gravity.

If we momentarily disregard the constraint  \eqref{constraintB} and consider $B$ as an arbitrary 2-form with values in the Lie algebra of $\mathrm{SL}(2,{\Bbb C})$, then the theory given by the action  \eqref{actiongravity} is nothing but a $BF$ theory with gauge group $\mathrm{SL}(2,{\Bbb C})$. Although its gauge group is non compact, it can be formally quantized in terms of spin foams using a discretization and writing the Dirac distribution as
\begin{equation}
\delta(g)=\sum_{j}\int_{\Bbb R}d\rho\,(\rho^{2}+j^{2})\,\mbox{Tr}\,_{V_{j,\rho}}(g)
\end{equation} 
where the sum runs over all representations of the principal series described in the last section \ref{unitary}. This would lead to a vertex amplitude of the $BF$ type as in  \eqref{BFvertex}.


Roughly speaking, taking into account the constraint  \eqref{constraintB} amounts to restrict the summation to  the representations that obey $\rho=\gamma j$ and project onto  the SU(2) subrepresentation of lowest spin in the decomposition  \eqref{SU(2)decomposition}. This amounts to replace the $\delta$ function on  $\mathrm{SL}(2,{\Bbb C})$ by 
\begin{equation}
\sum_{j}\int_{\mathrm{SU(2)}}\!\!\!dk\,\,d_{j}^{2}\,\mbox{Tr}\,_{V_{j}}(k)\,\mbox{Tr}\,_{V_{j,\gamma j}}(gk)=
\sum_{j}d_{j}\,\mbox{Tr}\,_{V_{j}}\big(Y_{j,\gamma j}^{\dagger}gY^{}_{j,\gamma j}\big)
 \end{equation}
with $Y_{j,\rho}:\,V_{j}\rightarrow V_{j,\rho}$ the embedding as SU(2) representation. The integration over $k$ implements a projection onto $V_{j}$, as follows from the Schur orthogonality relations  \eqref{Schur}. Therefore, the vertex reads
\begin{equation}
{\cal A}_{v}(h_{ab})=\int_{\mathrm{SL}(2,{\Bbb C})^{4}}\prod_{0\leq a\leq 4}d'g_{a}\prod_{0\leq a<b\leq 4}
\bigg\{
\sum_{j_{ab}}d_{j_{ab}}\mbox{Tr}\,_{V_{j_{ab}}}\big[h_{ab}\,Y_{j_{ab},\gamma j_{ab}}\,g^{}_{a}g_{b}^{-1}Y^{\dagger}_{j_{ab},\gamma j_{ab}}\big]
\bigg\}
\end{equation}
where $\prod_{0\leq a\leq 4}d'g_{a}$ means that, because of the gauge invariance $g_{a}\rightarrow g_{a}g$, one of the non compact $\mathrm{SL}(2,{\Bbb C})$ integration has to be dropped. Note that $h_{ab}$ commutes with $Y_{j_{ab},\gamma j_{ab}}$.

In this expression, we have used labels $a,b\in\left\{0,\dots,4\right\}$ for the tetrahedra on the boundary and the explicit expression of the vertex amplitude relies on the specific ordering from $a$ to $b$ with $a<b$. This defines a choice of an orientation of the faces or, equivalently, of the boundary links. Let us show that ${\cal A}(h_{l})$ does not depend on such a choice, provided we trade the corresponding link variable $h_{l}$ for its inverse. Our argument relies on the existence if an antilinear map $J:\,V_{j,\rho}\rightarrow V_{j,\rho}$ acting on unitary irreducible representations of the principal series of $\mathrm{SL}(2,{\Bbb C})$ that commutes with the group action and preserves the scalar product $\langle J\psi|J\chi\rangle=\langle \chi|\psi\rangle$, whose explicit form is given in \eqref{defJ}. Therefore, if $\psi_{i}$ is an orthornormal basis of $V_{j,\rho}$, so is $J\psi_{i}$ and we have
\begin{equation}
\mathrm{Tr}_{V_{j,\rho}}(g)=\sum_{i}\langle J\psi_{i}|gJ\psi_{i}\rangle=\sum_{i}\langle J\psi_{i}|Jg\psi_{i}\rangle
=\sum_{i}\langle g\psi_{i}|\psi_{i}\rangle
=\sum_{i}\langle \psi_{i}|g^{-1}\psi_{i}\rangle=\mathrm{Tr}_{V_{j,\rho}}(g^{-1})
\end{equation}
It is also interesting to notice that this amplitude differs from the $BF$ amplitude with gauge group SU(2) \eqref{BFvertex} only through the range of the integration over the variables $g_{a}$, $BF$ theory being recovered if we integrate over SU(2) instead of $\mathrm{SL}(2,{\Bbb C})$.  


Gluing together all the 4-simplexes, the amplitude for an arbitrary 2-complex can be written as
\begin{equation}
{\cal A}_{\cal C}(h_{l})=\int_{\mathrm{SL}(2,{\Bbb C})^{4V}}\prod_{e,v}d'g_{v,e}\quad\Big\{\prod_{f}{\cal A}_{f}(g_{v,e},h_{l})\Big\}
\label{EPRLspinfoam}
\end{equation}
where an $\mathrm{SL}(2,{\Bbb C)}$ has been assigned to each pair of a vertex $v$ which is not a node and an edge $e$ incident to $v$. For a face $f$ that does not meet the boundary of ${\cal C}$,  the face amplitude is defined by
\begin{equation}
{\cal A}_{f}(g_{v,e})=\sum_{j}d_{j}\,\mbox{Tr}\,_{V_{j}}\Big[\mathop{\prod}\limits^{\longrightarrow}_{v\in\partial f} Y_{j,\gamma j}^{\dagger}\,g_{v,e^{+}(v)}(g_{v,e^{-}(v)})^{-1} Y_{j,\gamma j}\Big]
\end{equation}
with $e^{+}(v)$ (resp. $e_{-}(v)$) the edge entering (resp. leaving) $v$ according to the orientation chosen for the face. If $f$ meets the boundary, an extra SU(2) variable $h_{l}$ has to be inserted on the link between two vertices. This amplitude does not depend on the choice of the face orientation provided we invert the link variable $h_{l}$. Moreover, for a complex without boundary, it is real thanks to the unitary of the representation,

The boundary of a 2-complex is a graph $\Gamma=\partial {\cal C}$ and the spin foam amplitude ${\cal A}_{\cal C}(h_{l})$ defines a wave function in
\begin{equation}
{\cal A}_{\cal C}(h_{l})\in{\cal H}=L^{2}(\mathrm{SU(2)}^{L}/\mathrm{SU(2)}^{N})
\end{equation}
where the node gauge transformations $g_{n}\in\mathrm{SU(2)}$ act on the link variables as $h_{l}\rightarrow g_{s(l)}h_{l\,}g_{t(l)}^{-1}$. This may be extended to gauge transformations with values in  $\mathrm{SL}(2,{\Bbb C})$ provided we use projected spin networks \cite{CarloSimone}.

The sums over spins in the EPRL/FK model are in general divergent and require some regularization. Trading the Lorentz group for its quantum group deformation \cite{Philippe} leads to a finite model  \cite{Muxin} and \cite{Winston}, which is expected to correspond to a non vanishing cosmological constant.

There is also a euclidian version of the EPRL/FK model based on the group $\mathrm{SU(2)}\times\mathrm{SU(2)}$ instead of $\mathrm{SL}(2,{\Bbb C})$. For $0<\gamma<1$, it involves the representations $V_{j^{+}}\otimes V_{j^{-}}$ with $j^{\pm}=\frac{1\pm\gamma}{2}$ and the subrepresentation $V_{j}$ of highest spin $j=j^{+}+j^{-}$  of the diagonal $\mathrm{SU(2)}$ subgroup.

Finally, let us conclude this section by emphasizing that the EPRL/FK model corrects two drawbacks of the older Barrett-Crane model, defined for Lorentzian geometries in \cite{BC}. First, its asymptotic expansion in the large spin limit reproduces Regge calculus, as shown in \cite{asymptoticvertex} for a single 4-simplex and in \cite{asymptotic2complex} for an entire 2-complex, using the stationary phase method. Second, its relation to the kinematical loop quantum gravity Hilbert space is transparent since it involves the Immirzi parameter and SU(2) boundary states \cite{CarloZakopane}.

\label{EPRLFK}
\label{euclidian}

\subsection{Group field theory formulation of the lorentzian EPRL/FK model}

Once the vertex amlitude of a spin foam model is known, the general techniques presented in section \ref{GFT} can be applied to construct the associated group field theory. In the case of the EPRL/FK vertex, this leads to the interaction
\begin{multline}
V(\Phi)=\frac{\lambda}{5}\int_{{\mathrm SU(2)}^{20}}\prod_{0\leq a\neq b\leq 4} dg_{a,b}
\int_{\mathrm{SL}(2,{\Bbb C})^{4}}\prod_{0\leq a\leq 4}d'g_{a}\cr
\bigg\{ \prod_{0\leq a<b\leq 4}\Big\{
\sum_{j_{ab}}d_{j_{ab}}\mbox{Tr}\,_{V_{j_{ab}}}\big[g^{-1}_{ab}\,Y^{\dagger}_{j_{ab},\gamma j_{ab}}g^{}_{a}g_{b}^{-1}Y_{j_{ab},\gamma j_{ab}}g_{ba}^{}\big]
\Big\}
\prod_{0\leq a\leq 4}\Phi\big(g_{a(a+1)},\dots,g_{a4},g_{a,0}\dots,g_{a(a-1)}\big)\bigg\}\label{interactionGFTEPRL}
\end{multline}
This interaction is adapted to colored model by adding a color index $\Phi^{a}\big(g_{a(a+1)},\dots,g_{a4},g_{a,0}\dots,g_{a(a-1)}\big)$. It is also invariant under gauge transformations  \eqref{gaugeGFT} and global translations  \eqref{globalGFT} since $Y_{j,\rho}$ commutes with the SU(2) action. The propagator can be chosen to be either the trivial one  \eqref{trivialpropagator} or the projector onto SU(2) gauge invariant states  \eqref{invariantpropagator}. Then, an arbitrary Feynman graph is evaluated by performing the SU(2) integration using Schur's orthogonality relations  \eqref{Schur} and yields the EPRL/FK spin foam amplitude  \eqref{EPRLspinfoam} for the partition function as a sum over graphs without external legs. 
The case of spin foam with non trivial boundary can be treated along the lines presented in section \ref{boundarysec}.

Using the canonical basis (see appendix \ref{unitary}), the interaction can be written in terms of coherent states (see appendix \ref{coherent}). Indeed, 
\begin{multline}
\sum_{j_{ab}}d_{j_{ab}} 
\mbox{Tr}\,_{V_{j_{ab}}}\big[g_{ba}\,Y^{\dagger}_{j_{ab},\gamma j_{ab}}g^{}_{a}g_{b}^{-1}Y_{j_{ab},\gamma j_{ab}}g_{ab}^{-1}\big]
=\\\sum_{j_{ab}}(d_{j_{ab}})^{2}\int_{\mathrm{S^{2}}}dn_{ab}\,
\frac{\big[\langle n_{ab}| (g_{ba}g_{a})^{\dagger}g_{ab} g_{b}|n_{ab}\rangle\big]^{2j_{ab}}}
{\Big[\langle n_{ab}|g_{a}^{\dagger}g_{a}|n_{ab}\rangle \Big]^{(\mathrm{i}\gamma_{j_{ab}}+1+j_{ab})}
\Big[\langle n_{ab}|g_{b}^{\dagger}g_{b}^{}|n_{ab}\rangle \Big]^{(-\mathrm{i}\gamma_{j_{ab}}+1+j_{ab})}}
\end{multline}
Replacing in the expression of the vertex \eqref{interactionGFTEPRL}, we get
\begin{multline}
V(\Phi)=\frac{\lambda}{5}\int_{{\mathrm SU(2)}^{20}}\prod_{0\leq a\neq b\leq 4} dg_{a,b}
\int_{\mathrm{SL}(2,{\Bbb C})^{4}}\prod_{0\leq a\leq 4}d'g_{a}\quad\sum_{j_{ab}}(d_{j_{ab}})^{2}\int_{\mathrm{S^{2}}}dn_{ab}\cr
\bigg\{ \prod_{0\leq a<b\leq 4}
\frac{\big[\langle n_{ab}| (g_{ba}g_{a})^{\dagger}g_{ab} g_{b}|n\rangle\big]^{2j_{ab}}}
{\Big[\langle n_{ab}|g_{a}^{\dagger}g_{a}|n_{ab}\rangle \Big]^{(\mathrm{i}\gamma_{j_{ab}}+1+j_{ab})}
\Big[\langle n_{ab}|g_{b}^{\dagger}g_{b}^{}|n_{ab}\rangle \Big]^{(-\mathrm{i}\gamma_{j_{ab}}+1+j_{ab})}}
\cr
\prod_{0\leq a\leq 4}\Phi\big(g_{a(a+1)},\dots,g_{a4},g_{a,0}\dots,g_{a(a-1)}\big)\bigg\}\label{interactionGFTEPRLcoherent}
\end{multline}

Using this representation of the vertex, the sum over spins can be performed explicitly. Indeed, starting with a geometrical series, we get
\begin{equation}
\sum_{2j\in{\Bbb N}}(2j+1)^{2}\xi^{2j}= \Big(\xi\frac{d}{d\xi}+1\Big)^{2}\frac{1}{1-\xi}=\frac{1+\xi}{(1-\xi)^{3}}
\end{equation}
which  is convergent for $|\xi|<1$ and becomes singular at $\xi=1$. Therefore, we set 
\begin{equation}
\xi_{ab}=\frac{\langle n| (g_{ab}g_{a})^{\dagger}g_{ab} g_{b}|n\rangle}
{\Big[\langle n|(g_{ba}g_{a})^{\dagger}g_{ba}g_{a}|n\rangle \Big]^{\frac{\mathrm{i}\gamma+1}{2}}
\Big[\langle n|(g_{ab}g_{b})^{\dagger}g_{ab}g_{b}|n\rangle \Big]^{\frac{-\mathrm{i}\gamma+1}{2}}}
\end{equation}
so that the sum over spins is expressed in terms of the previous function.

By the Cauchy-Schwartz inequality $|\langle\chi|\psi\rangle|^{2}\leq \langle\chi|\chi\rangle\langle\psi|\psi\rangle$, it turns out that we always have $|\xi_{an}|\leq 1$.  Moreover, the equality $|\langle\chi|\psi\rangle|^{2}= \langle\chi|\chi\rangle\langle\psi|\psi\rangle$ is reached only when the kets $|\chi\rangle$ and $|\psi\rangle$ are proportional. In our context, it amounts to
\begin{equation}
\frac{g_{ab} g_{a}|n_{ab}\rangle \langle n_{ab}|g_{a}^{\dagger}g_{ba}^{\dagger}}{\langle n_{ab}|g_{a}^{\dagger}g_{a}|n_{ab}\rangle}=
\frac{g_{ba} g_{b}|n_{ab}\rangle \langle n_{ab}|g_{b}^{\dagger}g_{ab}^{\dagger}}{\langle n_{ab}|g_{b}^{\dagger}g_{b}^{}|n_{ab}\rangle}
\end{equation}
which is precisely the gluing condition found in the analysis of the asymptotic behavior of the vertex \cite{asymptoticvertex}. Geometrically, it means that the bivectors defining the geometry of the triangles $ab$ and $ab$ in the four simplex agree. 

This leads us to an interesting analogy with the mass shell of ordinary quantum field theory, as first noticed by Pereira in his PhD thesis \cite{Roberto}. 
From the viewpoint of group field theory, the vertex becomes singular when the geometry is close to a real geometry. This is reminiscent of the mass shell singularity in quantum field theory, whose propagator is singular when the particles are real. It is particularly clear  if we use Schwinger's proper time, as introduced in section \ref{Schwingersec}
\begin{equation}
\frac{1}{k^{2}+m^{2}+\mathrm{i}\epsilon}=\int_{0}^{\infty}\!\!d\alpha\,\exp\mathrm{i}\big\{\alpha(k^{2}+m^{2}+\mathrm{i}\epsilon)\big\}
\end{equation}
In particular, we can also regulate the sum over spin by inserting a factor of $(\mathrm{e}^{-\epsilon})^{j_{ab}}$, with $\epsilon>0$. In this construction, the spins $j_{ab}$ play in group field theory a role similar to the Schwinger parameter in quantum field theory.

Note that a group field theory reproducing the EPRL/FK spin foam amplitude is by no means unique. We have chosen to work with fields defined on SU(2)$^{4}$ and put the vertex weight in the interaction. It could also be possible to split this weight into two parts and put them on the propagators, so that we would have a propagator of the type
\begin{equation}
{\cal K}(g_{1},\dots,g_{4};\tilde{g}_{1},\dots,\tilde{g}_{4})=
\int_{\mathrm{SL}(2,{\Bbb C})^{2}}dgd\tilde{g}
\prod_{0\leq a\leq 4}\Big\{
\sum_{j_{a}}d_{j_{a}}\mbox{Tr}\,_{V_{j_{a}.\gamma j_{a}}}\big[g_{a}(\tilde{g}_{a})^{-1}\,\tilde{g}Y^{\dagger}_{j_{a},\gamma j_{ab}}Y_{j_{a},\gamma j_{a}}{g}^{-1}\big]
\Big\}
\end{equation}
and an interaction of the $\mathrm{SL}(2,{\Bbb C})$ $BF$ type. This requires us to work with a group field $\Phi(g_{1},\dots,g_{4})$ defined on $\mathrm{SL}(2,{\Bbb C})^{4}$ and use the $\mathrm{SL}(2,{\Bbb C})$ analogue of Schur's orthogonality relations to recover the spin foam amplitude. Because of the non compact nature of the group, this is slightly more technical. Moreover, factoring out the gauge degrees of freedom by dropping one of the $\mathrm{SL}(2,{\Bbb C})$ integration at each vertex is no longer possible and requires more work. In the euclidian version of the model, $\mathrm{SL}(2,{\Bbb C})$ is replaced by $\mathrm{SU(2)}\times\mathrm{SU(2)}$ and does not create any difficulty. This is the formulation intialy proposed in \cite{FK} and further developed in \cite{Vincentetal1} and \cite{Vincentetal2}. 

The field $\Phi(g_{1},\dots,g_{4})$ obeying the SU(2) invariance property $\Phi(gg_{1},\dots,gg_{4})=\Phi(g_{1},\dots,g_{4})$ is related to a quantum tetrahedron \cite{quantumtetrahedron} embedded in three dimensional space, as shown by the discussion following \eqref{intertwiner}. Therefore, the external legs of a group field theory Feynman graph define the boundary tetrahedra. If we further include a boundary graph observable as defined in  \eqref{observable}, we recover the EPRL/FK spin foam amplitude for a 2-complex with boundary in terms of spin networks. In a four dimensional theory, it is more transparent to work with projective spin networks since they yield a manifestly covariant covariant formulation \cite{CarloSimone}.  This is achieved by introducing an extra variable $x$ in the coset $\mathrm{SL}(2,{\Bbb C})/\mathrm{SU(2)}$. It is represented by $g_{x}\in\mathrm{SL}(2,{\Bbb C})$, up to a multiplication on the right by $h\in\mathrm{SU(2)}$:$g_{x}$ and $g_{x}h$ represent the same $x$. Moreover, any $x\in\mathrm{SL}(2,{\Bbb C})/\mathrm{SU(2)}$ can be represented in a unique way as a positive define matrix $X=g_{x}g_{x}^{\dagger}$ of determinant one. Writing any such matrix as $X=X^{I}\sigma_{I}$ with $\sigma_{I}$ the identity and the Pauli matrices, we can identify $X$ with a normalized future-pointing time-like vector in the hyperboloid $H^{+}$. Then, we expand the field describing a tetrahedron in three dimensional space as in \eqref{expansionM}
\begin{equation}
\Phi(h_{1},\dots,h_{4})=\sum_{j_{1},\dots,j_{D},m_{1},\dots,m_{D}\atop\mbox{\tiny intertwiner between } j_{1},\dots,j_{4}}
M_{m_{1},\dots,m_{D}}^{j_{1},\dots,j_{4};\,i}\,
{\cal D}_{m_{1},m'_{1}}^{j_{1}}(h_{1})\cdots{\cal  D}_{m_{4},m'_{4}}^{j_{4}}(h_{4})\,i^{\ast}_{m'_{1},\dot,m'_{4}}
\end{equation}
with $h_{i}\in\mathrm{SU(2)}$. Its four dimensional cousin is expressed as
\begin{equation}
\widetilde{\Phi}(X,g_{1},\dots,g_{4})=\sum_{j_{1},\dots,j_{D},m_{1},\dots,m_{D}\atop\mbox{\tiny intertwiner between } j_{1},\dots,j_{4}}
M_{m_{1},\dots,m_{D}}^{j_{1},\dots,j_{4};\,i}\,
{\cal D}_{j_{1},m_{1},j_{1},m'_{1}}^{j_{1},\gamma j_{1}}(g_{1}g_{x})\cdots{\cal  D}_{j_{4},m_{4},j_{4},m'_{4}}^{j_{4},\gamma j_{4}}(g_{4}g_{x})\,i^{\ast}_{m'_{1},\dot,m'_{4}}\label{deftildephi}
\end{equation}
with $g_{i}\in\mathrm{SL}(2,{\Bbb C})$, $X\in H^{+}$ and
\begin{equation}
{\cal D}_{k,m,k'm'}^{\rho,j}(g)=_{\rho,j}\!\!\langle k,m|g|k',m'\rangle_{\rho,j}\qquad 
\end{equation}
the matrix element of $g\in\mathrm{SL}(2,{\Bbb C})$ in the unitary representation $V_{j,\rho}$ of the principal series (see appendix \ref{unitary}). Note that $\widetilde{\Phi}$ only depends on $X=g_{x}g_{x}^{\dagger}$ because of the SU(2) gauge invariance of $\Phi$.

The field $\widetilde{\Phi}$ is invariant under the $\mathrm{SL}(2,{\Bbb C})$ transformation
\begin{equation}
\widetilde{\Phi}(g^{-1}\triangleright X,g_{1}g,\dots,g_{4}g)=\widetilde{\Phi}(X,g_{1},\dots,g_{4}) 
\end{equation}
with $g\triangleright X= gX g{\dagger} $ the standard Lorentz transformation of $X$. The field $\Phi(h_{1},\dots,h_{4})$ is recovered by assuming that $X=\overline{X}=(1,0,0,0)$ and restricting the argument of the field to SU(2). In order not to introduce spurious degrees of freedom, we assume that the group elements in  \eqref{deftildephi} are such that $X=g\triangleright \overline{X}$. Therefore,  $\widetilde{\phi}$ does not contain more degrees of freedom than $\phi$, it merely provides a covariant version of the latter, with $X$ the four dimensional normal to the tetrahedron under consideration.

Finally, let us note that one can rewrite the interaction in terms of the new field $\widetilde{\Phi}$, trading the integration over the group elements $g_{a}$ in \eqref{interactionGFTEPRL} for an integration over the normals $X_{a}$, taking into account the properties of the $\mathrm{SL}(2,{\Bbb C})$ Haar measure recalled in Appendix \ref{unitary},
\begin{multline}
V(\widetilde{\Phi})=
\frac{\lambda'}{5}
\int_{(H^{+})^{4}}\prod_{0\leq a\leq 4}d'X_{a}
\int_{{{\mathrm SL}(2,{\Bbb C})}^{20},\,\, X_{a}=g_{ab}\triangleright \overline{X} }\quad\prod_{0\leq a\neq b\leq 4} d'g_{a,b}
\quad\sum_{j_{ab}}(d_{j_{ab}})^{2}\int_{\mathrm{S^{2}}}dn_{ab}\cr
\bigg\{ \prod_{0\leq a<b\leq 4}
\frac{\big[\langle n_{ab}| g_{ba}^{\dagger}g_{ab}^{}|n_{ab}\rangle\big]^{2j_{ab}}}
{\Big[\langle n_{ab}|X_{a}|n_{ab}\rangle \Big]^{(\mathrm{i}\gamma_{j_{ab}}+1+j_{ab})}
\Big[\langle n_{ab}|X_{b}|n_{ab}\rangle \Big]^{(-\mathrm{i}\gamma_{j_{ab}}+1+j_{ab})}}
\cr
\prod_{0\leq a\leq 4}\widetilde{\Phi}\big(X_{a},g_{a(a+1)},\dots,g_{a4},g_{a,0}\dots,g_{a(a-1)}\big)\bigg\}
\end{multline}
$\lambda'$  is a new coupling constant introduced in order to absorb normalization factors in the measure over $H^{+}$. Primed integration measure $d'X$ and $d'g$ signify that  global $\mathrm{SL(2,\mathbb{C})}$ invariance has to be factored, because of gauge invariance. This could be done, for instance, by imposing that one of the normals is fixed.




\section{Conclusion and outlook}

In these lecture notes we have presented group field theory as a generalization of matrix models of two dimensional quantum gravity in order to deal with higher dimensional quantum gravity. We first generalized matrices $M_{ij}$ to higher rank tensors $M_{i_{1}\dots i_{D}}$ to generate a summation over $D$-dimensional triangulations. Then, we promoted tensors to functions $\Phi(g_{1},\dots,g_{D})$ to account for the gravitational degrees of freedom as encoded in loop quantum gravity and spin foam models. In this case, the perturbative expansion of the group field theory path integral
\begin{equation} 
\langle{\cal O}_{\Gamma}[\Phi](h_{l})\rangle_{c}=\sum_{T\, \mbox{\tiny possibly disconnected triangulations with }\atop n\,\mbox{\tiny boundary ({\it D }-1)-simplexes }}\frac{1}{C_{T}}\,{\cal A}_{T/\Gamma}(h_{l})\label{observableexp}
\end{equation}

As shown in the last section, for a suitable choice of the action functional $S[\Phi]$, it reproduces the spin foam amplitudes of the Lorentzian EPRL/FK model. Therefore, it can be considered as a convenient generating functional for these amplitudes.

However, it is more than a mere generating functional: it provides a prescription to sum the various amplitudes. Its relation to spin foam models is similar to the relation of quantum field theory to particle physics, albeit in reverse logical order. Indeed, instead of starting with field quantization, particle physicists could have devised the Feynman rules for evaluating scattering processes between particles and only later realize that these follow from the perturbative expansion of a field theory. Needless to say that the quantum field viewpoint has been instrumental to understand various phenomena (for instance the relation to statistical physics through renormalization) and is the only way to related elementary particles physics to everyday life classical field theory.

Something similar may take place in quantum gravity. Loop quantum gravity and spin foam models provide a tentative description of the gravitational field at very short distance. The transition amplitudes they provide are the Feynman graph expansion of the group field theory path integral. Therefore, group field theory may shed a new light on loop quantum gravity through the use of quantum field theoretic techniques like semi-classical expansions, Schwinger-Dyson equations or Wilsonian renormalization.

However, in order for group field theory to qualify as a bona fide physical theory, a few problems have to be circumvented. Let us end these lectures by listing some of them as well as tentative solutions.

\begin{itemize}

\item{\bf Coupling to matter} The group field theory theories we have presented only describe pure gravity, Matter fields have to be included in order to provide a physically sensible theory. Inclusion of matter can be performed in a variety of ways. For instance, one can add extra arguments to the group field $\Phi$, in analogy with the coupling of the Potts model to two dimensional gravity \eqref{matrixmodelpotts}. Such a construction could be used to reproduce the spin foam amplitude of gravity coupled to Yang-Mills theory and fermions \cite{SpinFoamFermions}. Alternatively, matter fields can also arise as some excitations of the group field itself around a classical solution, as proposed by Di Mare and Oriti \cite{MatterEmergent}. The simplest for of matter is certainly the vacuum, in the form of a non zero cosmological constant. It is known that the inclusion of a cosmological constant can be achieved by trading SU(2) or $\mathrm{SL}(2,{\Bbb C})$ for their q-derformed counterpart. Since quantum group are Hopf algebras, it is natural to define group field theories based on Hopf algebras. Such a framework has been developed by Krasnov  in order to include matter fields of various sorts \cite{KrasnovHopf}.  For instance, in the case of the Boulatov-Ooguri models, the basic field $\Phi(g_{1},\dots, g_{D})$ is an element of ${\cal A}^{\otimes D}$ where ${\cal A}$ is the Hopf algebra of functions over SU(2). Then, the
the action can be expressed using the Haar measure, the coproduct and the antipode. 

\item{\bf Continuum theory through the  double scaling limit} Spin foam models provide a proposal for a quantum theory of gravity that relies on discretizations of space-time. It has to be  related to a continuum, long distance,  theory formulated in terms of fields living on a smooth manifold.
The double scaling limit is essential in establishing the link between matrix models and two dimensional gravity by enhancing the contribution of triangulation with many simplexes. Since group field theories are generalizations of matrix models, it is reasonable to attempt at a similar construction proceeding as follows. To begin with, we restrict the summation to spins $j\leq N$ and identify various classes of graphs  that correspond to a fixed topology and share the same power law behavior $N^{\alpha}$ as $N\rightarrow \infty$, possibly after a renormalization of the coupling and/or the field.  Next, let us assume that the summation over any such class of graph is convergent with $\lambda\in[0,\lambda_{c}]$, with $\lambda_{c}$ a critical coupling and that it exhibits a power law behavior of the type $(\lambda_{c}-\lambda)^{\beta}$ when $\lambda\rightarrow\lambda_{c}$. Note that such an algebraic singularity often results from the summation of tree-like graphs. Then,  we can define the double scaling limit by simultaneously taking $N\rightarrow\infty$ and $\lambda\rightarrow\lambda_{c}$ with $N^{\alpha}(\lambda-\lambda_{c})^{\beta}$ held fixed. 
Although this is far from being realized in the EPRL/FK model, two important steps in this direction have already been taken in simpler models based on colored tensor models. First, Gurau has identified the dominant terms in the large $N$ behavior of the Boulatov-Ooguri models in \cite{complete}. Second, in the case of tensor models without group arguments, Bonzom, Gurau, Riello and Rivasseau \cite{melons} have been able to sum the dominant class of graphs and show that it leads to a continuum limit.  Alternatively, one could also consider two dimensional gravity coupled Yang-Mills theory as proposed in \cite{equivariant}. This is an interesting playground because it possesses a non trivial spin foam formulation and is technically simple since it only involves ribbon graphs.

\item{\bf Schwinger-Dyson equations and constrains on boundary states} In order to understand various physically relevant situations, like early cosmology or black hole physics, it is necessary to relate group field theories to more conventional formulations of quantum gravity like loop quantum gravity. Since group field theory implements a prescription on the summation on spin foam amplitude contributing to a fixed boundary graph, it is interesting to translate such a prescription into an equation for the boundary state. This can be achieved through the Schwinger-Dyson equations \eqref{SchwingerDyson}. Indeed, choosing an external leg in a non necessarily connected $n$-point function, this leg can be either related to a vertex or to an other external leg. When implemented on the expectation value of a boundary graph, a given $(D\!-\!1)$-simplex can either split into $D$ $(D\!-\!1)$-simplexes or identified with an other $(D\!-\!1)$-simplex on the boundary, as noted by Freidel \cite{Freidel}. The resulting equations do not refer anymore to spin foams but simply enforce constraints on the expectation values of boundary states, which have been shown to provide a higher dimensional generalization of the Virasoro constraints of two dimensional quantum gravity in the context of colored tensor models by Gurau \cite{GurauSD} and by Bonzom to study the large $N$ limit of colored tensor models \cite{Bonzom}. Schwinger-Dyson equations have also been used by Ooguri to control the sum over topologies in the three dimensional case \cite{OoguriSDE}.  In the context of the EPRL/FK model, it would be interesting to study the possible relation of these constraints to loop quantum gravity, especially to the hamiltonian constraint. Alternatively, group field theory can also be used to defined a effective hamiltonian constraint through a semi-classical expansion, as proposed by Livine, Oriti and Ryan \cite{effectivehamiltonian}.

\end{itemize}

Obviously, all these topics are related: the divergences may be cured through a renormalization procedure which may shed a new light on the low energy limit of the theory while the inclusion of matter fields may alter the renormalization group equations.

\appendix

\section{Principal series of unitary irreducible repreentations of $SL(2,{\Bbb C})$}

\label{unitary}
In order to  give the construction of the Lorentzian  EPL/FK${\gamma}$ model, it is necessary to list a few facts concerning the unitary irreducible representations of $\mathrm{SL}(2,{\mathbb C})$ \cite{Ruhl}. We work with the group $\mathrm{SL}(2,{\mathbb C})$ of $2\times 2$ complex matrices of determinant 1 which is the universal cover of connected component of the identity of the Lorentz group SO(3,1), just as SU(2) is the universal cover of the rotation group SO(3). Unlike the latter, $\mathrm{SL}(2,{\mathbb C})$ is non compact so that its unitary representations are necessary infinite dimensional. When decomposing a function over $\mathrm{SL}(2,{\mathbb C})$ using the matrix elements of representations, only unitary irreducible representations of the principal series appear. These representations $V_{j,\rho}$ are labelled by a real number $\rho\in{\Bbb R}$ and a half-integer $j\in\frac{{\Bbb Z}}{2}$ and admit two equivalent descriptions.

In the spinorial formulation, let $V_{j,\rho}$ be the space of (not necessarily holomorphic) functions of $z=(z_{1},z_{2})\in{\Bbb C}^{2}$ obeying the homogeneity property
\begin{equation}
\psi(\lambda z)=\lambda^{-1+\mathrm{i}\rho+j}\overline{\lambda}^{-1+\mathrm{i}\rho-j}\psi(z)\label{homogeneity}
\end{equation}
It is a Hilbert space for the scalar product
\begin{equation}
\langle\chi|\psi\rangle=\int_{\Bbb{CP}^{1}}\!\!\!\Omega\,\,\overline{\chi}\psi
\end{equation}
The two form
\begin{equation}
\Omega=\frac{\mathrm{i}}{2}\big(z_{1}dz_{2}-z_{2}dz_{1}\big)\wedge\big(\overline{z}_{1}d\overline{z}_{2}-\overline{z}_{2}d\overline{z}_{1}\big)
\end{equation}
is such that $\overline{\chi}\psi\,\,\Omega$ is invariant under $z\rightarrow \lambda z$ and may integrated over the complex projective space $\Bbb{CP}^{1}$.
The action of $g\in\mathrm{SL}(2,{\mathbb C})$ on $\psi$ is
\begin{equation}
g\!\cdot\!\psi(z)=\psi({}^{t}gz)
\end{equation}
with ${}^{t}g$ the transpose of $g$.  

This presentation of $V_{j,\rho}$ allows us to define an antiunitary operator $J$ commuting the action of $g\in\mathrm{SL}(2,{\mathbb C})$ and generalizing to the principal unitary representations of $\mathrm{SL}(2,{\mathbb C})$ the map
\begin{equation}
\begin{pmatrix}z_{1}\cr z_{2}\end{pmatrix}\quad\rightarrow\quad\begin{pmatrix}-\overline{z}_{2}\cr \overline{z}_{1}\end{pmatrix}
=\begin{pmatrix}0&-1\cr1&0\end{pmatrix}\begin{pmatrix}z_{1}\cr z_{2}\end{pmatrix}
\end{equation}
commuting with the SU(2) action. On $V_{j,\rho}$, it is defined as
\begin{equation}
J\psi(z)={\textstyle \frac{\sqrt{j^{2}+\rho^{2}}}{\pi}}\int_{\Bbb{CP}^{1}}\!\!\!\Omega\,\,(w_{0}z_{1}-w_{1}z_{0})^{-1+j+\mathrm{i}\rho}\,(\overline{w_{0}z_{1}-w_{1}z_{0}})^{-1-j+\mathrm{i}\rho}\,
\overline{\psi}(w)\,\,\label{defJ}
\end{equation}
This map obeys $Jg=gJ$ and $J^{2}=(-1)^{2k}$. As its SU(2) cousin, it is antiunitary in the sense that $\langle J\chi|J\psi\rangle=\langle\psi|\chi\rangle$. It is in fact the composition of an intertwiner between the representations $V_{j,\rho}$ and $V_{-j,-\rho}$ and complex conjugation. Since the representations $V_{j,\rho}$ and $V_{-j,-\rho}$ are equivalent, we assume that $j$ is non negative. 

Alternatively, it is convenient in actual computations to define $V_{j,\rho}$ as the space of $L^{2}$ functions over SU(2)  obeying the covariance property
\begin{equation}
\psi(\mathrm{e}^{\mathrm{i}\phi\sigma_{3}} u)=\psi(u)\mathrm{e}^{2\mathrm{i}j\phi}\label{covariance}
\end{equation}
for all $u\in\mathrm{SU(2)}$ and $\phi\in{\Bbb R}$ with $\sigma_{3}$ the third Pauli matrix. 
To define the $\mathrm{SL}(2,{\mathbb C})$ action, recall that any $g\in\mathrm{SL}(2,{\mathbb C})$ can be uniquely factorized as 
\begin{equation}
g=kh\qquad\mathrm{with}\qquad h\in \mbox{SU(2)}\quad
k=\begin{pmatrix}\lambda^{-1}&\mu\\
0&\lambda
\end{pmatrix}
\qquad\lambda>0\quad\mu=u+\mathrm{i}v\in{\mathbb C}
\end{equation}
Under this decomposition, the Haar measure on $\mathrm{SL}(2,{\mathbb C})$ factorizes as $dg=C\,dk\,dh$ with $dg$ the Haar measure on SU(2), $dk=C'\lambda\, d\lambda\,du\,dv$ an $\mathrm{SL}(2,{\mathbb C})$-invariant measure on the $\mathrm{SL}(2,{\mathbb C})/\mathrm{SU(2)}$. $C$ and $C'$ are two normalization constants that can be found  in \cite{Ruhl}. For our purposes, it is convenient to identify the coset $\mathrm{SL}(2,{\mathbb C})/\mathrm{SU(2)}$ with the hyperboloid $H^{+}=\left\{X^{I}|X^{I}X_{I}=1\,,\, X^{0}>0\right\}$ using the relation $X=kk^{\dagger}$. Using this identification, the measure on the coset turns out to be proportional to the usual measure on the hyperboloid
\begin{equation}
dX=\frac{dX^{1}dX^{2}dX^{3}}{\sqrt{1+(X^{1})^{2}+(X^{2})^{2}+(X^{3})^{2}}}
\end{equation}

Taking $u\in\mathrm{SU(2)}$ and $g\in\mathrm{SL}(2,{\Bbb C})$, we factorize $ug\in\mathrm{SL}(2,{\Bbb C})$ as $ug=k_{g}(u)h_{g}(u)$ and define
\begin{equation}
g\!\cdot\!\psi(u)=\lambda_{g}(u)^{2\mathrm{i}\rho-2}\psi(h_{g}(u))
\end{equation}
The correspondence with the spinorial representation relies on the use of the homogeneity condition  \eqref{homogeneity} for positive real numbers  to assume  that $|z_{1}|^{2}+|z_{2}|^{2}=1$ and define
\begin{equation}
u=\begin{pmatrix}z_{1}&z_{2}\cr -\overline{z}_{2}&\overline{z}_{1}\end{pmatrix}
\end{equation}
Then, the covariance condition  \eqref{covariance} is equivalent to the homogeneity condition for complex numbers of modulus 1. 

Recall that the Peter-Weyl theorem theorem states that any $L^{2}$ function on SU(2) can be expanded over Wigner matrices $D_{m'm}^{k}(u)$ with $k\in\frac{{\Bbb N}}{2}$, $m,m'\in\left\{-k,-k+1,\dots,k\right\}$ and that the latter obey the Schur orthogonality relations  
\begin{equation}
\int_{\mathrm{SU(2)}}\!\!dg\,\, \overline{{\cal D}_{m,d}^{j}(g)}{\cal D}_{m',d}^{j'}(g)=\frac{\delta_{j,j'}\delta_{m,m'}\delta_{n,d}}{2j+1}\label{Schur}
\end{equation}
The covariance condition  \eqref{covariance} imposes $m'=j$ so that an orthonormal basis of $V_{j,\rho}$, called the canonical basis, is provided by the functions
\begin{equation} 
|k,m\rangle_{j,\rho}={\sqrt{2k+1}}\,D_{jm}^{k}(u)\qquad k-j\in{\Bbb N}\quad m\in\left\{-k,-k+1,\dots,k\right\}\label{canonical}
\end{equation}
Consequently, as a representation of $\mathrm{SU(2)}\subset\mathrm{SL}(2,{\Bbb C})$, it decomposes as 
\begin{equation}
V_{j,\rho}=\mathop{\oplus}\limits_{k-j\in {\mathbb N}}V_{k}\label{SU(2)decomposition}
\end{equation}
Let us denote $Y_{j,\rho}:\,V_{j}\rightarrow V_{j,\rho}$ the embedding as SU(2) representation, so that
\begin{equation}
Y_{j,\rho}|j,m\rangle=|j,m\rangle_{\rho,j}
\qquad
\mathrm{and} 
\qquad
Y^{\dagger}_{j,\rho}|k,m\rangle_{j,\rho}=\delta_{k,j}|j,m\rangle
\end{equation}
with $|j,m\rangle$ the standard basis of $V_{j}$.

\section{SU(2) coherent states and their use in $\mathrm{SL}(2,\mathbb{C})$ representation theory}

\label{coherent}

It is also useful to recall a few facts about SU(2) coherent states, see \cite{Perelomov} for a complete survey. Given a unit vector $n$, let $g_{n}\in\mathrm{SU(2)}$ be any rotation that takes the unit vector $n_{0}$ along the $z$ axis to $n$. The spin $j$ coherent state is defined by acting with $g_{n}$ on the highest spin vector $|j,j\rangle$
\begin{equation}
|j,n\rangle=g|j,j\rangle=\sum_{-j\leq m\leq j}D^{j}_{mj}(g)|j,m\rangle
\end{equation}
The rotation $g_{n}$ is defined up to a rotation around the $z$ axis so that the coherent state $|j,n\rangle$ is defined up to a phase. This phase is irrelevant as soon as the coherent state appear as a ket $|j,n\rangle$ and a bra $\langle j,n|$. To be definite, let us fix the rotation in such a way that its axis is $n_{0}\times n$ with a angle in $[0,\pi]$. Coherent states satisfy the following properties, which are very useful in our context.

The scalar product of two coherent states is
\begin{equation}
\langle j,n_{1}|j,n_{2}\rangle=\exp\{\mathrm{i}j\varphi(n_{0},n_{1},n_{1})\}\,\Big(\frac{1+n_{1}\cdot n_{2}}{2}\Big)^{j}
\end{equation}
where $\varphi(n_{0},n_{1},n_{1}$ is the algebraic area of the geodesic triangle on the unit sphere with vertices at $n_{0}$, $n_{1}$ and $n_{2}$. 
Although not orthogonal, they form an over complete basis of $V_{j}$ in the sense that
\begin{equation}
\frac{2j+1}{4\pi}\int_{S^{2}}dn\,|j,n\rangle\langle j,n|=1_{V_{j}}
\end{equation}
The action of $g\in\mathrm{SU(2)}$ on a coherent state is again a coherent state,
\begin{equation}
g|j,n\rangle=\exp\{\mathrm{i}j\varphi(n_{0},n,n_{g})\}\,|j,n_{g}\rangle
\end{equation}
with $n_{g}$ the transformed of $n_{0}$ under the rotation defined by $g$.

All these properties allow us to reduce any computation of matrix elements of $g$ (by convention $g$ acts on the ket on its right) between two coherent states to the spin $\frac{1}{2}$ representation
\begin{equation}
\langle j,n_{1}|g|j,n_{2}\rangle=\big(\langle n_{1}|g|n_{2}\rangle\big)^{2j}
\end{equation}
where we have dropped the label $j=\frac{1}{2}$ to alleviate the notation.

All these relations prove to be useful in order to compute the following trace, with $g_{a},g_{b}\in\mathrm{SL}(2,{\Bbb C})$, in terms of coherent states 
\begin{align}
\mathrm{Tr}_{V_{j}}\big(Y_{\rho,j}^{\dagger}g_{a}^{-1}g^{}_{b}Y_{\rho,j}\big)&=\sum_{-j\leq m\leq j}{}_{j,\rho}\langle j,m|g_{a}^{-1}g^{}_{b}|j,m\rangle_{j,\rho}\\
&=\sum_{-j\leq m\leq j}(2j+1)\int_{\mathrm{SU(2)}}du\,\lambda_{g_{a}}(u)^{-2\mathrm{i}\rho-2}\lambda_{g_{b}}(u)^{2\mathrm{i}\rho-2}\overline{D}^{j}_{jm}\big(h_{g_{a}}(u)\big){D}^{j}_{jm}\big(h_{g_{b}}(u)\big)\\
&=\sum_{-j\leq m\leq j}(2j+1)\int_{\mathrm{SU(2)}}du\,\lambda_{g_{a}}(u)^{-2\mathrm{i}\rho-2}\lambda_{g_{b}}(u)^{2\mathrm{i}\rho-2}\langle j,m|h^{-1}_{g_{a}}(u)|j,j\rangle\langle j,j|h_{g_{b}}(u)|j,m\rangle\\
&=(2j+1)\int_{\mathrm{SU(2)}}du\,\lambda_{g_{a}}(u)^{-2\mathrm{i}\rho-2}\lambda_{g_{b}}(u)^{2\mathrm{i}\rho-2}\langle j,j|h_{g_{b}}(u)h^{-1}_{g_{a}}(u)|j,j\rangle\\
\end{align}
To express the face amplitude in terms of coherent states, we have to compute the matrix element
\begin{align}
\langle j,j|h_{g_{a}}(u)h^{-1}_{g_{b}}(u)|j,j\rangle=
\langle \textstyle{\frac{1}{2},\frac{1}{2}}|h_{g_{a}}(u)h^{-1}_{g_{b}}(u)|\textstyle{\frac{1}{2},\frac{1}{2}}\rangle^{2j}
\end{align}
with all group elements considered as $2\times 2$ matrices evaluated in the fundamental representation and $|\frac{1}{2},\frac{1}{2}\rangle=\begin{pmatrix}1\cr0\end{pmatrix}$.

Since $h_{g}(u)=k_{g}^{-1}(u)ug$ is defined by the decomposition $ug=k_{g}(u)h_{g}(u)$, we have
\begin{equation}
\big(h_{g}(u)\big)^{-1}=\big(h_{g}(u)\big)^{\dagger}=(g^{-1})u^{\dagger}k_{g}(u)
\end{equation}
and
\begin{equation}
h_{g}(u)=k_{g}^{\dagger}(u)u(g^{-1})^{\dagger}
\end{equation}
Recall that
\begin{equation}
k_{g}(u)=\begin{pmatrix}\lambda_{g}^{-1}(u)&\mu_{g}(u)\cr 0&\lambda_{g}(u)\end{pmatrix},
\end{equation}
so that $|\frac{1}{2},\frac{1}{2}\rangle$ is an eigenvector of $k_{g}(u)$ with eigenvalue $\lambda^{-1}_{g}(u)$. Thus the matrix element reads
\begin{align}
\langle \textstyle{\frac{1}{2},\frac{1}{2}}|h_{g_{a}}(u)h^{-1}_{g_{b}}(u)|\textstyle{\frac{1}{2},\frac{1}{2}}\rangle=
\lambda_{g_{a}}^{-1}(u)\lambda_{g_{b}}^{-1}(u)\langle \textstyle{\frac{1}{2},\frac{1}{2}}|u (g_{a}^{-1})^{\dagger} g_{b}^{-1}u^{\dagger}|\textstyle{\frac{1}{2},\frac{1}{2}}\rangle
\end{align}
Taking all terms into account, we have
\begin{align}
\mathrm{Tr}_{V_{j}}\big(Y_{\rho,j}^{\dagger}g_{a}^{-1}g^{}_{b}Y_{\rho,j}\big)&=(2j+1)\int_{\mathrm{SU(2)}}du\,\lambda_{g_{a}}(u)^{2(-\mathrm{i}\rho-1-j)}\lambda_{g_{b}}(u)^{2(\mathrm{i}\rho-1-j)}\big[\langle \textstyle{\frac{1}{2},\frac{1}{2}}|u (g_{a}^{-1})^{\dagger} g_{b}^{-1}u^{\dagger}|\textstyle{\frac{1}{2},\frac{1}{2}}\rangle\big]^{2j}
\end{align}
To compute explicitly $\lambda_{g}(u)>0$, notice $\lambda_{g}^{2}(u)$ is nothing but the upper left corner of the matrix  $(k_{g}(u)^{-1})^{\dagger}k_{g}^{-1}(u)=u(g^{-1})^{\dagger}g^{-1}u^{\dagger}$,  
\begin{equation}
\lambda_{g}(u)=\Big[\langle\textstyle{\frac{1}{2},\frac{1}{2}}|u(g^{-1})^{\dagger}g^{-1}u^{\dagger}|\textstyle{\frac{1}{2},\frac{1}{2}}\rangle \Big]^{\frac{1}{2}}
\end{equation}
Therefore,
\begin{multline}
\mathrm{Tr}_{V_{j}}\big(Y_{\rho,j}^{\dagger}g_{a}^{-1}g^{}_{b}Y_{\rho,j}\big)=
(2j+1)\int_{\mathrm{SU(2)}}du\cr
\frac{\big[\langle \textstyle{\frac{1}{2},\frac{1}{2}}|u (g_{a}^{-1})^{\dagger} g_{b}^{-1}u^{\dagger}|\textstyle{\frac{1}{2},\frac{1}{2}}\rangle\big]^{2j}}
{\Big[\langle\textstyle{\frac{1}{2},\frac{1}{2}}|u(g_{a}^{-1})^{\dagger}g_{a}^{-1}u^{\dagger}|\textstyle{\frac{1}{2},\frac{1}{2}}\rangle \Big]^{(+\mathrm{i}\rho+1+j)}\Big[\langle\textstyle{\frac{1}{2},\frac{1}{2}}|u(g_{b}^{-1})^{\dagger}g_{b}^{-1}u^{\dagger}|\textstyle{\frac{1}{2},\frac{1}{2}}\rangle \Big]^{(-\mathrm{i}\rho+1+j)}}
\end{multline}
Finally, we identify $u^{\dagger}|\textstyle{\frac{1}{2},\frac{1}{2}}\rangle=|n\rangle$ with a coherent state in the spin $\frac{1}{2}$ repsentation. Therefore, up to a factor of $\pi$, the integral can be written as an integral over $S^{2}$
\begin{align}
\mathrm{Tr}_{V_{j}}\big(Y_{\rho,j}^{\dagger}g_{a}^{-1}g^{}_{b}Y_{\rho,j}\big)&=
(2j+1)\int_{\mathrm{S^{2}}}dn\,
\frac{\big[\langle n| (g_{a}^{-1})^{\dagger} g_{b}^{-1}|n\rangle\big]^{2j}}
{\Big[\langle n|(g_{a}^{-1})^{\dagger}g_{a}^{-1}|n\rangle \Big]^{(\mathrm{i}\rho+1+j)}\Big[\langle n|(g_{b}^{-1})^{\dagger}g_{b}^{-1}|n\rangle \Big]^{(-\mathrm{i}\rho+1+j)}}
\end{align}


\bigskip
\bigskip
{\bf Acknowledgement:} The author is indebted to Joseph Ben Geloun, Razvan Gurau, Vincent Rivasseau, Carlo Rovelli and Simone Speziale for sharing their insights into the various topics presented here. He also thanks the organizers of the "Third Quantum Gravity and Quantum Geometry School" that took place in Zakopane in february 2011. Thanks also to Franck Hellmann, Editor of the Proceedings, who patiently waited for this contribution.

\end{document}